\documentclass[
,aps%
,pra%
,10pt%
,showpacs%
,showkeys%
,superscriptaddress%
,preprint%
,preprintnumbers%
,nofootinbib%
,amsmath%
,amssymb%
,floatfix%
]{revtex4-2}

\usepackage[utf8]{inputenc}
\usepackage{graphicx}
\usepackage{tikz}

\usepackage{amsfonts}
\usepackage{amsmath}
\usepackage{amssymb}
\usepackage{bm, bbm}
\usepackage{mathrsfs}
\usepackage{mathtools}

\usepackage[caption=false]{subfig}
\usepackage{float}
\usepackage[utf8]{inputenc}

\usepackage{hyperref} 

\begin{document}

\title{Multi-scale reconstruction of large supply networks}

\author{Leonardo Niccol\`{o} Ialongo}
\email[To whom correspondence should be addressed: ]{ialongo@csh.ac.at}
\affiliation{Complexity Science Hub, Vienna, 1030, Austria}
\author{Sylvain Bangma}
\affiliation{University of Leiden, Lorentz Institute for Theoretical Physics (LION), Leiden, 2333 CA, The Netherlands}
\affiliation{ING Bank N.V., Amsterdam, 1102 CT, The Netherlands}
\author{Fabian Jansen}
\affiliation{ING Bank N.V., Amsterdam, 1102 CT, The Netherlands}
\author{Diego Garlaschelli}
\affiliation{University of Leiden, Lorentz Institute for Theoretical Physics (LION), Leiden, 2333 CA, The Netherlands}
\affiliation{IMT School for Advanced Studies, Lucca, 55100, Italy}

\begin{abstract}
The structure of the supply chain network has important implications for modelling economic systems, from growth trajectories to responses to shocks or natural disasters. However, reconstructing firm-to-firm networks from available information poses several practical and theoretical challenges: the lack of publicly available data, the complexity of meso-scale structures, and the high level of heterogeneity of firms. With this work we contribute to the literature on economic network reconstruction by proposing a novel methodology based on a recently developed multi-scale model. This approach has three main advantages over other methods: its parameters are defined to maintain statistical consistency at different scales of node aggregation, it can be applied in a multi-scale setting, and it is computationally more tractable for very large graphs. The consistency at different scales of aggregation, inherent to the model definition, is preserved for any hierarchy of coarse-grainings The arbitrariness of the aggregation allows us to work across different scales, making it possible to estimate model parameters even when node information is inconsistent, such as when some nodes are firms while others are countries or regions. Finally, the model can be fitted at an aggregate scale with lower computational requirements, since the parameters are invariant to the grouping of nodes. We assess the advantages and limitations of this approach by testing it on two complementary datasets of Dutch firms constructed from inter-client transactions on the bank accounts of two major Dutch banking institutions. We show that the model reliably predicts important topological properties of the observed network in several scenarios of practical interest and is therefore a suitable candidate for reconstructing firm-to-firm networks at scale.
\end{abstract}

\keywords{Complex Networks, Economic Systems, Financial Systems}
\pacs{89.75.Fb; 02.50.Tt; 89.65.Gh}

\maketitle
\newpage

\section{Introduction}
Supply-chains have received a lot of attention in recent times as their importance in the global economy has become more evident. This has revived interest from academia in understanding the systemic importance of these networks on the global economy \cite{schweitzer2009economic,bernanke2018real}. Several studies have shown how the network structure can amplify shocks from natural disasters \cite{carvalho2021jap,henriet2012firm}, from idiosyncratic fluctuations \cite{acemoglu2012net,contreras2014propagation}, from the labour restrictions during the Covid pandemic \cite{pichler2022fore}, or due to the reduction in liquidity during a financial crisis \cite{huremovic2023production}. The structure of these networks has also been shown to play a crucial role in growth dynamics \cite{mcnerney2022production,klimek2019quantifying} and has been used in order to study the systemic risk induced by the supply-chains' structure \cite{diem2021quantifying,colon2021criticality}.

While many studies have been focused on the supply-chains observed at the industrial level, we now know that in many cases this can severely underestimate the amplification effects due to the firm-level interactions \cite{diem2023macro,moran2019may}. Several studies have now been performed on firm-level data in many countries and from different data source \cite{lafond2023firm}: data from the Turkish production network has been used to document preferential attachment due to skill matching \cite{demir2024ring}; Japanese large commercial datasets on self-reported suppliers and customer, from Tokyo Shoko Research Ltd. and from Teikoku Databank Ltd., have been used in numerous studies \cite{fujiwara2010large,mizuno2014structure,inoue2019firm,carvalho2021jap}; Hungarian VAT tax reporting has been used to develop a model of systemic risk \cite{diem2021quantifying,diem2023macro}; Japanese payment data has been used to build a national supply-chain network \cite{fujiwara2021money}; in the US, the SEC filings have extensively been studied \cite{atalay2011network}; finally the national network of Belgium firms, built from VAT data, has been investigated in many publications \cite{dhyne2021trade,bernard2019production}. This list is by no means exhaustive as many more articles have come out in recent years. For a complete overview of firm-level network data we refer to \cite{lafond2023firm}.

Unfortunately analysis of firm-level networks is still very limited in scope. Most complete datasets are available at a national level only and cannot be shared or integrated easily. The few global data sources that are available, report data on a very small fraction of the firms that exist worldwide \cite{lafond2023firm}. The limited availability of this data and the need to work across regions has been pushing for better reconstruction methodologies to be developed from the available public information. Several approaches have been developed to this end which have been surveyed in \cite{squartini2018reconstruction} and specifically for supply-chains in \cite{mungo2023reconstructing}. Among these, we have methods based on inferring connections from the correlation of observed time-series data \cite{campajola2021equivalence,mungo2023revealing}, from the data on calls between employees \cite{reisch2021inferring}, and many others based on firm-level information \cite{mungo2022reconstructing,ialongo2022recon}. From a methodological point of view we can see a clear distinction between methods that focus on link prediction \cite{brintrup2018predicting,kosasih2021machine,mungo2022reconstructing}, where the objective is to identify the true network with highest probability, and maximum-entropy ensemble methods \cite{ialongo2022recon,bacilieri2023recon} where instead the aim is to find the constrained set of graphs that share similar characteristics with the true one. 

Our contribution belongs to this second family of models as it builds a probabilistic ensemble of graphs meant to preserve important characteristics of the empirical one. This is because we are most interested in the ability to correctly determine the ensemble of possible networks that are functionally equivalent to the real network, rather than having a high confidence of finding the ``true'' graph. This stems from the knowledge that while we may observe a given network at a given time this is by no means a stable configuration \cite{moran2019may} and what the true network is might change rapidly over time. As we expect that many rewiring events would not lead to a change in the properties of the network or in a different macro-behaviour of the system, we hope to build ensembles that contain all graphs that satisfy this functional equivalence relation with the true network. 

Differently from what is done in \cite{ialongo2022recon}, the model we propose here is not based on maximum entropy but rather on a recently proposed method that originates from the principle of invariance to arbitrary node partitions \cite{garuccio2020multiscale, lalli2024geometry}. We do so as we want to highlight a clear yet so far understudied problem in reconstruction methods for production networks, that is the multi-scale nature of the system under study. Indeed, while it may seem clear what we mean by a firm-level network, definition of firms can vary from the legal perspective to the plant-level point of view. Large corporations may span multiple countries and production lines making it difficult to develop a methodology that applies well to multinationals and your local bakery alike. Furthermore, while data on firms is increasingly available, for many regions of the world this level of detail is not yet achievable. As such, we would like a model that can correctly handle data at the firm, industry, region, or country level. 

In this work we present a principled approach to modelling large supply chain networks in a multi-scale environment. Our methodology, based on the work of Garuccio et al. \cite{garuccio2020multiscale}, provides a simple yet sound  theoretical background for us to satisfy the challenges inherent to supply chains. In the next section, we outline the theory at the heart of the approach and adapt it to the needs of our application. We then test the model on a real firm-to-firm network constructed from Dutch payment data in order to assess its performance in a couple of scenarios intended to highlight the properties of this model. Our goal is to highlight the importance of modelling production networks in a multi-scale environment and, in doing so, establish a benchmark for its performance.

\subsection*{Modelling production networks}
Production networks are striking in the complexity of features that they present at various scales \cite{lafond2023firm}. From the non-trivial meso-structure to the high level of node heterogeneity, it is difficult to identify sufficient statistics to correctly construct maximum-entropy network ensembles. One of the salient features of these networks is the functional structure due to the technological constraints to production. This has been shown to favour certain networks motifs both at the firm level \cite{mattsson2021functional} and at the industry level \cite{di2024commodity}. In particular at the industry level it seems that this structure effectively determines a ``fingerprint'' of the sector \cite{di2024commodity}. In the economic literature the technology that a firm employs is usually captured in a production function whose functional form is assumed to be known. The production function will determine both which connections are allowed and the weight of these edges. Production functions also tell us how the weights of the links are expected to change due to changes in prices and demand. Usually however they do not give information on the probability of link formation, as they are more focused in determining the intensive margin for change.

A simple approach to use the logic of production functions for network reconstruction has been introduced in \cite{ialongo2022recon}. In this model the probability of a link forming between two nodes is proportional to the relative size of the two firms in the production and consumption of a specific good. A similar principle has also been used in \cite{bernard2022sparse} to develop sparse production networks starting from a random allocation model. Indeed it can be shown that the model we propose is a more general approach that contains the allocation model of Bernard and Zi \cite{bernard2022sparse} as a special case. The main advantages of our model with respect to \cite{bernard2022sparse} is that in our approach the density of connections is an exogenous parameter that we are free to change, that we have a consistent model at all scales, and that we take the functional structure of production into account. Further details on the relation between the models can be found in the supplementary materials. 

A way to rationalize this approach with a geometric understanding is to represent a firm using two vectors, the in- and out- embedding of the firm, respectively $\bm{x}_i$ and $\bm{y}_i$. The probability of two firms being connected will depend on the similarity between their out- and in- embedding in this space, such that we may write 
\begin{equation}
    P(a_{ij} = 1) \coloneqq p_{ij} = f(\langle \bm{y}_i, \bm{x}_j\rangle) \ .
\end{equation}
We note that each firm is represented by two vectors, as this is necessary to introduce direction in a way that allows the probability of buying from a firm to be substantially different from selling to it. Furthermore, the similarity in out-embedding should represent similar production capacity, while a small distance in the in-embedding space signals that similar products are used in production. While this embedding could be learned \cite{milocco2024multi}, here we want to test the simplest possible implementation of this approach that satisfies our multi-scale environment. As such, we choose as in \cite{ialongo2022recon} to represent each firm by its production and consumption quantities by product. Therefore, our embedding space is $S \subseteq \mathbb{R}^D$ where $D$ is the number of traded products. This is consistent with the allocation model proposed by Bernard and Zi with explicit differentiation of the product markets \cite{bernard2022sparse}.

\begin{figure}[thp]
    \centering
    \includegraphics[width=0.75\linewidth]{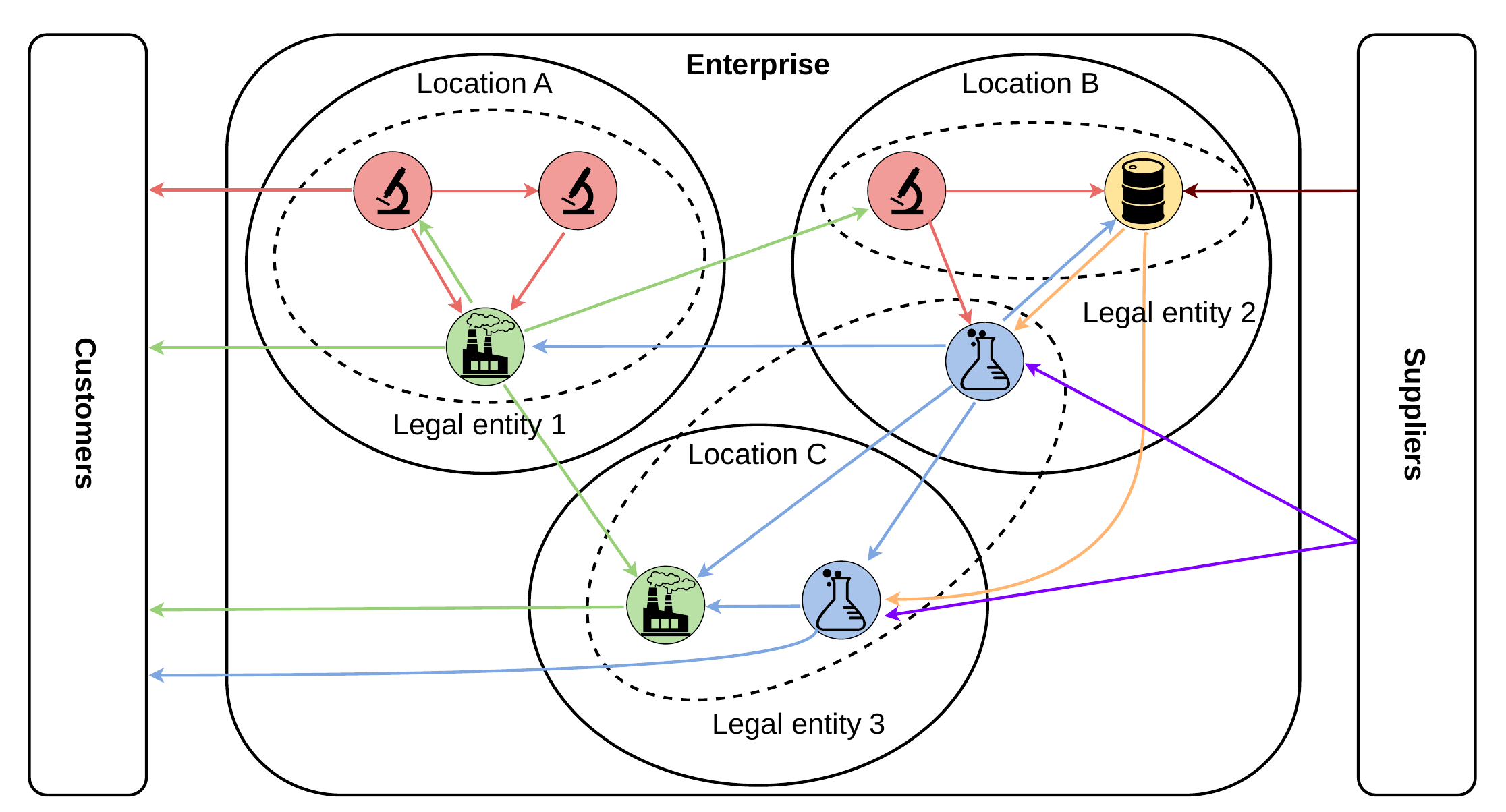}
    \caption{Schematic representation of an example business enterprise with its input and output relations coloured by product type. The diagram shows the three different production locations in the circles and the legal entities in the dashed ovals. Official statistics will divide the enterprise according to the Kind of Activity Units (KAU), which in our diagram are the set of nodes of the same colour which share the same NACE 4-digit code. Data reporting on firms can follow geographical location, KAUs, or legal entity, making the description potentially inconsistent. }
    \label{fig:kau_diag}
\end{figure}

The definition we have for the technological embedding of the firm poses an interesting challenge from an accounting point of view. Firms can be comprised of multiple production units in potentially different locations, sometimes involving more than one legal entity, and trade a variety of goods. The boundaries of a firm are never exactly defined and it is difficult to find the atomic\footnote{In the sense that it is not divisible.} component of an enterprise. Statistical offices define these elementary constituents as kind-of-activity units (KAUs), which are accounting units that group together ``all the offices, production facilities, etc. of an enterprise, which contribute to the performance of a specific economic activity defined at class level (four digits) of the European classification of economic activities'' \cite{kau1993}. Each KAU can be further divided into single establishments (local KAU). In figure \ref{fig:kau_diag} we have drawn an example enterprise divided in its establishments, the coloured circle, and represented with arrows the products they exchange. We have highlighted with different colours the NACE classification of each unit and therefore of the services exchanged. As can be seen from the diagram each establishment has its input-output relations, taking multiple products as inputs to generate a single type of output. It should be clear therefore that our probability could be defined at the establishment level, for each location, for each legal entity, or for the enterprise as a whole. Given that the data we might have might not always be for the same level of description, the challenge is to define an approach that maintains consistency at all possible scales and that can work in a multi-scale environment. 

The technological embedding we proposed, a vector description of production and consumption by product, has the strong advantage of being easy to define at any scale by observing the input-output relations or by summing the appropriate embedding of the constituent elements at lower scales. Our embedding is therefore additive under coarse graining. We further note that by definition KAUs will have the output embedding in the product space being zero for all dimensions except the one associated to their production activity. The input embedding however will be different from zero for those products which are necessary for production. Because of the sparsity of these vectors many will be mutually orthogonal ensuring that connections can only exist between compatible units. As we aggregate more and more of these KAUs we will get a less sparse representation and the problem will become less constrained. Our objective is to formulate a model capable of handling this representation effectively at all possible observation scales. 

The scale-invariant model proposed by \cite{garuccio2020multiscale} can be adapted to fit the needs our application. The model has a functional form that ensures that if the node embedding is additive under coarse-graining, then the functional form of the probability is the same for any scale. The general form of this probability functional is given by
\begin{equation}\label{eq:p_inv_gen}
    p_{i_{l}j_{l}} = 1 - e^{-{\bm{\theta}_{i_{l}}}^T \bm{B} \bm{\theta}_{j_{l}}} \quad .
\end{equation}
A summary derivation of this form can be found in the supplementary materials. A simple way to impose local production constraints in the equation above is by a specific selection of the parameters $\bm{\Theta}^{(l)}$ and matrix $\bm{B}$. Our production embedding can be directly used as parameters for the model. In particular, if we denote the total expenses of firm $i$ for product $\alpha$ as $s^{\text{in}}_{i, \alpha}$, and the corresponding income for that product as $s^{\text{out}}_{i, \alpha}$, we can now construct our parameter vector as $\bm{\theta}_{i_{l}} \coloneqq \begin{bmatrix} \bm{s}^{\text{out}}_i \\ \bm{s}^{\text{in}}_i \end{bmatrix}$. Here we have denoted as $\bm{s}^{\text{out}}_i$ and  $\bm{s}^{\text{in}}_i$ the column vectors with elements $\alpha$ given by $s^{\text{out}}_{i, \alpha}$ and $s^{\text{in}}_{i, \alpha}$ respectively. We can now set the matrix $\bm{B} \coloneqq \begin{bmatrix} \bm{0} & \text{diag}(\bm{\delta}) \\ \bm{0} & \bm{0} \end{bmatrix}$ in order to obtain the following simplified functional form:
\begin{equation} \label{eq:p_inv_stripe}
    p_{i_{l}j_{l}} (\bm{\delta}) = 1 - e^{-\sum_{\alpha} \delta_\alpha s^{\text{out}}_{i_l, \alpha} s^{\text{in}}_{j_l, \alpha}}
\end{equation}
where $D$ is the number of products in the economy and $\text{diag}(\bm{\delta})$ is a diagonal matrix of size $D$ with free parameters as diagonal elements used to fit the density of each product layer. Note that this is a methodological choice that we have done for the purpose of this analysis, in general one can make different choices if it better suits the problem at hand.

We note that the problem can be divided in independent layers defined by the product $\alpha$ where the link exists with probability
\begin{equation} \label{eq:p_layer}
    p_{i_{l}j_{l}}^{\alpha} (\delta_\alpha) = 1 - e^{- \delta_\alpha s^{\text{out}}_{i_l, \alpha} s^{\text{in}}_{j_l, \alpha}}
\end{equation}
and where we can recover the probability of the original graph using an aggregation similar to \eqref{eq:agg_edge} given by
\begin{equation}
    1 - p_{i_{l}j_{l}} (\bm{\delta}) = e^{-\sum_{\alpha} \delta_\alpha s^{\text{out}}_{i_l, \alpha} s^{\text{in}}_{j_l, \alpha}} = \prod_\alpha (1 - p_{i_{l}j_{l}}^{\alpha} (\delta_\alpha) )
\end{equation}

This equation can be further generalised as is done in \cite{garuccio2020multiscale} to incorporate dyadic factors such as geographic distances. In this case equation \eqref{eq:p_inv_stripe} becomes
\begin{equation} \label{eq:p_stripe_dyad}
    p_{i_{l}j_{l}} (\bm{\delta}) = 1 - e^{-\sum_{\alpha} \delta_\alpha s^{\text{out}}_{i_l, \alpha} s^{\text{in}}_{j_l, \alpha} f(d_{i_{l}j_{l}}^\alpha)} \quad .
\end{equation}
For the invariance to hold we must require that the dyadic component aggregates following
\begin{equation} \label{eq:dyad_inv}
    f(d_{i_{l}+1j_{l+1}}^\alpha) = \frac{ \sum_{i_l \in i_{l+1}}  \sum_{j_l \in j_{l+1}} s^{\text{out}}_{i_l, \alpha} s^{\text{in}}_{j_l, \alpha} f(d_{i_{l}j_{l}}^\alpha)}{\sum_{i_l \in i_{l+1}} s^{\text{out}}_{i_l, \alpha} \sum_{j_l \in j_{l+1}} s^{\text{in}}_{j_l, \alpha}} \quad .
\end{equation}
In this work however we will not be using dyadic components, so unless otherwise specified the multi-scale model will refer to equation \eqref{eq:p_inv_stripe} with the added simplification that we consider a single parameter $\delta$ equal for all product layers.

It is important to note here a substantial theoretical difference between the multi-scale model and the maximum entropy ensembles. In the multi-scale model, as we are no longer building the ensemble from the constraints of the desired properties of the network, we cannot define the sufficient statistics of the max-entropy approach. This implies that the identity between maximising the likelihood and matching the constrained ensemble measures is no longer true. In practical terms for the multi-scale method we must choose between maximising the likelihood and matching the empirical density as done in \cite{ialongo2022recon}. Empirically we have found that the imbalanced nature of the dataset, due to the sparsity of the network, means that maximising the likelihood to fit the $\delta$ parameter yields ensembles that are much more sparse than the empirical one. For this reason, in this work we will calibrate the parameter $\delta$ to ensure that the expected link density of the ensemble is equal to the observed one. Note that we are not claiming that maximum likelihood should never be performed with this model. On the contrary, we expect this result to be different if the fitnesses of each node are left as parameters to be estimated.

\section*{Results}
Our aim is to show how this model can be used in practice in a series of cases that are typical given the partial nature of data that is available. We hope here to provide a useful benchmark that is easily applicable in most scenarios, is theoretically self-consistent at different scales, and can be further improved to use the information available. Our main objective is to establish a benchmark performance in a series of scenarios of practical interest. In particular we will address three scenarios: first we assess the performance of the model at the firm-level to establish its reconstruction accuracy with respect to stripe-corrected Gravity Model (scGM) presented in \cite{ialongo2022recon}; in the second scenario we will discuss how the model allows us to correctly incorporate knowledge about the unobserved rest-of-the-world node (ROW) to improve the modelling at any scale; finally we will look at how the model performs at different aggregation levels by calibrating the model on an intermediate level and assessing the predictions at both coarser and finer grains.

\subsection*{Comparing performance at a single scale}
The theoretical advantages of the model in terms of its scale-invariance are only useful insofar as the model can perform adequately at any scale. In order to appropriately test this, we compare the reconstruction performance of the model on a series of structural properties of the empirical network at firm-level. As a benchmark for the quality of the model we will compare it to the density-corrected Gravity Model (dcGM) and stripe-corrected Gravity Model (scGM) analysed in \cite{ialongo2022recon}. For a correct comparison we discuss two formulations of the invariant model such that they differ from the dcGM and scGM only in functional form and not in the fitnesses used: the density-corrected Invariant model (dcIN) is obtained by using the total in- and out- strength of each node as fitnesses, while the stripe-corrected Invariant model (scIN) is the result of using the strength by sector.  

\begin{figure*}[p]
    \centering
    \subfloat[\label{fig:icdf_out_degree}]{%
    \includegraphics[width=0.49\textwidth]{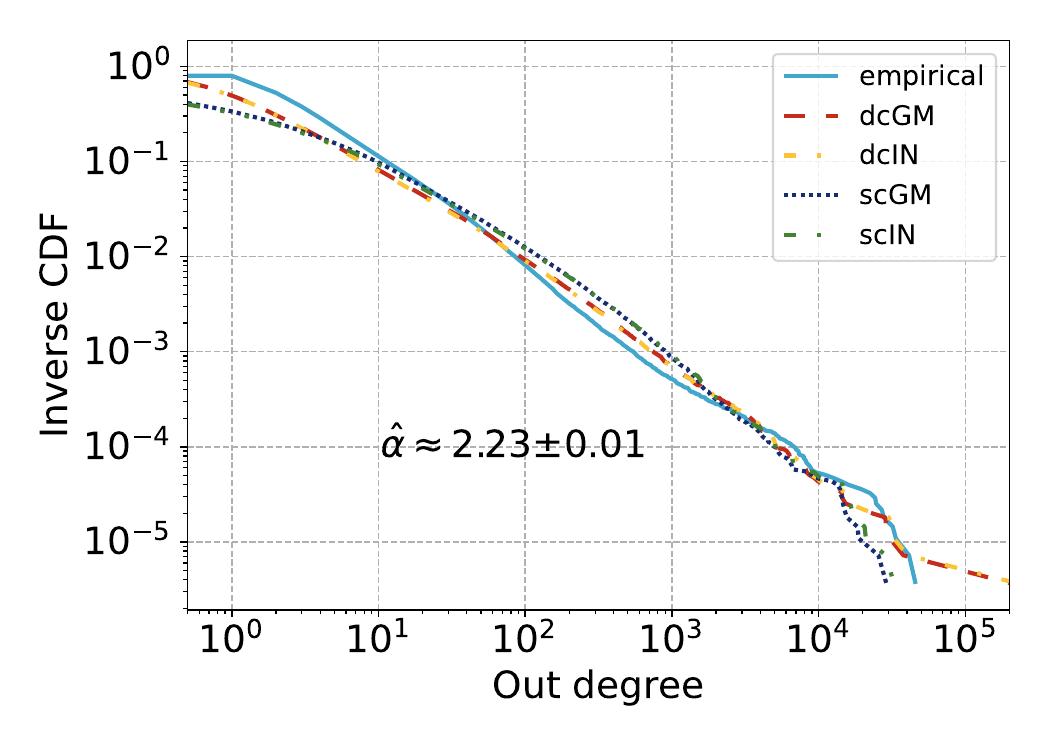}
    }\hfil 
    \subfloat[\label{fig:icdf_in_degree}]{%
    \includegraphics[width=0.49\textwidth]{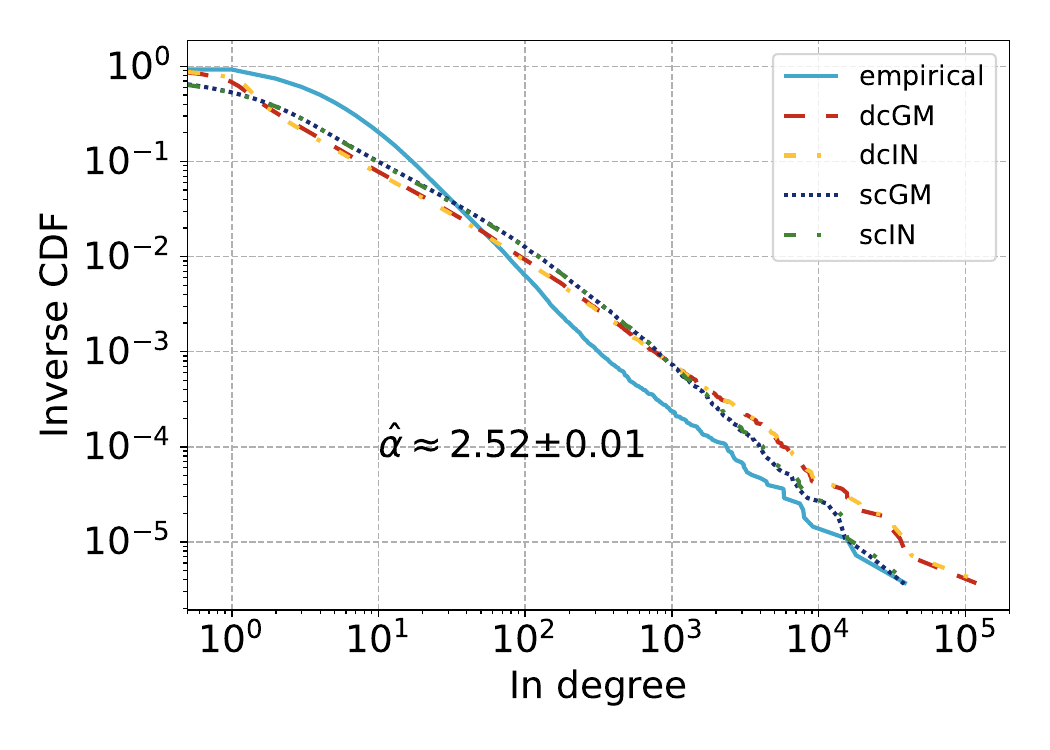}%
    }\hfil \vspace{-1em}
    \subfloat[\label{fig:knn_1scale_oo}]{%
    \includegraphics[width=0.49\textwidth]{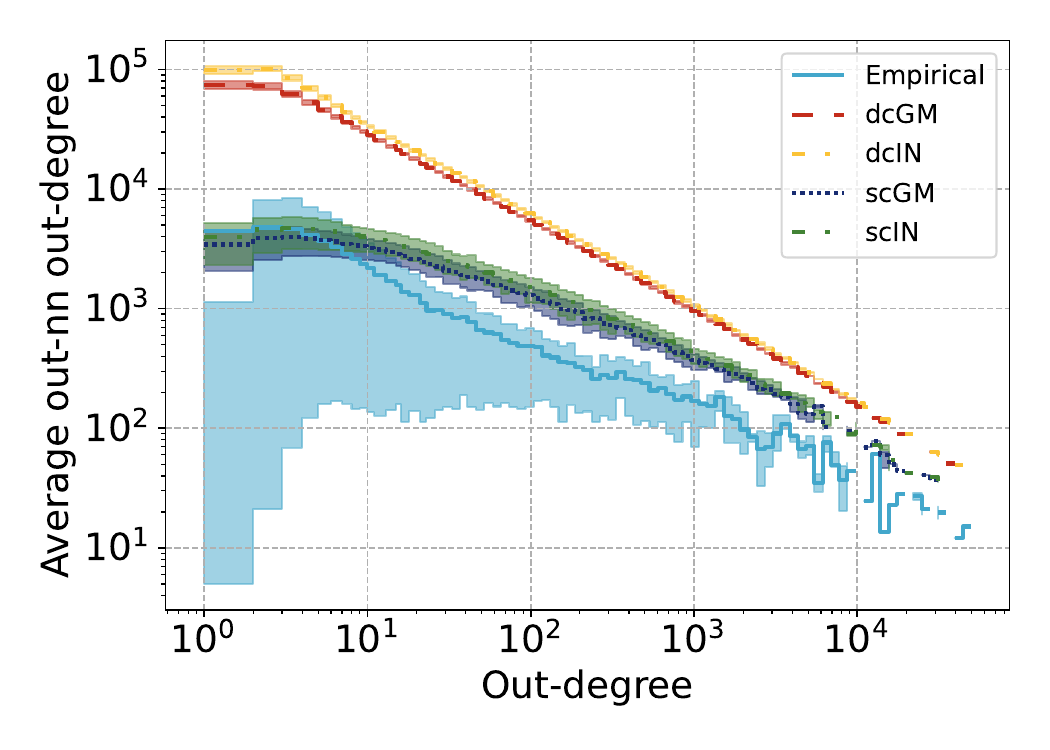}
    }\hfil
    \subfloat[\label{fig:knn_1scale_ii}]{%
    \includegraphics[width=0.49\textwidth]{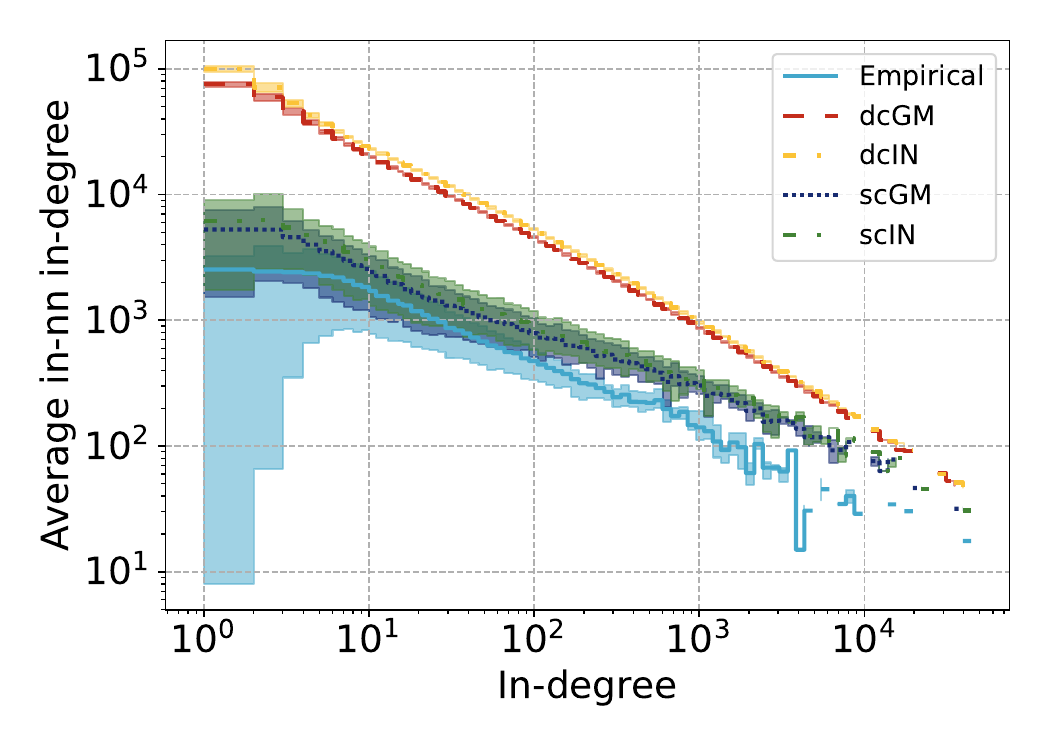}%
    }\hfil \vspace{-1em}
    \subfloat[\label{fig:roc_1scale}]{%
    \includegraphics[width=0.49\textwidth]{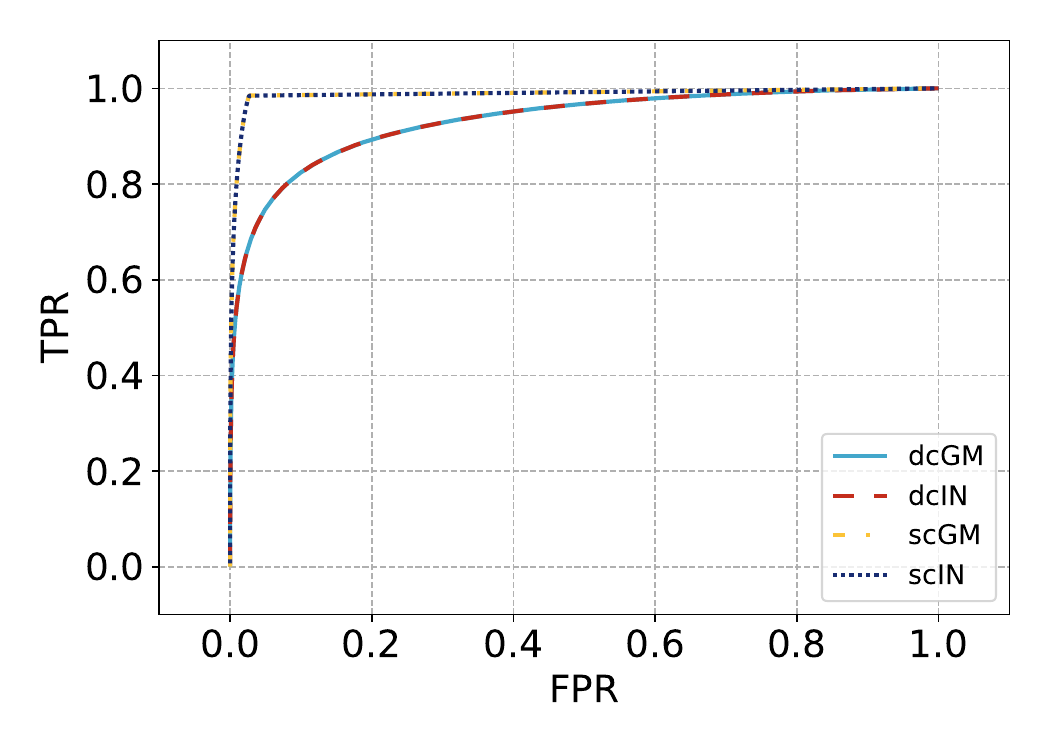}
    }\hfil
    \subfloat[\label{fig:prec_1scale}]{%
    \includegraphics[width=0.49\textwidth]{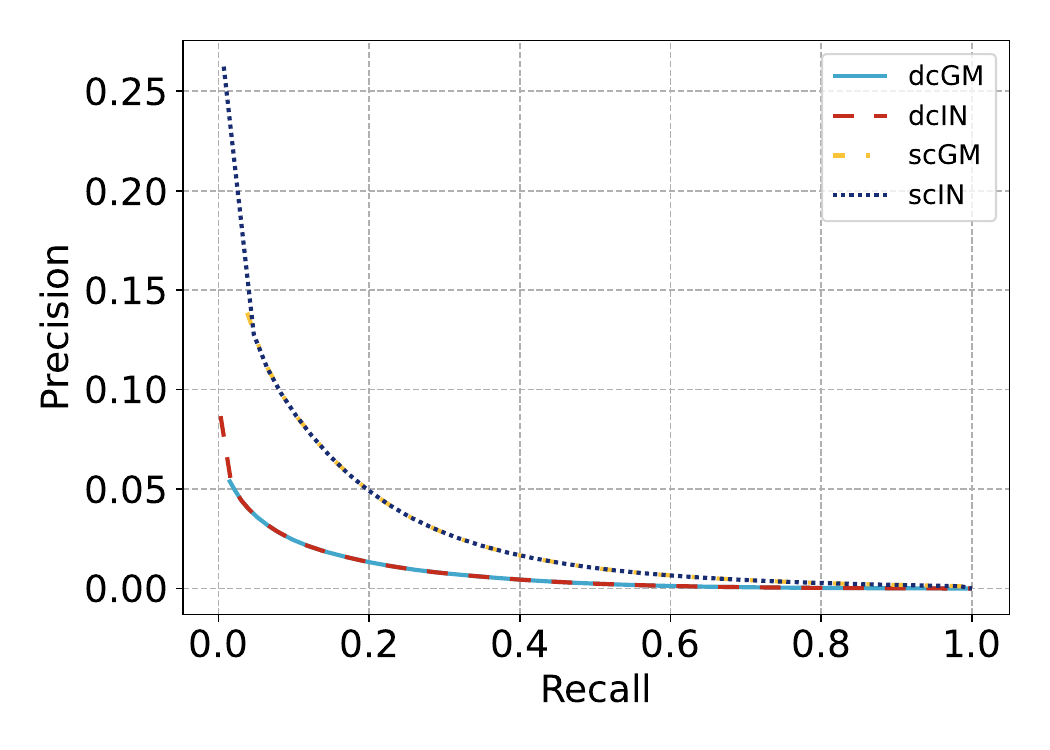}%
    }
    \caption{\scriptsize In panel (a) and (b) the inverse cumulative distributions (ICDF) of the empirical and the expected in and our degree sequences of the benchmark and multi-scale models. In text the estimated exponent of the power law fitted on the tail of the empirical distribution. In panels (c) and (d) the average nearest neighbour degrees as a function of degree. The plots are obtained by sampling 100 times from the ensembles and pooling the results. The average line and the shaded interquartile ranges are constructed by binning the degree on a logarithmic scale. Finally in panels (e) and (f) the receiver operating curve (ROC) and the precision-recall curve for the specified models at firm-level. }\label{fig:single_scale}
\end{figure*} 

As a first measure of ensemble quality we look at the distribution of the in- and out- degrees of each node. From figures \ref{fig:icdf_out_degree} and \ref{fig:icdf_in_degree} we can see that all models perform similarly and adequately in the reconstruction of the degree distribution. We note that as was found in \cite{ialongo2022recon} the models perform better for the out-degree than for the in-degree with the stripe models doing a slightly better job with the tails of the distribution. The major discrepancy between the models and the empirical network are due to the error in the low-degree range were the model tends to assign degrees lower than what is observed empirically. From the figure it is however clear that the two models are indistinguishable in their results at the single scale. This is not surprising as the two models have been shown to approximate to the Chung-Lu model \cite{chung2002connected} for very sparse networks \cite{garuccio2020multiscale}.

While the quality of reconstruction of the degree distribution is similar between the stripe- and density- corrected models this is not the case when looking at higher order properties such as the average nearest-neighbour degree. In figures \ref{fig:knn_1scale_oo} and \ref{fig:knn_1scale_ii} we can see once again that the invariant and gravity models perform similarly but there is a clear improvement in using the technological embedding of the stripes. The added information is in fact key in constraining the mesoscopic structure of the production network and it allows for much better reconstruction of the neighbourhood of each node. This is also true for weighted properties such as the average nearest-neighbour strengths as shown in \cite{ialongo2022recon}. It is clear from the figures \ref{fig:knn_1scale_oo} and \ref{fig:knn_1scale_ii} that the stripe models are much more realistic ensembles as compared to the density-corrected versions. 

As a final measure of the quality of fit at a single scale we report here two standard measures form the link prediction literature. In figure \ref{fig:roc_1scale} we report the Receiver Operating Curve (ROC) and in figure \ref{fig:prec_1scale} the precision vs recall. From these figures we can see that the stripe models clearly outperform their density counterparts with both curves being much higher. It is important to remember here that by construction these models are meant to generate ensembles that are as random as possible while replicating the chosen constraints. As such it is a desirable property of these methods to perform poorly in terms of link prediction metrics, so a comparison of these results is only reasonable when applied to methods that have a similar approach. In the context of this analysis we use these figures to highlight how the models perform relatively to each other and in the ranking of the observed links. Indeed the clearly improved performance of the stripe models can be interpreted in its ability to assign higher probabilities to the links that are observed in the network resulting in a higher precision for a given recall. We are confident therefore that the added constraints of the stripe models are consistent with the observed graph.

\subsection*{Handling the rest of the world}
\begin{figure}[t]
    \centering
    \includegraphics[width=0.55\linewidth]{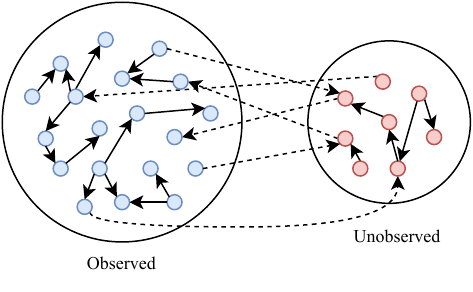}
    \caption{Schematic representation of the rest-of-the-world node.}
    \label{fig:row_diag}
\end{figure}

Typically when working with firm-level data the network that we observe is a sub-graph of the whole production network. In most cases however we still have some information on the unobserved rest-of-the world (ROW). This may be because we observe the partial links between firms in our data and other countries, or because we have data but at a different resolution level, for example in the form of a supply and use table. One of the advantages of the multi-scale model is our ability to be able to incorporate this information in the modelling in a coherent multi-scale framework. To highlight how this can be done in practice we devise here the following scenario shown in figure \ref{fig:row_diag}. As it can be seen in the diagram, we split our nodes in two sets, one observed and one unobserved one, comprising 70\% and 30\% of the vertices respectively. We perform 25 random selections of this separation and compare two approaches: first we estimate the model only on the observed sub-graph ignoring the rest-of-the-world (internal case), and in the second case we instead group the rest-of-the-world into a single node (ROW) and use any available information on it. 

\begin{figure}[t]
    \centering
    \includegraphics[width=0.55\linewidth]{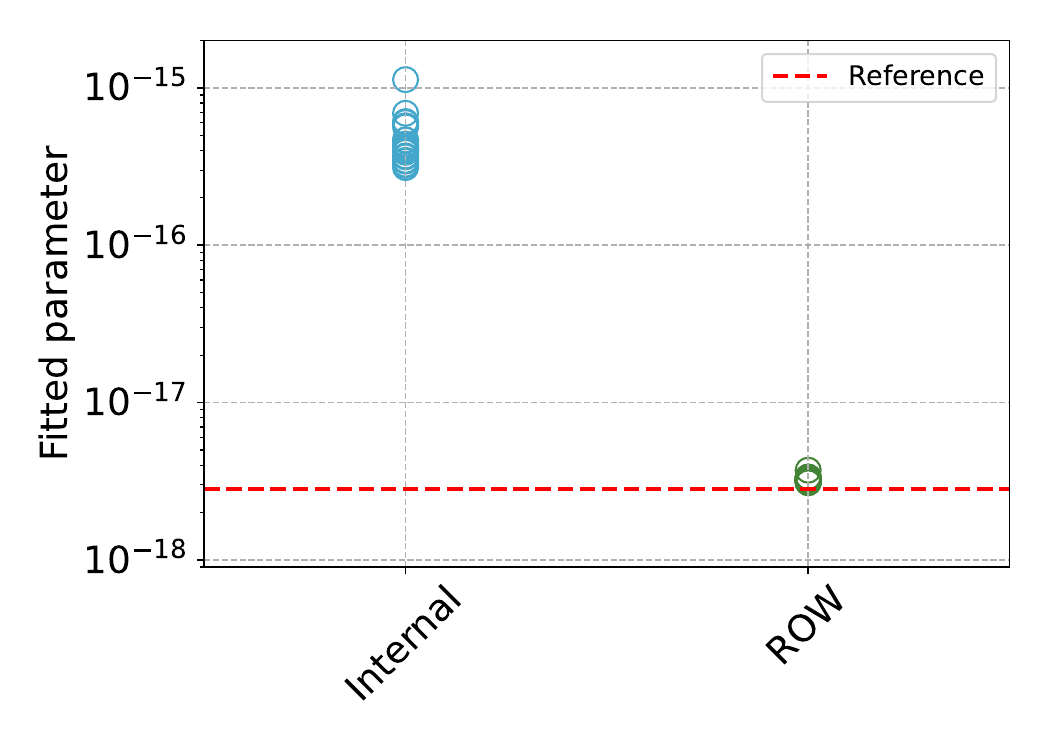}
    \caption{Estimation of the $\delta$ parameter in the internal and rest-of-the-world (ROW) scenario. The reference line represents the parameter estimated on the whole dataset.}\label{fig:row_params}
\end{figure}

We then test the estimation of the model in these two cases. From figure \ref{fig:row_params} we can see that the while the estimation of the density parameter performed including the ROW node is consistent with what would be estimated on the complete graph, the internal estimation is significantly different. This stems from the fact that ignoring the information on the strengths of the nodes towards the ROW node results in a biased estimation of the fitnesses of the firms and hence in a wrong calibration of the parameter. The effect of this is further highlighted by measuring the error in the implied density estimation for the complete graph using the estimation methods outlined above. While the row estimates keep the maximum absolute error in the range of 50-60\%, the internal estimation leads to a consistent overestimation of the density of the complete graph in the order of a 2000 \%. Furthermore, the ROW estimates, although they are subject to significant fluctuations, are much more closely centred around zero. The performance of the model is somewhat disappointing, however the scenario we are testing it in is quite challenging since the random selection of the nodes in the observed sub-graph can by itself lead to significant fluctuations in the observed density which result in the spread of predicted values we have shown. Further work is necessary in order to characterise the expected variance of the density such that we may incorporate this information in a more stable estimation of the parameter. 

The main message we are trying to convey with this scenario is that when observing a sub-graph of the true network we are implicitly working in a multi-scale environment as only seldom do we not have any information on the rest-of-the-world. The model we have proposed is able to correctly handle this information and doing so improves the reconstruction quality of the model. This is perhaps even more evident when applied to our own case. The ABN and ING networks are but a small part of the Dutch graph. Taking into consideration the information of the unobserved component can improve our results. Indeed in the supplementary material we show that taking into consideration the ROW in the computation of the node embeddings translates to a better reconstruction accuracy overall.

\subsection*{Working with aggregate data}\label{sec:agg_data}
\begin{figure}[t]
    \centering
    \includegraphics[width=0.55\linewidth]{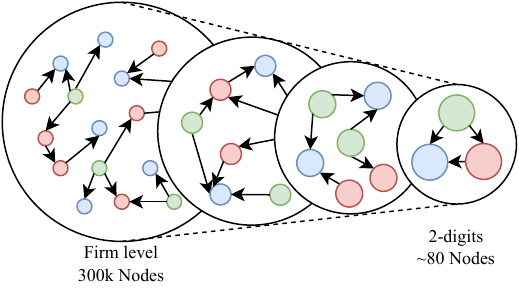}
    \caption{Schematic representation of the coarse graining procedure applied in this scenario.}
    \label{fig:agg_diag}
\end{figure}

In this last scenario we look at how well our methodology allows us to work at different levels of aggregation as shown in figure \ref{fig:agg_diag}. This is the most common case we can find, where we have access to a network at the industry-level but only partial information on the firms. Our objective is to demonstrate how well we can pass from an aggregate network to a disaggregated one at firm level. We will assume that only aggregate data is available in the form of an input-output table at a given level of aggregation implied by the number of digits used for the industrial classification of the firms. The model will then be fitted to the aggregate graph and then we will test its predictive ability on the properties of the firm-level graph. To make our data consistent we will use an implied input-output table from our payment datasets since using the data from the statistical offices would imply introducing two significant sources of potential error. First our data is based on payments, not VAT taxes, and as such we would need some logic to transform one into the other, this is anything but trivial. Second, since our network is only a part of the full graph, understanding how the strengths we observe relate to the national IO table could once again introduce a major element of error. For this reason, and given the experimental nature of these exercises, we build the aggregate network starting from our firm-level graph.  

\begin{figure*}[tp]
    \centering
    \subfloat[\label{fig:d_err_digits}]{%
    \includegraphics[width=0.49\textwidth]{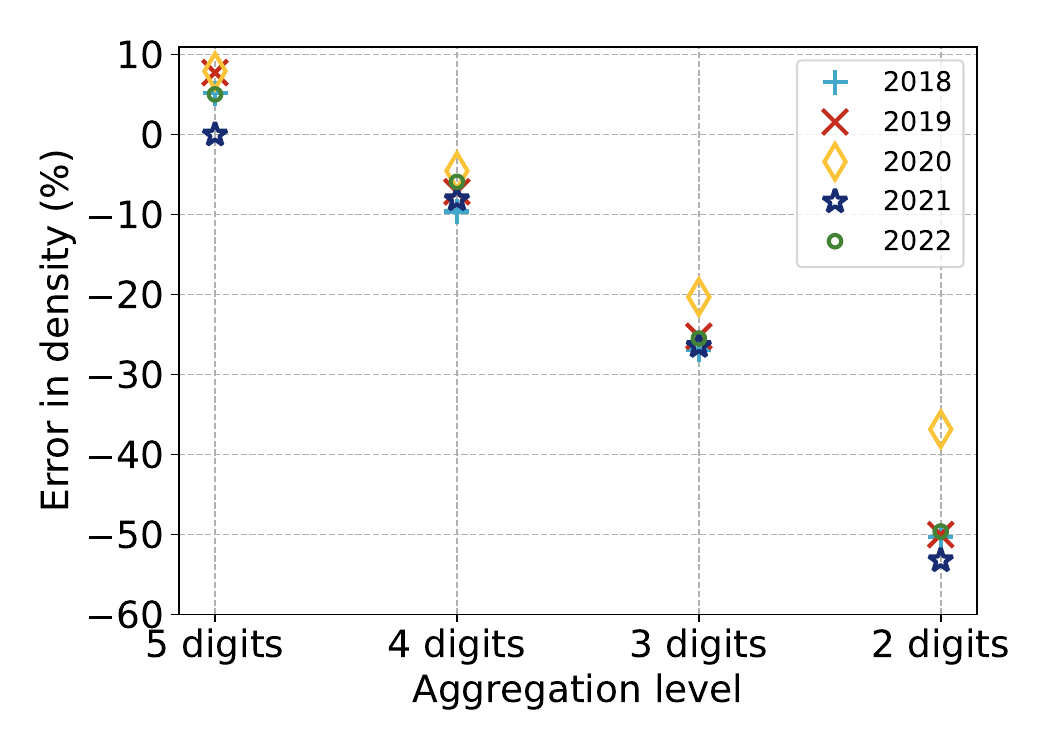}
    }\hfil
    \subfloat[\label{fig:ks_out_digits}]{%
    \includegraphics[width=0.49\textwidth]{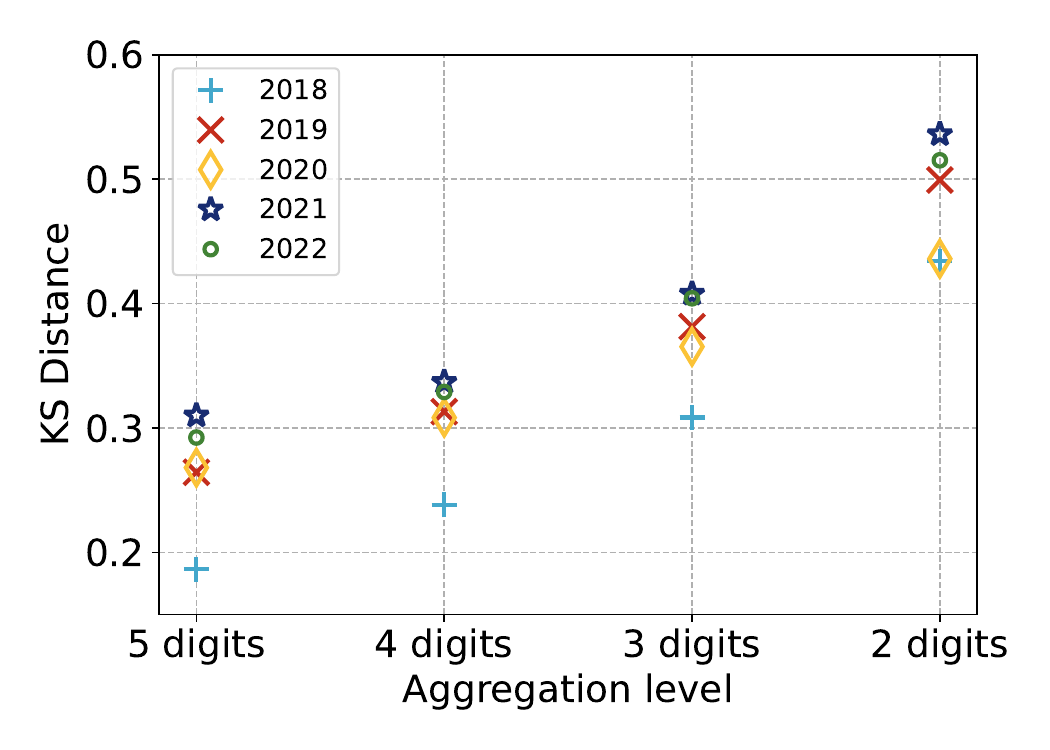}%
    }\hfil
    \subfloat[\label{fig:d_agg_lvl}]{%
    \includegraphics[width=0.49\textwidth]{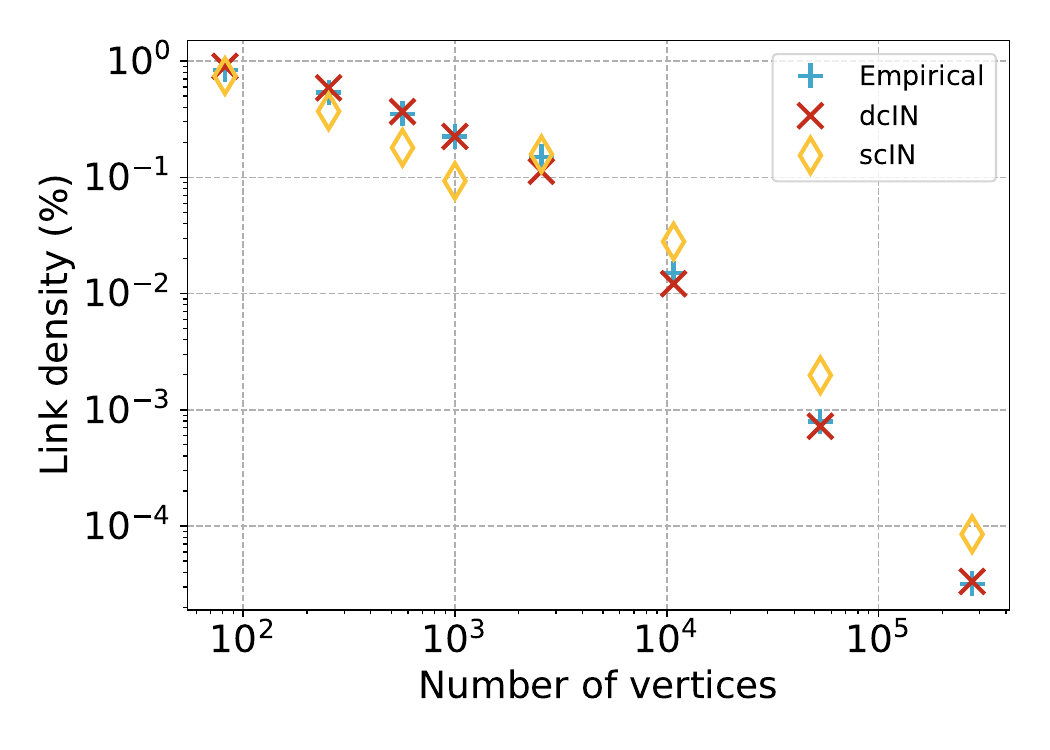}
    }\hfil
    \subfloat[\label{fig:agg_param}]{%
    \includegraphics[width=0.49\textwidth]{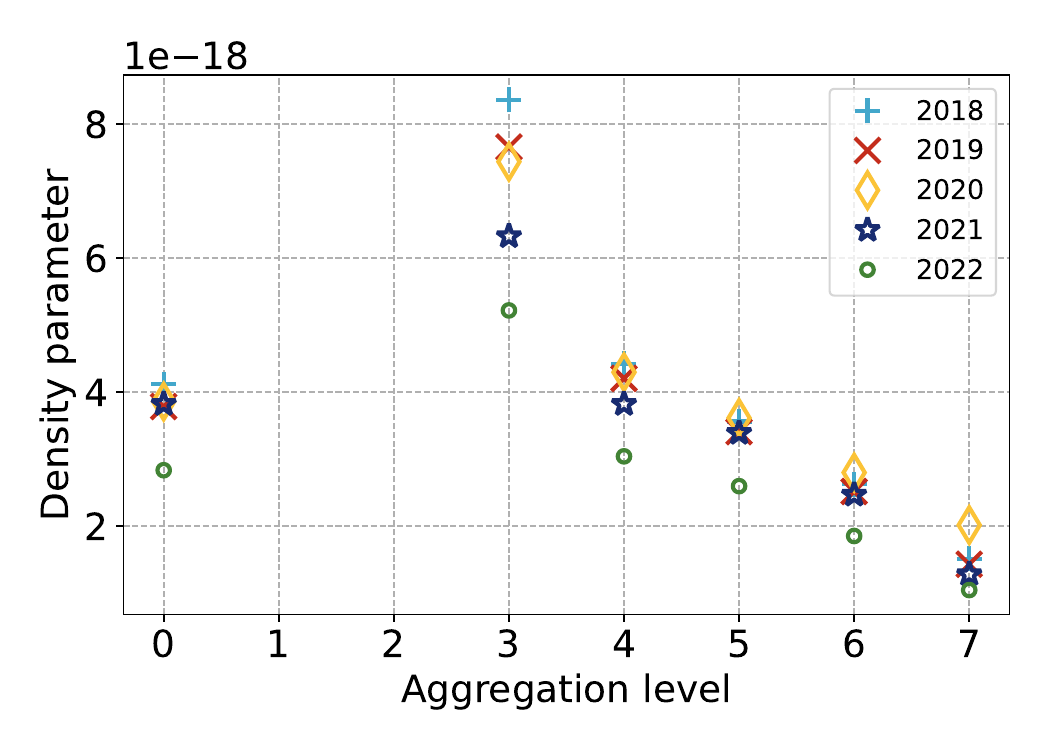}%
    }
    \caption{Reconstruction quality in terms of error in estimating the link density (a) and out-degree distribution (b) at firm-level as a function of the number of digits used to construct the aggregated level at which the model is fitted. In panel (c) the comparison between the expected link density under fine and coarse graining of our models and the empirical one, and in (d) the value of the parameter fitted at most of the scales for all the years available in our data.}
\end{figure*}

From the definition of our model in equation \eqref{eq:p_inv_stripe} it should be clear that there are two elements that determine the probability of the connections between firms. The first element is the $\delta$ parameter which is fitted to match the density of the empirical graph. The second is the node fitnesses which in our case we are taking to be equal to the strength of the firm, either by sector or the total one. As a first exercise we focus on the estimation of the global parameter to see how this scales with the aggregation level we have access to. In figure \ref{fig:d_err_digits} we plot the error in predicting the number of links at the firm level given the IO network defined for a given number of digits. We can clearly see that as the number of digits is increased from two to five the error reduces substantially for all the years we have tested this in. Note that this is not trivial, since we are passing from a very dense small graph at the two-digit level to a very sparse large network at firm-level. That a single parameter is able to correctly capture this trend with relatively small error is quite surprising. The effect of this improvement can also be seen in figure \ref{fig:ks_out_digits} where we are looking at the Kolmogorov-Smirnoff distance between the CDFs of the empirical and reconstructed network degree distributions. 

Note that the estimation of the $\delta$ parameter is done here using the dcIN model and not the scIN one. This is due to the fact that the estimation of the scIN model is ill-posed if the observed network coincides with the stripe definitions. We further discuss this in the supplementary information, however to understand the point it is simpler to imagine estimating $\delta$ on a fully connected graph. It should be clear that the only way to guarantee a link density of one is for the parameter to tend to infinity. Computationally this would mean that after a certain point all values of the parameter are equivalent and as such it is not possible to identify the optimal parameter. This is probably the reason why as the network becomes more dense, our estimation of $\delta$ becomes more unreliable as seen in figure \ref{fig:d_err_digits}.

The model we have presented here is by construction invariant to any partition of the graph and as such we expect that fitting the parameters at any scale would result in an equally accurate prediction at any other scale. Of course this is not true in practice for two reasons: first, it is not necessarily true that the data is well approximated by this model, and second, that the fitnesses we have constructed ensure this relationship holds. To assess this we fit the model at an intermediate scale (5-digits level) and observe how the model performs in predicting the density at all other scales. We build seven graphs that specify seven increasing levels of aggregation, starting from the firm-level graph (level 0) with roughly $3\times10^5$ nodes, we go to levels 4, 5, 6, and 7 that are built aggregating according to the firms' NACE classification at 5, 4, 3, and 2 digits respectively. Given that this would imply jumping from $3\times10^5$ to roughly 900 nodes in level 4, we construct artificially some intermediate levels 1, 2 and 3, such that we have a more continuous plot in the number of firms. These levels are however built by forming random groups of similar size within a 5-digit industry, and as we will see this has an impact on the result. 

In figure \ref{fig:d_agg_lvl} we can see that the density of the empirical network is well reproduced by the dcIN at all scales while the scIN, although still able to follow the pattern, it performs more poorly. This is because the scIN has to be fitted on level three since, as we discussed previously, fitting it at the 5-digit level is not possible. It seems that the randomization that we perform somehow breaks the pattern of the non-random grouping due to the industrial classification. This is perhaps more evident when looking at figure \ref{fig:agg_param}: here we can see that when fitting the parameter on the aggregated level three, this gives us an outlier with respect to the much smoother change between all the other levels. This suggests that while the model is scale-invariant by design, when applied to real data the kind of partitioning that is applied might matter. Further investigation is necessary on this last issue.

From the discussion above it should be clear what is another advantage of the multi-scale functional. In figure \ref{fig:agg_param} we have seen that we can correctly estimate the parameter at the 5-digit level with a very small error. This implies that instead of having to compute the expected density for a $3\times10^5$ graph we can fit the model on the 900 nodes graph and use the estimated parameter successfully. If we were estimating the fitnesses using maximum likelihood then this could be even more of an advantage. Here we applied this logic to computing the parameter but the same idea could be applied for an efficient sampling method. We could first sample a graph at a given desired aggregation level to obtain an aggregated adjacency matrix. It is then possible to go one step deeper on a more disaggregated level with the knowledge that for any ``macro'' link that we did not sample before there can be no ``micro'' link. We can now sample a new more fine-grained adjacency matrix by simply conditioning on the previous matrix. This conditioning, which in general is time consuming, is achievable thanks to the specific form of the invariant functional. Iterating this process, depending on the sparsity of the sampled matrix, can mean reducing significantly the complexity of the sampling process. If the graph is sparse, and we take a clever level of aggregation, this iterative process can yield a much lower computational load than the expected $N^2$. Although beyond the scope of this work, we thought it worth mentioning as another advantage of the invariant model.

So far we have discussed the impact of the aggregation level on the estimation of $\delta$ however the other source of uncertainty is due to the fitnesses of the node. In general when observing the aggregated data we have the aggregate flows and hence the coarse-grained node fitness. Given the additive properties of the fitnesses we know that the sum of the firm fitnesses must be equal to the ones we infer from the aggregate data, but we might not know how. If we do not know the true values of the strengths per sector of the firms we can try several strategies to infer them. To understand this issue better, we compare five cases with increasing level of detail: first we assume to know the number of firms per sector but not their size (Uniform), then we assume to have estimated a log-normal distribution on the true sizes of firms and use it to generate realistic firm sizes (Distribution), finally we assume to know the real size of the firms' in and out flows but not their by sector distribution as in the dcIN (Total). We compare these cases with two stipe scenarios: in the first we assume to know the true stripes (Stripe) while in the second we construct the stripes homogeneously from the IO flows such that all firms in a given sector have the same percentage of flows coming from the other sectors. We leave a more complete discussion of the various cases to the supplementary informations, but we note here that we unfortunately find that the heterogeneous information given by the true stripes is a clear advantage in terms of reconstruction accuracy. 

\section*{Discussion}
The contribution of this work has been of adapting the promising multi-scale model proposed by \cite{garuccio2020multiscale} to production networks. This novel methodology is particularly suited to supply-chain reconstruction on a large scale because it recognises the inherent multi-scale nature of the problem. The necessity of integrating data from different sources at various aggregation levels fits perfectly with the theoretical properties of the model. Furthermore, as it allows for arbitrary node partitions, it can be used to model sub-graphs of interest without losing track of the complexity of the whole. This is a good starting point to be able to work towards a more comprehensive picture of supply-chains at a global scale, while keeping it computationally more tractable. 

As we have shown the invariant formulation performs just as well as a the maximum entropy model proposed in \cite{ialongo2022recon} if given the same fitness vector. However differently from the latter its parameters are defined as to maintain consistency at different scales under an additive scaling rule. The tests we have performed, under two different practical scenarios, have highlighted the importance of including any knowledge of the rest-of-the-world as this can strongly improve the performance of the reconstruction. This principle has often been underestimated in previous work\footnote{Including our own!} as the information on the unobserved graph might seem difficult to incorporate into the model. Fortunately the multi-scale model presents a simple and principled way to work in this multi-scale environment. 

In the second scenario we have attempted to illustrate how the model can be used when limited aggregate information is present. The ability of the model to maintain the quality of fit across scales allow for its parameters to be efficiently estimated on coarse-grained graphs at a fraction of the usual computational cost. Furthermore, this provides a simple way to estimate the parameters of the model on available public information such as input-output tables. However there are several limitations in this approach. The main limitation here is that although it seems that with the right firm-level information we can perform reasonably well, this information might be difficult to obtain. Note that when fitting the model on a coarse-grained graph, we have so far assumed that the only interesting parameter to estimate is the one controlling for the density. Of course this can be relaxed allowing for the estimation of the whole node vector fitness. This in principle could bring advantages, since we can now extract more information from the aggregate scale. However the additive nature of the parameters does not give us a unique way to obtain the firm-level fitnesses from the estimated aggregate ones. How to solve this in practice is the subject for future work. 

Note that our analysis does not incorporate geographic distance, although we are convinced it plays an important role even for a small country like the Netherlands. Fortunately, any such information can be incorporated in a straightforward manner into the model. The objective of the paper was to present a proof of concept for this model to highlight its major advantages in a controlled yet realistic environment based on real data. We hope that this will provide a useful benchmark in developing better models for reconstructing production networks at the global scale.

\section*{Data and code availability}
The datasets on transactions used in the paper are highly confidential and cannot be made public. The code is freely available as the \texttt{graph-ensembles} python package.

\section*{Acknowledgements}
We thank ABN AMRO Bank N.V. for their support and active collaboration. A special thanks to the FR\&R team at ABN AMRO for their advice that helped shape this research.
We acknowledge support from Stichting Econophysics (Leiden, The Netherlands).
This work is supported by the European Union - NextGenerationEU - National Recovery and Resilience Plan (Piano Nazionale di Ripresa e Resilienza, PNRR), project `SoBigData.it - Strengthening the Italian RI for Social Mining and Big Data Analytics' - Grant IR0000013 (n. 3264, 28/12/2021) (\url{https://pnrr.sobigdata.it/}).
It is also supported by the project ``Reconstruction, Resilience and Recovery of Socio-Economic Networks'' RECON-NET EP\_FAIR\_005 - PE0000013 ``FAIR'' - PNRR M4C2 Investment 1.3, financed by the European Union – NextGenerationEU.

\section*{Author contributions statement}
L.N.I. and D.G. designed the research and methodology; L.N.I and S.B. performed the data analysis on the two datasets separately; all author wrote and approved the paper.

\section*{Additional information}
The authors declare no competing interests.

\bibliography{scenarios}

\begin{thebibliography}{42}%
\makeatletter
\providecommand \@ifxundefined [1]{%
 \@ifx{#1\undefined}
}%
\providecommand \@ifnum [1]{%
 \ifnum #1\expandafter \@firstoftwo
 \else \expandafter \@secondoftwo
 \fi
}%
\providecommand \@ifx [1]{%
 \ifx #1\expandafter \@firstoftwo
 \else \expandafter \@secondoftwo
 \fi
}%
\providecommand \natexlab [1]{#1}%
\providecommand \enquote  [1]{``#1''}%
\providecommand \bibnamefont  [1]{#1}%
\providecommand \bibfnamefont [1]{#1}%
\providecommand \citenamefont [1]{#1}%
\providecommand \href@noop [0]{\@secondoftwo}%
\providecommand \href [0]{\begingroup \@sanitize@url \@href}%
\providecommand \@href[1]{\@@startlink{#1}\@@href}%
\providecommand \@@href[1]{\endgroup#1\@@endlink}%
\providecommand \@sanitize@url [0]{\catcode `\\12\catcode `\$12\catcode
  `\&12\catcode `\#12\catcode `\^12\catcode `\_12\catcode `\%12\relax}%
\providecommand \@@startlink[1]{}%
\providecommand \@@endlink[0]{}%
\providecommand \url  [0]{\begingroup\@sanitize@url \@url }%
\providecommand \@url [1]{\endgroup\@href {#1}{\urlprefix }}%
\providecommand \urlprefix  [0]{URL }%
\providecommand \Eprint [0]{\href }%
\providecommand \doibase [0]{https://doi.org/}%
\providecommand \selectlanguage [0]{\@gobble}%
\providecommand \bibinfo  [0]{\@secondoftwo}%
\providecommand \bibfield  [0]{\@secondoftwo}%
\providecommand \translation [1]{[#1]}%
\providecommand \BibitemOpen [0]{}%
\providecommand \bibitemStop [0]{}%
\providecommand \bibitemNoStop [0]{.\EOS\space}%
\providecommand \EOS [0]{\spacefactor3000\relax}%
\providecommand \BibitemShut  [1]{\csname bibitem#1\endcsname}%
\let\auto@bib@innerbib\@empty
\bibitem [{\citenamefont {Schweitzer}\ \emph {et~al.}(2009)\citenamefont
  {Schweitzer}, \citenamefont {Fagiolo}, \citenamefont {Sornette},
  \citenamefont {Vega-Redondo}, \citenamefont {Vespignani},\ and\ \citenamefont
  {White}}]{schweitzer2009economic}%
  \BibitemOpen
  \bibfield  {author} {\bibinfo {author} {\bibfnamefont {F.}~\bibnamefont
  {Schweitzer}}, \bibinfo {author} {\bibfnamefont {G.}~\bibnamefont {Fagiolo}},
  \bibinfo {author} {\bibfnamefont {D.}~\bibnamefont {Sornette}}, \bibinfo
  {author} {\bibfnamefont {F.}~\bibnamefont {Vega-Redondo}}, \bibinfo {author}
  {\bibfnamefont {A.}~\bibnamefont {Vespignani}},\ and\ \bibinfo {author}
  {\bibfnamefont {D.~R.}\ \bibnamefont {White}},\ }\href@noop {} {\bibfield
  {journal} {\bibinfo  {journal} {science}\ }\textbf {\bibinfo {volume}
  {325}},\ \bibinfo {pages} {422} (\bibinfo {year} {2009})}\BibitemShut
  {NoStop}%
\bibitem [{\citenamefont {Bernanke}(2018)}]{bernanke2018real}%
  \BibitemOpen
  \bibfield  {author} {\bibinfo {author} {\bibfnamefont {B.~S.}\ \bibnamefont
  {Bernanke}},\ }\href@noop {} {\bibfield  {journal} {\bibinfo  {journal}
  {Brookings Papers on Economic Activity}\ }\textbf {\bibinfo {volume}
  {2018}},\ \bibinfo {pages} {251} (\bibinfo {year} {2018})}\BibitemShut
  {NoStop}%
\bibitem [{\citenamefont {Carvalho}\ \emph {et~al.}(2021)\citenamefont
  {Carvalho}, \citenamefont {Nirei}, \citenamefont {Saito},\ and\ \citenamefont
  {Tahbaz-Salehi}}]{carvalho2021jap}%
  \BibitemOpen
  \bibfield  {author} {\bibinfo {author} {\bibfnamefont {V.~M.}\ \bibnamefont
  {Carvalho}}, \bibinfo {author} {\bibfnamefont {M.}~\bibnamefont {Nirei}},
  \bibinfo {author} {\bibfnamefont {Y.~U.}\ \bibnamefont {Saito}},\ and\
  \bibinfo {author} {\bibfnamefont {A.}~\bibnamefont {Tahbaz-Salehi}},\
  }\href@noop {} {\bibfield  {journal} {\bibinfo  {journal} {The Quarterly
  Journal of Economics}\ }\textbf {\bibinfo {volume} {136}},\ \bibinfo {pages}
  {1255} (\bibinfo {year} {2021})}\BibitemShut {NoStop}%
\bibitem [{\citenamefont {Henriet}\ \emph {et~al.}(2012)\citenamefont
  {Henriet}, \citenamefont {Hallegatte},\ and\ \citenamefont
  {Tabourier}}]{henriet2012firm}%
  \BibitemOpen
  \bibfield  {author} {\bibinfo {author} {\bibfnamefont {F.}~\bibnamefont
  {Henriet}}, \bibinfo {author} {\bibfnamefont {S.}~\bibnamefont
  {Hallegatte}},\ and\ \bibinfo {author} {\bibfnamefont {L.}~\bibnamefont
  {Tabourier}},\ }\href@noop {} {\bibfield  {journal} {\bibinfo  {journal}
  {Journal of Economic Dynamics and Control}\ }\textbf {\bibinfo {volume}
  {36}},\ \bibinfo {pages} {150} (\bibinfo {year} {2012})}\BibitemShut
  {NoStop}%
\bibitem [{\citenamefont {Acemoglu}\ \emph {et~al.}(2012)\citenamefont
  {Acemoglu}, \citenamefont {Carvalho}, \citenamefont {Ozdaglar},\ and\
  \citenamefont {Tahbaz-Salehi}}]{acemoglu2012net}%
  \BibitemOpen
  \bibfield  {author} {\bibinfo {author} {\bibfnamefont {D.}~\bibnamefont
  {Acemoglu}}, \bibinfo {author} {\bibfnamefont {V.~M.}\ \bibnamefont
  {Carvalho}}, \bibinfo {author} {\bibfnamefont {A.}~\bibnamefont {Ozdaglar}},\
  and\ \bibinfo {author} {\bibfnamefont {A.}~\bibnamefont {Tahbaz-Salehi}},\
  }\href@noop {} {\bibfield  {journal} {\bibinfo  {journal} {Econometrica}\
  }\textbf {\bibinfo {volume} {80}},\ \bibinfo {pages} {1977} (\bibinfo {year}
  {2012})}\BibitemShut {NoStop}%
\bibitem [{\citenamefont {Contreras}\ and\ \citenamefont
  {Fagiolo}(2014)}]{contreras2014propagation}%
  \BibitemOpen
  \bibfield  {author} {\bibinfo {author} {\bibfnamefont {M.~G.~A.}\
  \bibnamefont {Contreras}}\ and\ \bibinfo {author} {\bibfnamefont
  {G.}~\bibnamefont {Fagiolo}},\ }\href@noop {} {\bibfield  {journal} {\bibinfo
   {journal} {Physical Review E}\ }\textbf {\bibinfo {volume} {90}},\ \bibinfo
  {pages} {062812} (\bibinfo {year} {2014})}\BibitemShut {NoStop}%
\bibitem [{\citenamefont {Pichler}\ \emph {et~al.}(2022)\citenamefont
  {Pichler}, \citenamefont {Pangallo}, \citenamefont {del Rio-Chanona},
  \citenamefont {Lafond},\ and\ \citenamefont {Farmer}}]{pichler2022fore}%
  \BibitemOpen
  \bibfield  {author} {\bibinfo {author} {\bibfnamefont {A.}~\bibnamefont
  {Pichler}}, \bibinfo {author} {\bibfnamefont {M.}~\bibnamefont {Pangallo}},
  \bibinfo {author} {\bibfnamefont {R.~M.}\ \bibnamefont {del Rio-Chanona}},
  \bibinfo {author} {\bibfnamefont {F.}~\bibnamefont {Lafond}},\ and\ \bibinfo
  {author} {\bibfnamefont {J.~D.}\ \bibnamefont {Farmer}},\ }\href@noop {}
  {\bibfield  {journal} {\bibinfo  {journal} {Journal of Economic Dynamics and
  Control}\ }\textbf {\bibinfo {volume} {144}},\ \bibinfo {pages} {104527}
  (\bibinfo {year} {2022})}\BibitemShut {NoStop}%
\bibitem [{\citenamefont {Huremovic}\ \emph {et~al.}(2023)\citenamefont
  {Huremovic}, \citenamefont {Jim{\'e}nez}, \citenamefont {Moral-Benito},
  \citenamefont {Peydr{\'o}},\ and\ \citenamefont
  {Vega-Redondo}}]{huremovic2023production}%
  \BibitemOpen
  \bibfield  {author} {\bibinfo {author} {\bibfnamefont {K.}~\bibnamefont
  {Huremovic}}, \bibinfo {author} {\bibfnamefont {G.}~\bibnamefont
  {Jim{\'e}nez}}, \bibinfo {author} {\bibfnamefont {E.}~\bibnamefont
  {Moral-Benito}}, \bibinfo {author} {\bibfnamefont {J.-L.}\ \bibnamefont
  {Peydr{\'o}}},\ and\ \bibinfo {author} {\bibfnamefont {F.}~\bibnamefont
  {Vega-Redondo}},\ }\href@noop {} {\bibfield  {journal} {\bibinfo  {journal}
  {Available at SSRN 4657236}\ } (\bibinfo {year} {2023})}\BibitemShut
  {NoStop}%
\bibitem [{\citenamefont {McNerney}\ \emph {et~al.}(2022)\citenamefont
  {McNerney}, \citenamefont {Savoie}, \citenamefont {Caravelli}, \citenamefont
  {Carvalho},\ and\ \citenamefont {Farmer}}]{mcnerney2022production}%
  \BibitemOpen
  \bibfield  {author} {\bibinfo {author} {\bibfnamefont {J.}~\bibnamefont
  {McNerney}}, \bibinfo {author} {\bibfnamefont {C.}~\bibnamefont {Savoie}},
  \bibinfo {author} {\bibfnamefont {F.}~\bibnamefont {Caravelli}}, \bibinfo
  {author} {\bibfnamefont {V.~M.}\ \bibnamefont {Carvalho}},\ and\ \bibinfo
  {author} {\bibfnamefont {J.~D.}\ \bibnamefont {Farmer}},\ }\href@noop {}
  {\bibfield  {journal} {\bibinfo  {journal} {Proceedings of the National
  Academy of Sciences}\ }\textbf {\bibinfo {volume} {119}},\ \bibinfo {pages}
  {e2106031118} (\bibinfo {year} {2022})}\BibitemShut {NoStop}%
\bibitem [{\citenamefont {Klimek}\ \emph {et~al.}(2019)\citenamefont {Klimek},
  \citenamefont {Poledna},\ and\ \citenamefont
  {Thurner}}]{klimek2019quantifying}%
  \BibitemOpen
  \bibfield  {author} {\bibinfo {author} {\bibfnamefont {P.}~\bibnamefont
  {Klimek}}, \bibinfo {author} {\bibfnamefont {S.}~\bibnamefont {Poledna}},\
  and\ \bibinfo {author} {\bibfnamefont {S.}~\bibnamefont {Thurner}},\
  }\href@noop {} {\bibfield  {journal} {\bibinfo  {journal} {Nature
  communications}\ }\textbf {\bibinfo {volume} {10}},\ \bibinfo {pages} {1677}
  (\bibinfo {year} {2019})}\BibitemShut {NoStop}%
\bibitem [{\citenamefont {Diem}\ \emph {et~al.}(2022)\citenamefont {Diem},
  \citenamefont {Borsos}, \citenamefont {Reisch}, \citenamefont {Kert{\'e}sz},\
  and\ \citenamefont {Thurner}}]{diem2021quantifying}%
  \BibitemOpen
  \bibfield  {author} {\bibinfo {author} {\bibfnamefont {C.}~\bibnamefont
  {Diem}}, \bibinfo {author} {\bibfnamefont {A.}~\bibnamefont {Borsos}},
  \bibinfo {author} {\bibfnamefont {T.}~\bibnamefont {Reisch}}, \bibinfo
  {author} {\bibfnamefont {J.}~\bibnamefont {Kert{\'e}sz}},\ and\ \bibinfo
  {author} {\bibfnamefont {S.}~\bibnamefont {Thurner}},\ }\href@noop {}
  {\bibfield  {journal} {\bibinfo  {journal} {Scientific reports}\ }\textbf
  {\bibinfo {volume} {12}},\ \bibinfo {pages} {7719} (\bibinfo {year}
  {2022})}\BibitemShut {NoStop}%
\bibitem [{\citenamefont {Colon}\ \emph {et~al.}(2021)\citenamefont {Colon},
  \citenamefont {Hallegatte},\ and\ \citenamefont
  {Rozenberg}}]{colon2021criticality}%
  \BibitemOpen
  \bibfield  {author} {\bibinfo {author} {\bibfnamefont {C.}~\bibnamefont
  {Colon}}, \bibinfo {author} {\bibfnamefont {S.}~\bibnamefont {Hallegatte}},\
  and\ \bibinfo {author} {\bibfnamefont {J.}~\bibnamefont {Rozenberg}},\
  }\href@noop {} {\bibfield  {journal} {\bibinfo  {journal} {Nature
  Sustainability}\ }\textbf {\bibinfo {volume} {4}},\ \bibinfo {pages} {209}
  (\bibinfo {year} {2021})}\BibitemShut {NoStop}%
\bibitem [{\citenamefont {Diem}\ \emph {et~al.}(2024)\citenamefont {Diem},
  \citenamefont {Borsos}, \citenamefont {Reisch}, \citenamefont {Kert{\'e}sz},\
  and\ \citenamefont {Thurner}}]{diem2023macro}%
  \BibitemOpen
  \bibfield  {author} {\bibinfo {author} {\bibfnamefont {C.}~\bibnamefont
  {Diem}}, \bibinfo {author} {\bibfnamefont {A.}~\bibnamefont {Borsos}},
  \bibinfo {author} {\bibfnamefont {T.}~\bibnamefont {Reisch}}, \bibinfo
  {author} {\bibfnamefont {J.}~\bibnamefont {Kert{\'e}sz}},\ and\ \bibinfo
  {author} {\bibfnamefont {S.}~\bibnamefont {Thurner}},\ }\href@noop {}
  {\bibfield  {journal} {\bibinfo  {journal} {PNAS nexus}\ }\textbf {\bibinfo
  {volume} {3}},\ \bibinfo {pages} {pgae064} (\bibinfo {year}
  {2024})}\BibitemShut {NoStop}%
\bibitem [{\citenamefont {Moran}\ and\ \citenamefont
  {Bouchaud}(2019)}]{moran2019may}%
  \BibitemOpen
  \bibfield  {author} {\bibinfo {author} {\bibfnamefont {J.}~\bibnamefont
  {Moran}}\ and\ \bibinfo {author} {\bibfnamefont {J.-P.}\ \bibnamefont
  {Bouchaud}},\ }\href@noop {} {\bibfield  {journal} {\bibinfo  {journal}
  {Physical Review E}\ }\textbf {\bibinfo {volume} {100}},\ \bibinfo {pages}
  {032307} (\bibinfo {year} {2019})}\BibitemShut {NoStop}%
\bibitem [{\citenamefont {Lafond}\ \emph {et~al.}(2023)\citenamefont {Lafond},
  \citenamefont {Astudillo-Est{\'e}vez}, \citenamefont {Bacilieri},\ and\
  \citenamefont {Borsos}}]{lafond2023firm}%
  \BibitemOpen
  \bibfield  {author} {\bibinfo {author} {\bibfnamefont {F.}~\bibnamefont
  {Lafond}}, \bibinfo {author} {\bibfnamefont {P.}~\bibnamefont
  {Astudillo-Est{\'e}vez}}, \bibinfo {author} {\bibfnamefont {A.}~\bibnamefont
  {Bacilieri}},\ and\ \bibinfo {author} {\bibfnamefont {A.}~\bibnamefont
  {Borsos}},\ }\href@noop {} {\emph {\bibinfo {title} {Firm-level production
  networks: what do we (really) know?}}},\ \bibinfo {type} {Tech. Rep.}\
  (\bibinfo  {institution} {INET Oxford Working Paper},\ \bibinfo {year}
  {2023})\BibitemShut {NoStop}%
\bibitem [{\citenamefont {Demir}\ \emph {et~al.}(2024)\citenamefont {Demir},
  \citenamefont {Fieler}, \citenamefont {Xu},\ and\ \citenamefont
  {Yang}}]{demir2024ring}%
  \BibitemOpen
  \bibfield  {author} {\bibinfo {author} {\bibfnamefont {B.}~\bibnamefont
  {Demir}}, \bibinfo {author} {\bibfnamefont {A.~C.}\ \bibnamefont {Fieler}},
  \bibinfo {author} {\bibfnamefont {D.~Y.}\ \bibnamefont {Xu}},\ and\ \bibinfo
  {author} {\bibfnamefont {K.~K.}\ \bibnamefont {Yang}},\ }\href@noop {}
  {\bibfield  {journal} {\bibinfo  {journal} {Journal of Political Economy}\
  }\textbf {\bibinfo {volume} {132}},\ \bibinfo {pages} {200} (\bibinfo {year}
  {2024})}\BibitemShut {NoStop}%
\bibitem [{\citenamefont {Fujiwara}\ and\ \citenamefont
  {Aoyama}(2010)}]{fujiwara2010large}%
  \BibitemOpen
  \bibfield  {author} {\bibinfo {author} {\bibfnamefont {Y.}~\bibnamefont
  {Fujiwara}}\ and\ \bibinfo {author} {\bibfnamefont {H.}~\bibnamefont
  {Aoyama}},\ }\href@noop {} {\bibfield  {journal} {\bibinfo  {journal} {The
  European Physical Journal B}\ }\textbf {\bibinfo {volume} {77}},\ \bibinfo
  {pages} {565} (\bibinfo {year} {2010})}\BibitemShut {NoStop}%
\bibitem [{\citenamefont {Mizuno}\ \emph {et~al.}(2014)\citenamefont {Mizuno},
  \citenamefont {Souma},\ and\ \citenamefont {Watanabe}}]{mizuno2014structure}%
  \BibitemOpen
  \bibfield  {author} {\bibinfo {author} {\bibfnamefont {T.}~\bibnamefont
  {Mizuno}}, \bibinfo {author} {\bibfnamefont {W.}~\bibnamefont {Souma}},\ and\
  \bibinfo {author} {\bibfnamefont {T.}~\bibnamefont {Watanabe}},\ }\href@noop
  {} {\bibfield  {journal} {\bibinfo  {journal} {Plos one}\ }\textbf {\bibinfo
  {volume} {9}},\ \bibinfo {pages} {e100712} (\bibinfo {year}
  {2014})}\BibitemShut {NoStop}%
\bibitem [{\citenamefont {Inoue}\ and\ \citenamefont
  {Todo}(2019)}]{inoue2019firm}%
  \BibitemOpen
  \bibfield  {author} {\bibinfo {author} {\bibfnamefont {H.}~\bibnamefont
  {Inoue}}\ and\ \bibinfo {author} {\bibfnamefont {Y.}~\bibnamefont {Todo}},\
  }\href@noop {} {\bibfield  {journal} {\bibinfo  {journal} {Nature
  Sustainability}\ }\textbf {\bibinfo {volume} {2}},\ \bibinfo {pages} {841}
  (\bibinfo {year} {2019})}\BibitemShut {NoStop}%
\bibitem [{\citenamefont {Fujiwara}\ \emph {et~al.}(2021)\citenamefont
  {Fujiwara}, \citenamefont {Inoue}, \citenamefont {Yamaguchi}, \citenamefont
  {Aoyama}, \citenamefont {Tanaka},\ and\ \citenamefont
  {Kikuchi}}]{fujiwara2021money}%
  \BibitemOpen
  \bibfield  {author} {\bibinfo {author} {\bibfnamefont {Y.}~\bibnamefont
  {Fujiwara}}, \bibinfo {author} {\bibfnamefont {H.}~\bibnamefont {Inoue}},
  \bibinfo {author} {\bibfnamefont {T.}~\bibnamefont {Yamaguchi}}, \bibinfo
  {author} {\bibfnamefont {H.}~\bibnamefont {Aoyama}}, \bibinfo {author}
  {\bibfnamefont {T.}~\bibnamefont {Tanaka}},\ and\ \bibinfo {author}
  {\bibfnamefont {K.}~\bibnamefont {Kikuchi}},\ }\href@noop {} {\bibfield
  {journal} {\bibinfo  {journal} {EPJ data science}\ }\textbf {\bibinfo
  {volume} {10}},\ \bibinfo {pages} {19} (\bibinfo {year} {2021})}\BibitemShut
  {NoStop}%
\bibitem [{\citenamefont {Atalay}\ \emph {et~al.}(2011)\citenamefont {Atalay},
  \citenamefont {Hortacsu}, \citenamefont {Roberts},\ and\ \citenamefont
  {Syverson}}]{atalay2011network}%
  \BibitemOpen
  \bibfield  {author} {\bibinfo {author} {\bibfnamefont {E.}~\bibnamefont
  {Atalay}}, \bibinfo {author} {\bibfnamefont {A.}~\bibnamefont {Hortacsu}},
  \bibinfo {author} {\bibfnamefont {J.}~\bibnamefont {Roberts}},\ and\ \bibinfo
  {author} {\bibfnamefont {C.}~\bibnamefont {Syverson}},\ }\href@noop {}
  {\bibfield  {journal} {\bibinfo  {journal} {Proceedings of the National
  Academy of Sciences}\ }\textbf {\bibinfo {volume} {108}},\ \bibinfo {pages}
  {5199} (\bibinfo {year} {2011})}\BibitemShut {NoStop}%
\bibitem [{\citenamefont {Dhyne}\ \emph {et~al.}(2021)\citenamefont {Dhyne},
  \citenamefont {Kikkawa}, \citenamefont {Mogstad},\ and\ \citenamefont
  {Tintelnot}}]{dhyne2021trade}%
  \BibitemOpen
  \bibfield  {author} {\bibinfo {author} {\bibfnamefont {E.}~\bibnamefont
  {Dhyne}}, \bibinfo {author} {\bibfnamefont {A.~K.}\ \bibnamefont {Kikkawa}},
  \bibinfo {author} {\bibfnamefont {M.}~\bibnamefont {Mogstad}},\ and\ \bibinfo
  {author} {\bibfnamefont {F.}~\bibnamefont {Tintelnot}},\ }\href@noop {}
  {\bibfield  {journal} {\bibinfo  {journal} {The Review of Economic Studies}\
  }\textbf {\bibinfo {volume} {88}},\ \bibinfo {pages} {643} (\bibinfo {year}
  {2021})}\BibitemShut {NoStop}%
\bibitem [{\citenamefont {Bernard}\ \emph {et~al.}(2019)\citenamefont
  {Bernard}, \citenamefont {Moxnes},\ and\ \citenamefont
  {Saito}}]{bernard2019production}%
  \BibitemOpen
  \bibfield  {author} {\bibinfo {author} {\bibfnamefont {A.~B.}\ \bibnamefont
  {Bernard}}, \bibinfo {author} {\bibfnamefont {A.}~\bibnamefont {Moxnes}},\
  and\ \bibinfo {author} {\bibfnamefont {Y.~U.}\ \bibnamefont {Saito}},\
  }\href@noop {} {\bibfield  {journal} {\bibinfo  {journal} {Journal of
  Political Economy}\ }\textbf {\bibinfo {volume} {127}},\ \bibinfo {pages}
  {639} (\bibinfo {year} {2019})}\BibitemShut {NoStop}%
\bibitem [{\citenamefont {Squartini}\ \emph {et~al.}(2018)\citenamefont
  {Squartini}, \citenamefont {Caldarelli}, \citenamefont {Cimini},
  \citenamefont {Gabrielli},\ and\ \citenamefont
  {Garlaschelli}}]{squartini2018reconstruction}%
  \BibitemOpen
  \bibfield  {author} {\bibinfo {author} {\bibfnamefont {T.}~\bibnamefont
  {Squartini}}, \bibinfo {author} {\bibfnamefont {G.}~\bibnamefont
  {Caldarelli}}, \bibinfo {author} {\bibfnamefont {G.}~\bibnamefont {Cimini}},
  \bibinfo {author} {\bibfnamefont {A.}~\bibnamefont {Gabrielli}},\ and\
  \bibinfo {author} {\bibfnamefont {D.}~\bibnamefont {Garlaschelli}},\
  }\href@noop {} {\bibfield  {journal} {\bibinfo  {journal} {Physics reports}\
  }\textbf {\bibinfo {volume} {757}},\ \bibinfo {pages} {1} (\bibinfo {year}
  {2018})}\BibitemShut {NoStop}%
\bibitem [{\citenamefont {Mungo}\ \emph {et~al.}(2024)\citenamefont {Mungo},
  \citenamefont {Brintrup}, \citenamefont {Garlaschelli},\ and\ \citenamefont
  {Lafond}}]{mungo2023reconstructing}%
  \BibitemOpen
  \bibfield  {author} {\bibinfo {author} {\bibfnamefont {L.}~\bibnamefont
  {Mungo}}, \bibinfo {author} {\bibfnamefont {A.}~\bibnamefont {Brintrup}},
  \bibinfo {author} {\bibfnamefont {D.}~\bibnamefont {Garlaschelli}},\ and\
  \bibinfo {author} {\bibfnamefont {F.}~\bibnamefont {Lafond}},\ }\href@noop {}
  {\bibfield  {journal} {\bibinfo  {journal} {Journal of Physics: Complexity}\
  }\textbf {\bibinfo {volume} {5}},\ \bibinfo {pages} {012001} (\bibinfo {year}
  {2024})}\BibitemShut {NoStop}%
\bibitem [{\citenamefont {Campajola}\ \emph {et~al.}(2021)\citenamefont
  {Campajola}, \citenamefont {Lillo}, \citenamefont {Mazzarisi},\ and\
  \citenamefont {Tantari}}]{campajola2021equivalence}%
  \BibitemOpen
  \bibfield  {author} {\bibinfo {author} {\bibfnamefont {C.}~\bibnamefont
  {Campajola}}, \bibinfo {author} {\bibfnamefont {F.}~\bibnamefont {Lillo}},
  \bibinfo {author} {\bibfnamefont {P.}~\bibnamefont {Mazzarisi}},\ and\
  \bibinfo {author} {\bibfnamefont {D.}~\bibnamefont {Tantari}},\ }\href@noop
  {} {\bibfield  {journal} {\bibinfo  {journal} {Journal of Statistical
  Mechanics: Theory and Experiment}\ }\textbf {\bibinfo {volume} {2021}},\
  \bibinfo {pages} {033412} (\bibinfo {year} {2021})}\BibitemShut {NoStop}%
\bibitem [{\citenamefont {Mungo}\ and\ \citenamefont
  {Moran}(2023)}]{mungo2023revealing}%
  \BibitemOpen
  \bibfield  {author} {\bibinfo {author} {\bibfnamefont {L.}~\bibnamefont
  {Mungo}}\ and\ \bibinfo {author} {\bibfnamefont {J.}~\bibnamefont {Moran}},\
  }\href@noop {} {\bibfield  {journal} {\bibinfo  {journal} {arXiv preprint
  arXiv:2302.09906}\ } (\bibinfo {year} {2023})}\BibitemShut {NoStop}%
\bibitem [{\citenamefont {Reisch}\ \emph {et~al.}(2021)\citenamefont {Reisch},
  \citenamefont {Heiler}, \citenamefont {Diem},\ and\ \citenamefont
  {Thurner}}]{reisch2021inferring}%
  \BibitemOpen
  \bibfield  {author} {\bibinfo {author} {\bibfnamefont {T.}~\bibnamefont
  {Reisch}}, \bibinfo {author} {\bibfnamefont {G.}~\bibnamefont {Heiler}},
  \bibinfo {author} {\bibfnamefont {C.}~\bibnamefont {Diem}},\ and\ \bibinfo
  {author} {\bibfnamefont {S.}~\bibnamefont {Thurner}},\ }\href@noop {}
  {\bibfield  {journal} {\bibinfo  {journal} {arXiv preprint arXiv:2110.05625}\
  } (\bibinfo {year} {2021})}\BibitemShut {NoStop}%
\bibitem [{\citenamefont {Mungo}\ \emph {et~al.}(2023)\citenamefont {Mungo},
  \citenamefont {Lafond}, \citenamefont {Astudillo-Est{\'e}vez},\ and\
  \citenamefont {Farmer}}]{mungo2022reconstructing}%
  \BibitemOpen
  \bibfield  {author} {\bibinfo {author} {\bibfnamefont {L.}~\bibnamefont
  {Mungo}}, \bibinfo {author} {\bibfnamefont {F.}~\bibnamefont {Lafond}},
  \bibinfo {author} {\bibfnamefont {P.}~\bibnamefont {Astudillo-Est{\'e}vez}},\
  and\ \bibinfo {author} {\bibfnamefont {J.~D.}\ \bibnamefont {Farmer}},\
  }\href@noop {} {\bibfield  {journal} {\bibinfo  {journal} {Journal of
  Economic Dynamics and Control}\ }\textbf {\bibinfo {volume} {148}},\ \bibinfo
  {pages} {104607} (\bibinfo {year} {2023})}\BibitemShut {NoStop}%
\bibitem [{\citenamefont {Ialongo}\ \emph {et~al.}(2022)\citenamefont
  {Ialongo}, \citenamefont {de~Valk}, \citenamefont {Marchese}, \citenamefont
  {Jansen}, \citenamefont {Zmarrou}, \citenamefont {Squartini},\ and\
  \citenamefont {Garlaschelli}}]{ialongo2022recon}%
  \BibitemOpen
  \bibfield  {author} {\bibinfo {author} {\bibfnamefont {L.~N.}\ \bibnamefont
  {Ialongo}}, \bibinfo {author} {\bibfnamefont {C.}~\bibnamefont {de~Valk}},
  \bibinfo {author} {\bibfnamefont {E.}~\bibnamefont {Marchese}}, \bibinfo
  {author} {\bibfnamefont {F.}~\bibnamefont {Jansen}}, \bibinfo {author}
  {\bibfnamefont {H.}~\bibnamefont {Zmarrou}}, \bibinfo {author} {\bibfnamefont
  {T.}~\bibnamefont {Squartini}},\ and\ \bibinfo {author} {\bibfnamefont
  {D.}~\bibnamefont {Garlaschelli}},\ }\href@noop {} {\bibfield  {journal}
  {\bibinfo  {journal} {Scientific Reports}\ }\textbf {\bibinfo {volume}
  {12}},\ \bibinfo {pages} {1} (\bibinfo {year} {2022})}\BibitemShut {NoStop}%
\bibitem [{\citenamefont {Brintrup}\ \emph {et~al.}(2018)\citenamefont
  {Brintrup}, \citenamefont {Wichmann}, \citenamefont {Woodall}, \citenamefont
  {McFarlane}, \citenamefont {Nicks},\ and\ \citenamefont
  {Krechel}}]{brintrup2018predicting}%
  \BibitemOpen
  \bibfield  {author} {\bibinfo {author} {\bibfnamefont {A.}~\bibnamefont
  {Brintrup}}, \bibinfo {author} {\bibfnamefont {P.}~\bibnamefont {Wichmann}},
  \bibinfo {author} {\bibfnamefont {P.}~\bibnamefont {Woodall}}, \bibinfo
  {author} {\bibfnamefont {D.}~\bibnamefont {McFarlane}}, \bibinfo {author}
  {\bibfnamefont {E.}~\bibnamefont {Nicks}},\ and\ \bibinfo {author}
  {\bibfnamefont {W.}~\bibnamefont {Krechel}},\ }\href@noop {} {\bibfield
  {journal} {\bibinfo  {journal} {Complexity}\ }\textbf {\bibinfo {volume}
  {2018}} (\bibinfo {year} {2018})}\BibitemShut {NoStop}%
\bibitem [{\citenamefont {Kosasih}\ and\ \citenamefont
  {Brintrup}(2021)}]{kosasih2021machine}%
  \BibitemOpen
  \bibfield  {author} {\bibinfo {author} {\bibfnamefont {E.~E.}\ \bibnamefont
  {Kosasih}}\ and\ \bibinfo {author} {\bibfnamefont {A.}~\bibnamefont
  {Brintrup}},\ }\href@noop {} {\bibfield  {journal} {\bibinfo  {journal}
  {International Journal of Production Research}\ ,\ \bibinfo {pages} {1}}
  (\bibinfo {year} {2021})}\BibitemShut {NoStop}%
\bibitem [{\citenamefont {Bacilieri}\ and\ \citenamefont
  {Austudillo-Estevez}(2023)}]{bacilieri2023recon}%
  \BibitemOpen
  \bibfield  {author} {\bibinfo {author} {\bibfnamefont {A.}~\bibnamefont
  {Bacilieri}}\ and\ \bibinfo {author} {\bibfnamefont {P.}~\bibnamefont
  {Austudillo-Estevez}},\ }\href@noop {} {\bibfield  {journal} {\bibinfo
  {journal} {arXiv preprint arXiv:2304.00081}\ } (\bibinfo {year}
  {2023})}\BibitemShut {NoStop}%
\bibitem [{\citenamefont {Garuccio}\ \emph {et~al.}(2023)\citenamefont
  {Garuccio}, \citenamefont {Lalli},\ and\ \citenamefont
  {Garlaschelli}}]{garuccio2020multiscale}%
  \BibitemOpen
  \bibfield  {author} {\bibinfo {author} {\bibfnamefont {E.}~\bibnamefont
  {Garuccio}}, \bibinfo {author} {\bibfnamefont {M.}~\bibnamefont {Lalli}},\
  and\ \bibinfo {author} {\bibfnamefont {D.}~\bibnamefont {Garlaschelli}},\
  }\href@noop {} {\bibfield  {journal} {\bibinfo  {journal} {Physical Review
  Research}\ }\textbf {\bibinfo {volume} {5}},\ \bibinfo {pages} {043101}
  (\bibinfo {year} {2023})}\BibitemShut {NoStop}%
\bibitem [{\citenamefont {Lalli}\ and\ \citenamefont
  {Garlaschelli}(2024)}]{lalli2024geometry}%
  \BibitemOpen
  \bibfield  {author} {\bibinfo {author} {\bibfnamefont {M.}~\bibnamefont
  {Lalli}}\ and\ \bibinfo {author} {\bibfnamefont {D.}~\bibnamefont
  {Garlaschelli}},\ }\href@noop {} {\bibfield  {journal} {\bibinfo  {journal}
  {arXiv preprint arXiv:2403.00235}\ } (\bibinfo {year} {2024})}\BibitemShut
  {NoStop}%
\bibitem [{\citenamefont {Mattsson}\ \emph {et~al.}(2021)\citenamefont
  {Mattsson}, \citenamefont {Takes}, \citenamefont {Heemskerk}, \citenamefont
  {Diks}, \citenamefont {Buiten}, \citenamefont {Faber},\ and\ \citenamefont
  {Sloot}}]{mattsson2021functional}%
  \BibitemOpen
  \bibfield  {author} {\bibinfo {author} {\bibfnamefont {C.~E.}\ \bibnamefont
  {Mattsson}}, \bibinfo {author} {\bibfnamefont {F.~W.}\ \bibnamefont {Takes}},
  \bibinfo {author} {\bibfnamefont {E.~M.}\ \bibnamefont {Heemskerk}}, \bibinfo
  {author} {\bibfnamefont {C.}~\bibnamefont {Diks}}, \bibinfo {author}
  {\bibfnamefont {G.}~\bibnamefont {Buiten}}, \bibinfo {author} {\bibfnamefont
  {A.}~\bibnamefont {Faber}},\ and\ \bibinfo {author} {\bibfnamefont {P.~M.}\
  \bibnamefont {Sloot}},\ }\href@noop {} {\bibfield  {journal} {\bibinfo
  {journal} {Frontiers in big Data}\ }\textbf {\bibinfo {volume} {4}},\
  \bibinfo {pages} {666712} (\bibinfo {year} {2021})}\BibitemShut {NoStop}%
\bibitem [{\citenamefont {Di~Vece}\ \emph {et~al.}(2024)\citenamefont
  {Di~Vece}, \citenamefont {Pijpers},\ and\ \citenamefont
  {Garlaschelli}}]{di2024commodity}%
  \BibitemOpen
  \bibfield  {author} {\bibinfo {author} {\bibfnamefont {M.}~\bibnamefont
  {Di~Vece}}, \bibinfo {author} {\bibfnamefont {F.~P.}\ \bibnamefont
  {Pijpers}},\ and\ \bibinfo {author} {\bibfnamefont {D.}~\bibnamefont
  {Garlaschelli}},\ }\href@noop {} {\bibfield  {journal} {\bibinfo  {journal}
  {Scientific Reports}\ }\textbf {\bibinfo {volume} {14}},\ \bibinfo {pages}
  {3625} (\bibinfo {year} {2024})}\BibitemShut {NoStop}%
\bibitem [{\citenamefont {Bernard}\ and\ \citenamefont
  {Zi}(2022)}]{bernard2022sparse}%
  \BibitemOpen
  \bibfield  {author} {\bibinfo {author} {\bibfnamefont {A.~B.}\ \bibnamefont
  {Bernard}}\ and\ \bibinfo {author} {\bibfnamefont {Y.}~\bibnamefont {Zi}},\
  }\href@noop {} {\emph {\bibinfo {title} {Sparse production networks}}},\
  \bibinfo {type} {Tech. Rep.}\ (\bibinfo  {institution} {National Bureau of
  Economic Research},\ \bibinfo {year} {2022})\BibitemShut {NoStop}%
\bibitem [{\citenamefont {Milocco}\ \emph {et~al.}(2024)\citenamefont
  {Milocco}, \citenamefont {Jansen},\ and\ \citenamefont
  {Garlaschelli}}]{milocco2024multi}%
  \BibitemOpen
  \bibfield  {author} {\bibinfo {author} {\bibfnamefont {R.}~\bibnamefont
  {Milocco}}, \bibinfo {author} {\bibfnamefont {F.}~\bibnamefont {Jansen}},\
  and\ \bibinfo {author} {\bibfnamefont {D.}~\bibnamefont {Garlaschelli}},\
  }\href@noop {} {\bibfield  {journal} {\bibinfo  {journal} {arXiv preprint
  arXiv:2412.04354}\ } (\bibinfo {year} {2024})}\BibitemShut {NoStop}%
\bibitem [{\citenamefont {{Council of European Union}}(1993)}]{kau1993}%
  \BibitemOpen
  \bibfield  {author} {\bibinfo {author} {\bibnamefont {{Council of European
  Union}}},\ }\href@noop {} {\bibinfo {title} {Regulation (ec) no 696/1993 of
  15 march 1993 on the statistical units for the observation and analysis of
  the production system in the community}} (\bibinfo {year} {1993})\BibitemShut
  {NoStop}%
\bibitem [{\citenamefont {Chung}\ and\ \citenamefont
  {Lu}(2002)}]{chung2002connected}%
  \BibitemOpen
  \bibfield  {author} {\bibinfo {author} {\bibfnamefont {F.}~\bibnamefont
  {Chung}}\ and\ \bibinfo {author} {\bibfnamefont {L.}~\bibnamefont {Lu}},\
  }\href@noop {} {\bibfield  {journal} {\bibinfo  {journal} {Annals of
  combinatorics}\ }\textbf {\bibinfo {volume} {6}},\ \bibinfo {pages} {125}
  (\bibinfo {year} {2002})}\BibitemShut {NoStop}%
\bibitem [{\citenamefont {Carvalho}(2014)}]{carvalho2014micro}%
  \BibitemOpen
  \bibfield  {author} {\bibinfo {author} {\bibfnamefont {V.~M.}\ \bibnamefont
  {Carvalho}},\ }\href@noop {} {\bibfield  {journal} {\bibinfo  {journal}
  {Journal of Economic Perspectives}\ }\textbf {\bibinfo {volume} {28}},\
  \bibinfo {pages} {23} (\bibinfo {year} {2014})}\BibitemShut {NoStop}%
\end{thebibliography}%

\newpage
\appendix
\section{Supplementary Information}
\subsection{Data}
The analysis has been conducted on production networks at firm level extracted from payments data of two major Dutch financial institutions. The data was provided to us by ABN AMRO Bank N.V. (ABN) and ING Bank N.V. (ING). Note that for privacy reasons the data was not shared but the analysis was conducted in parallel on the two datasets with only the resulting figures being shared for the writing of this paper. The data is mostly composed of SEPA transactions between accounts of the clients of the banks. These transactions are then used to construct a directed network with the total flows for the year 2022. For clarity we will mostly show results concerning the ABN dataset and refer to the Supplementary Information for the duplicate result when useful. Note that the direction of the connections has been chosen opposite the flow of money to reflect instead the movement of goods. In this datasets we do not unfortunately have access to details on the products being exchanged between firms. Therefore, we are using the NACE industrial classification of the producing node as a proxy for the product classification of each link. Therefore to construct the embedding of each firm we use the in and out strength by NACE code. This is of course a much coarser representation of firms then could be obtained from the knowledge of sales and use grouped by CPA category or more refined product classifications. The approach we have developed is however more general and can be adapted easily to the available data. For further details on the construction of the networks we refer to \cite{ialongo2022recon}. 

\subsection{Multi-scale model derivation}
Given a network with $N$ nodes and $L$ edges between them, we define an aggregated node as a collection of these original nodes. An edge will exist at a given aggregation level $l$ if and only if at least one edge existed from the set of nodes belonging to aggregate node $i$ going to any vertex belonging to aggregate node $j$. Formally,
\begin{equation} \label{eq:agg_edge}
    a_{i_{l+1}j_{l+1}}^{(l+1)}= 1 - \prod_{i_l \in i_{l+1}}\prod_{j_l \in j_{l+1}} (1 - a_{i_lj_l}^{(l)}) \quad .
\end{equation}
The multi-scale model is obtained from the invariance requirement which demands that we may generate an aggregated adjacency matrix $\bm{A}^{(l+1)}$ either directly given the probability $P(\bm{A}^{(l+1)} | \bm{\Theta}^{(l+1)})$ or indirectly by first generating a graph at a lower level of aggregation $l$ according to probability $P(\bm{A}^{(l)} | \bm{\Theta}^{(l)})$ and then aggregating according to equation \eqref{eq:agg_edge}. Note that $\bm{\Theta}^{(l)}$ denotes the parameters of the model at the specified aggregation level $(l)$. Two assumptions are necessary in order to obtain the functional form of the model. First we assume that the edges are independent of each other. We can now simplify notation by writing $p_{i_{l}j_{l}}^{(l)} \coloneqq  P(a_{i_{l}j_{l}}^{(l)} = 1 | \bm{\Theta}^{(l)})$. Our invariance requirement can now be formulated as
\begin{equation} \label{eq:inv_requirement}
    1 - p_{i_{l+1}j_{l+1}}^{(l+1)} = \prod_{i_l \in i_{l+1}} \prod_{j_l \in j_{l+1}} (1 - p_{i_l j_l}^{(l)}) \quad .
\end{equation}
The second assumption we require is the additivity of the parameters. For a general number of node specific parameters, it can be formulated as such:
\begin{align}
    \ln{(1 - p_{i_{l+1}j_{l+1}}^{(l+1)})} &= f \left( \bm{\theta}_{i_{l+1}}, \bm{\theta}_{j_{l+1}} \right) \\
    \ln{(1 - p_{i_lj_l}^{(l)})} &= f \left( \bm{\theta}_{i_{l}}, \bm{\theta}_{j_{l+1}} \right) \\
    \bm{\theta}_{i_{l+1}} &= \sum_{i_l \in i_{l+1}} \bm{\theta}_{i_l} \ , \quad \quad \forall i
\end{align}
where $\bm{\theta}_{i_l}$ is a $M$ dimensional vector. Using equation \eqref{eq:inv_requirement} we thus obtain that the functional $f$ must satisfy
\begin{equation}
    f \left( \sum_{i_l \in i_{l+1}} \bm{\theta}_{i_l}, \sum_{j_l \in j_{l+1}} \bm{\theta}_{j_l} \right) = \sum_{i_l \in i_{l+1}} \sum_{j_l \in j_{l+1}} f \left( \bm{\theta}_{i_l}, \bm{\theta}_{j_l} \right)
\end{equation}

This implies that $f$ must be bilinear in its arguments and we may write it in its matrix form as
\begin{equation}
    f(\bm{x}, \bm{y}) = \bm{x}^T \bm{B} \bm{y}
\end{equation}
where $B$ is a $M\times M$ matrix. We can see that the scale-invariance requirement under the additivity of the parameters implies that the functional form of the probability of each edge must be given by
\begin{equation}
    \ln{(1 - p_{i_{l}j_{l}})} = -{\bm{\theta}_{i_{l+1}}}^T \bm{B} \bm{\theta}_{j_{l+1}}
\end{equation}
where we have chosen to add the minus sign such that the constraint $p_{ij} \in [0,1]$ ensures that the parameters are all positive. This implies 
\begin{equation}
    p_{i_{l}j_{l}} = 1 - e^{-{\bm{\theta}_{i_{l}}}^T \bm{B} \bm{\theta}_{j_{l}}} 
\end{equation}
is the functional form for the probability of connection for any aggregation. Note that the aggregation is entirely arbitrary, that is we did not place any restriction on equation \eqref{eq:agg_edge} on which nodes must belong where other than the requirement that each edge belongs only to one group at each level $(l)$.

\subsection{Random allocation model}
Let us consider the multi-scale formula with a single dimension given by the size of the company. We can approximate the value by taking the first element of the Taylor series expansion of the exponential function, giving us the following:
$$
p_{ij} = 1 -e^{- \delta s_i^\text{out} s_j^\text{in}} = 1 - \left(e^{- \alpha s_i^\text{out}} \right) ^ {\beta s_j^\text{in}} \approx 
1 - \left(1 - \alpha s_i^\text{out} \right) ^ {\beta s_j^\text{in}}
$$
provided that $\alpha s_i^\text{out}$ is sufficiently small. Note that we have set $\alpha \beta = \delta$. In the random allocation model of \cite{bernard2022sparse}, the probability of connection between two nodes has the same functional form as the above approximation of the multi-scale model. The difference is given only by the values of $\alpha$ and $\beta$: in the random allocation model, $\alpha = \beta$ are equal to the one over the total size of all the market, such that the terms become the relative size of the seller and buyer. Indeed the model is given by $p_{ij}^{RA}=1-\left(1-s_i\right)^{s_j}$ where $s_i$ is the percentage of the sales of $i$ with respect to the total and $s_j$ is the same metric for the buyer. 
We note then a few important differences between this approach and our own. First, while our approach is well defined for any level of link density, the balls and bins inspired model by \cite{bernard2022sparse} is only defined at one particular value. We also note that this means that while our model can be adapted to any size distribution and still obtain any value of density, this is not the case for the RA model. Of course, if one wants to obtain an endogenous density as in Bernard and Zi's work, this can still be recovered by applying the same normalizing constant $\delta = \frac{1}{\left(\sum_i s_i^\text{out}\right)^2} = \frac{1}{\left(\sum_i s_i^\text{in}\right)^2}$.

\subsection{Rest of the world embeddings performance}
\begin{figure}
    \centering
    \subfloat[\label{fig:row_s_icdf}]{%
    \includegraphics[width=0.49\linewidth]{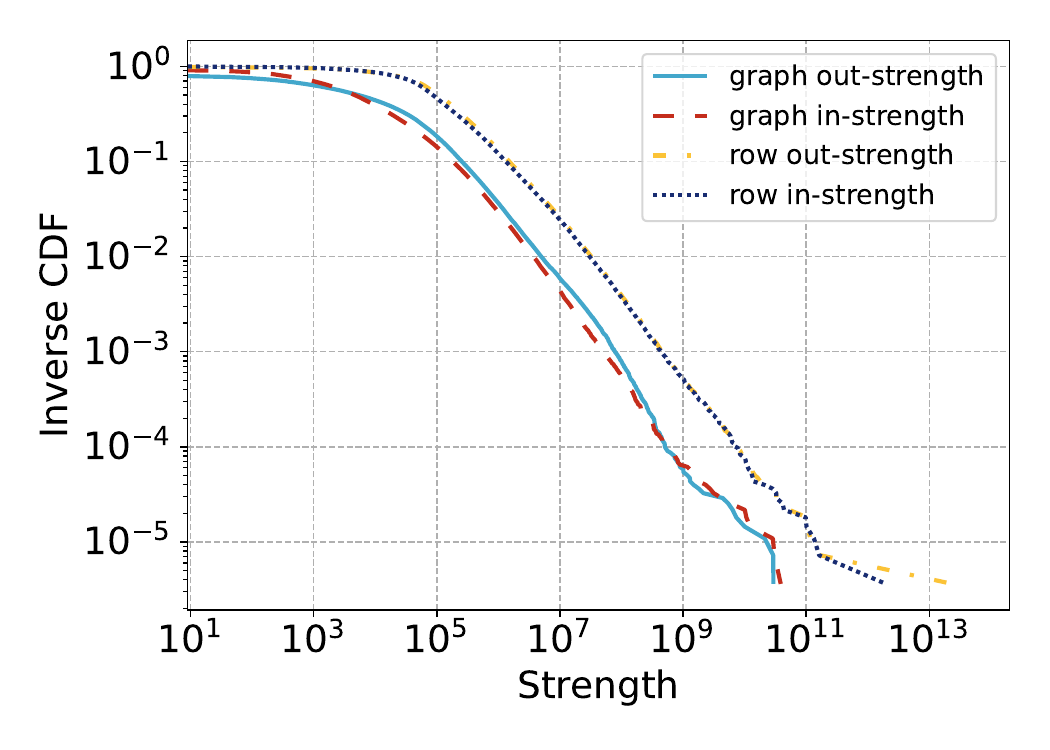}
    }\hfil
    \subfloat[\label{fig:row_ks}]{%
    \includegraphics[width=0.49\textwidth]{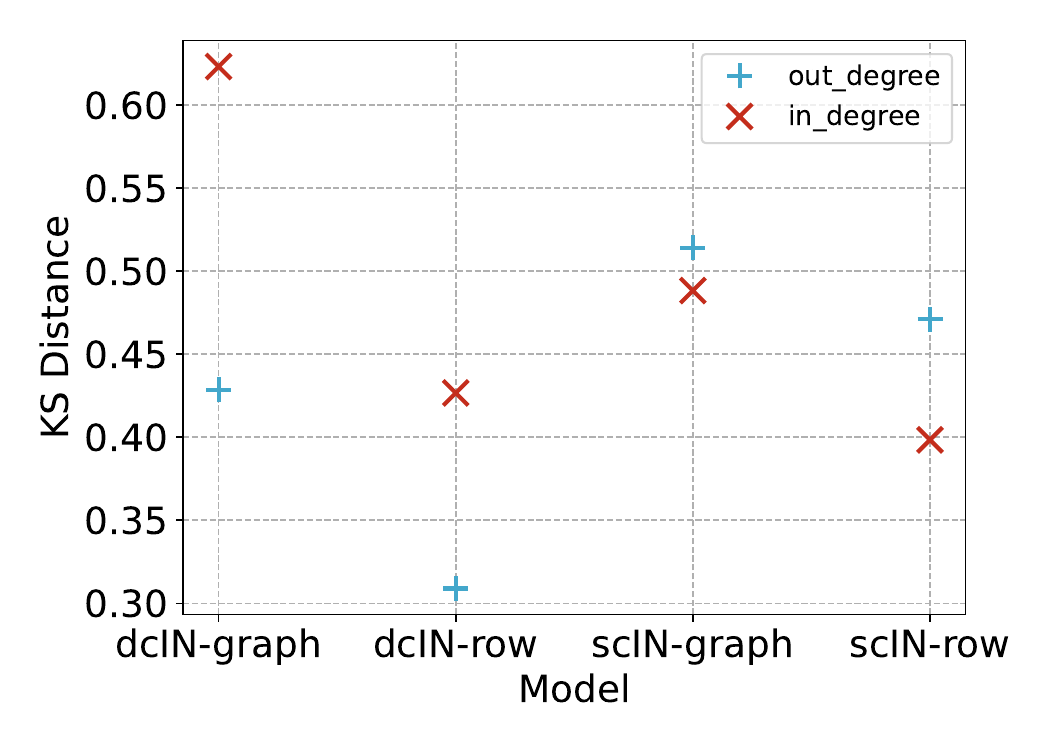}
    }\hfil \vspace{-1em}
    \subfloat[\label{fig:row_abs_err}]{%
    \includegraphics[width=0.49\textwidth]{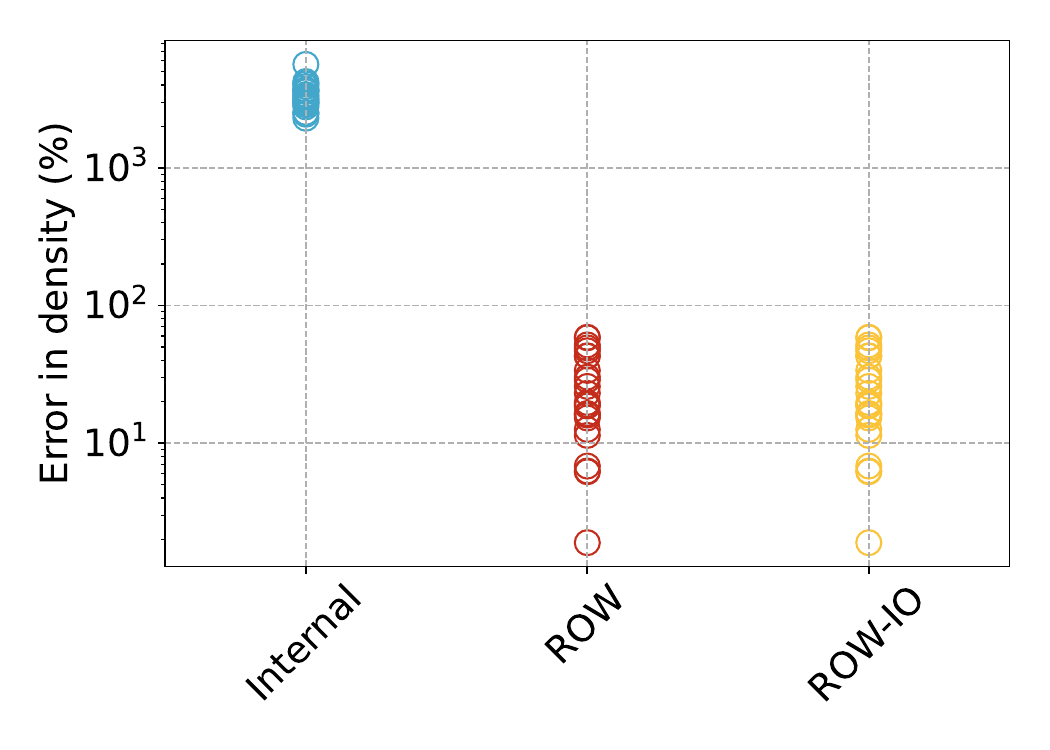}
    }\hfil
    \subfloat[\label{fig:row_d_err}]{%
    \includegraphics[width=0.49\textwidth]{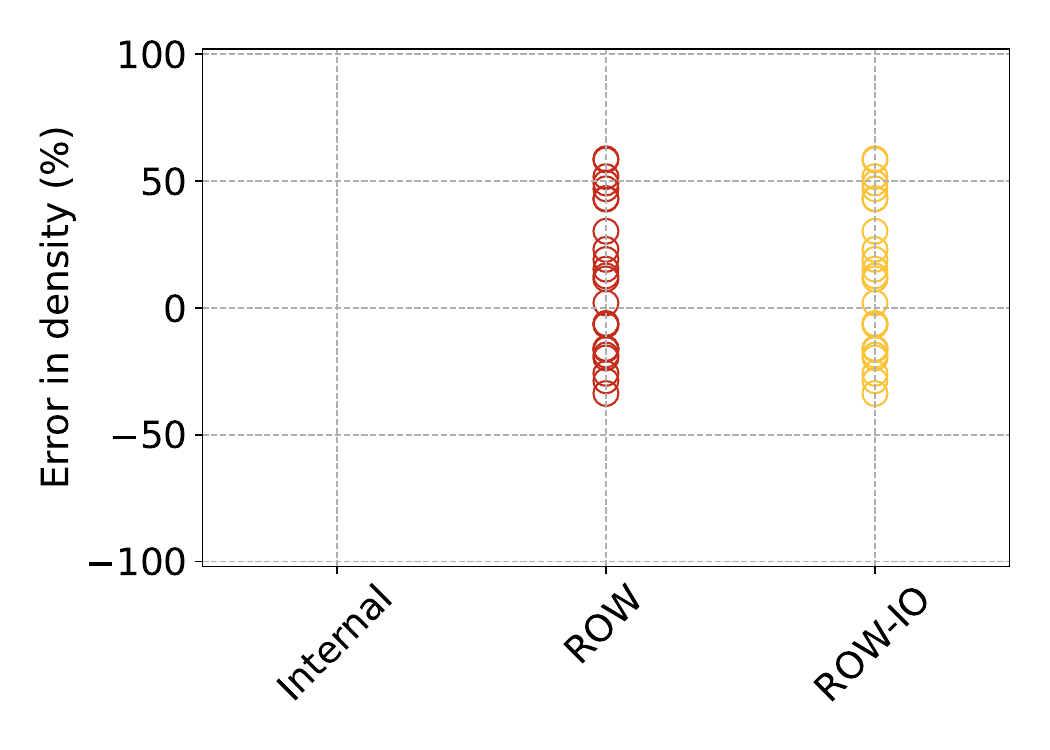}%
    }\hfil \vspace{-1em}
    \subfloat[\label{fig:params_b}]{%
    \includegraphics[width=0.49\textwidth]{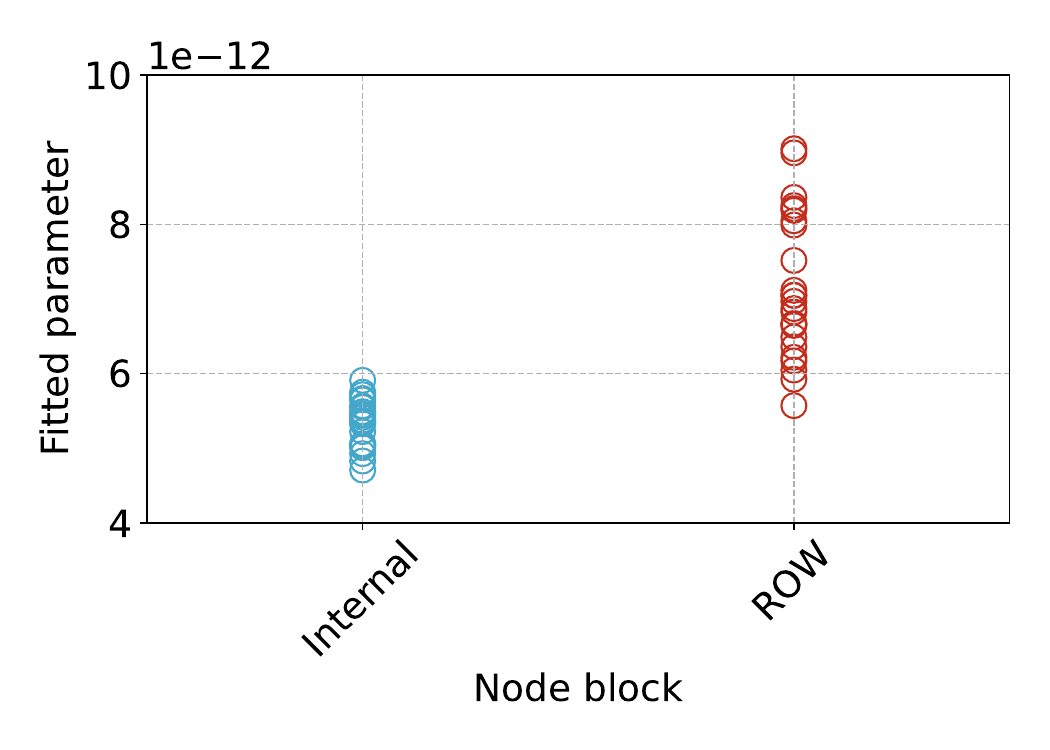}
    }\hfil
    \subfloat[\label{fig:row_d_err_b}]{%
    \includegraphics[width=0.49\textwidth]{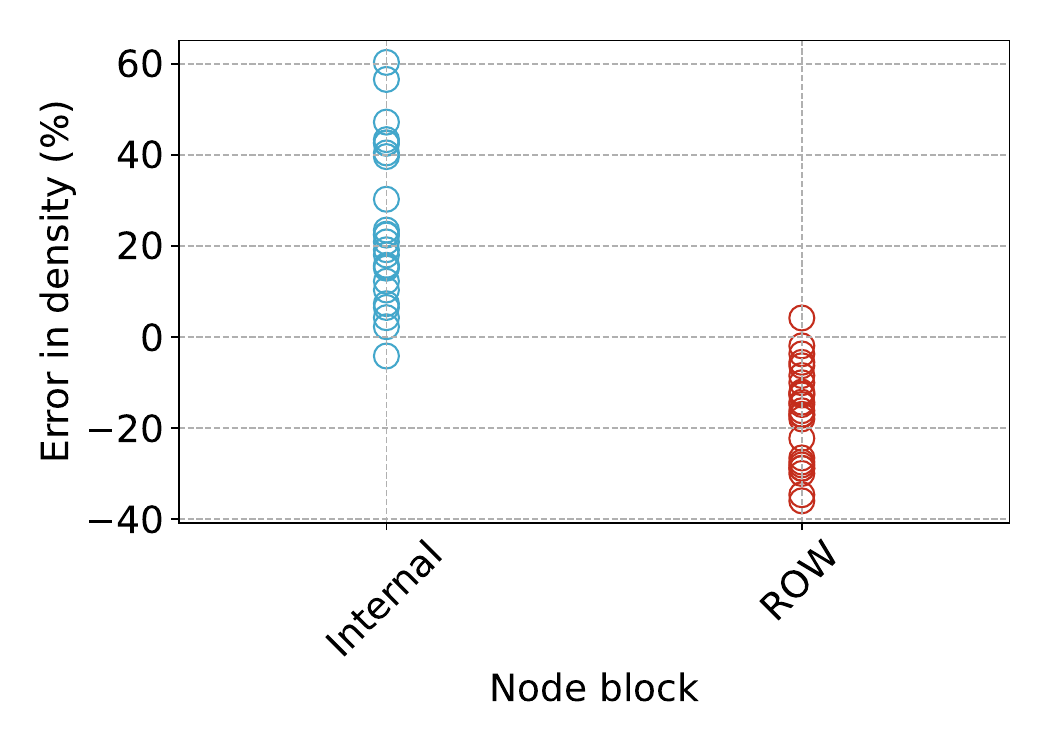}%
    }
    \caption{\scriptsize Complementary cumulative distribution of the in- and out-strengths including or excluding the ROW node (a) and the Kolmogorov-Smirnoff distance between the empirical and reconstructed degree distributions computed with or without the ROW (b). Percentage error in the reconstructed link density (d) and its absolute value (c) for the complete graph with the given the estimation method. Note that the internal value is not visible as it is outside of the 100\% range. We report here also a similar result for the ING dataset. In panel (e) we show the estimation of the parameter in the two cases, while in (f) we highlight the effect on the percentage error in the reconstructed link density.}
\end{figure} 
We show here the improvement in reconstruction performance obtained from using the full information available in our datasets about the firm embeddings. In figure \ref{fig:row_s_icdf} we plot the complementary cumulative distribution of the in- and out-strengths if we include or exclude the ROW node. We can see that although the tails have a similar slope, the curves are significantly shifted to the right. What this translates to is a better reconstruction accuracy especially for the low strength nodes. This can be seen quantitatively in the improvement of the Kolmogorov-Smirnoff distance between the empirical and expected degree distributions obtained including or excluding the ROW node as seen in figure \ref{fig:row_ks}. Including the ROW node improves both the out- and in-degree reconstruction for both the stripe and non-stripe versions of the model.

We report here the reconstruction accuracy in terms of link density of the ROW vs Internal estimation. We have introduced here a further case where we assume to have for the unobserved component an aggregated version of the graph based on the industrial classification of the nodes. This scenario is simulating the case when for our ROW node we have access to an input-output table (IO). The added information is used to constrain the possible connections that may exist between firms in the rest-of-the-world node such that they are consistent with the information available in the IO table. The detailed derivation of this constraining are outlined in the supplementary information section \ref{sec:conditional}. For the purposes of this comparison it is important to stress here that although this constraining is not limited to the multi-scale model, it is significantly easier to implement. In figures \ref{fig:row_abs_err} and \ref{fig:row_d_err} we can see that the estimation performance is greatly improved from using the ROW node. We note however that this still results in a significant error. As discussed in the main text, this is to be expected from the fluctuations in density that derive from extracting a subgraph from a network with power law degree distribution. We further note that in the ING dataset the error of the internal estimation is smaller, this is likely due to the fact that the ING dataset contains significantly more nodes. This could result in more stable strengths under different partitions.

\subsection{Firm level information effects}
\begin{figure}
    \centering
    \subfloat[\label{fig:info_dout_icdf}]{%
    \includegraphics[width=0.49\textwidth]{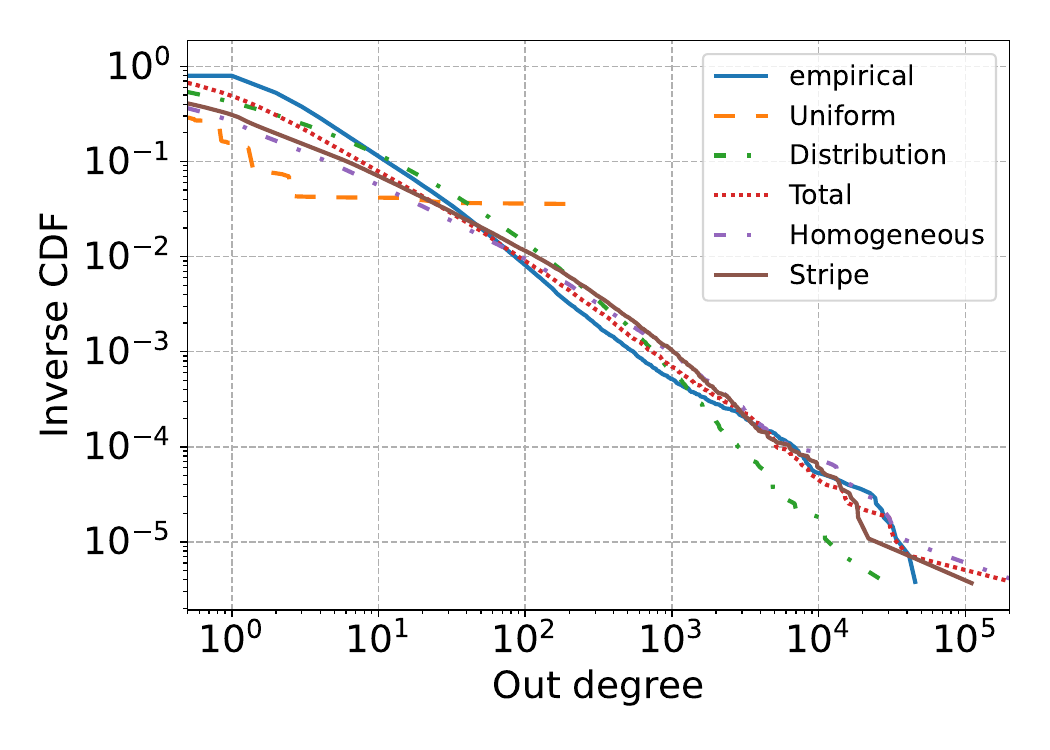}
    }\hfil
    \subfloat[\label{fig:info_din_icdf}]{%
    \includegraphics[width=0.49\textwidth]{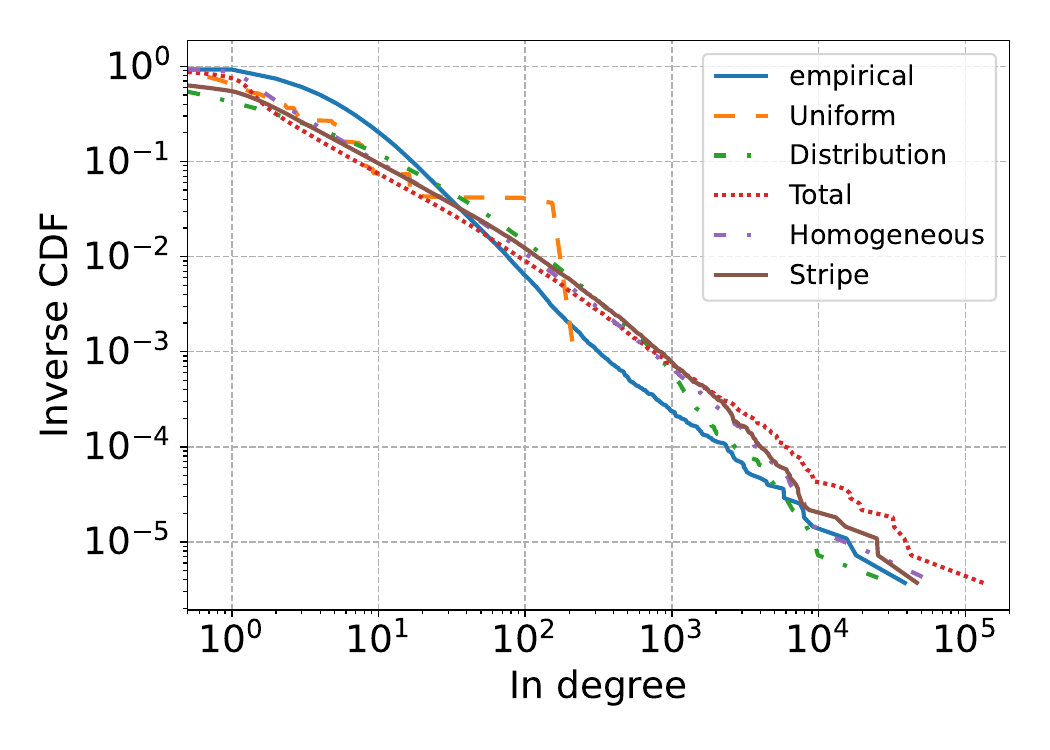}%
    }\hfil \vspace{-1em}
    \subfloat[\label{fig:info_knn_oo}]{%
    \includegraphics[width=0.49\textwidth]{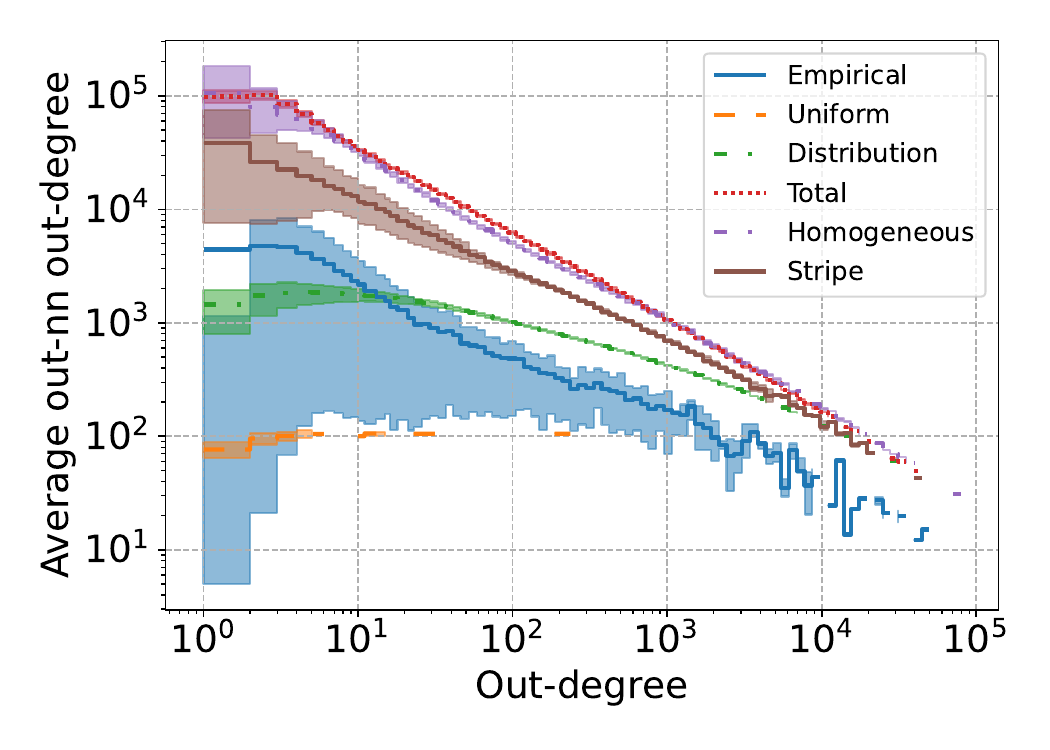}
    }\hfil
    \subfloat[\label{fig:info_knn_ii}]{%
    \includegraphics[width=0.49\textwidth]{sc3_info_knn_oo_mean.pdf}%
    }\hfil \vspace{-1em}
    \subfloat[\label{fig:info_roc}]{%
    \includegraphics[width=0.49\textwidth]{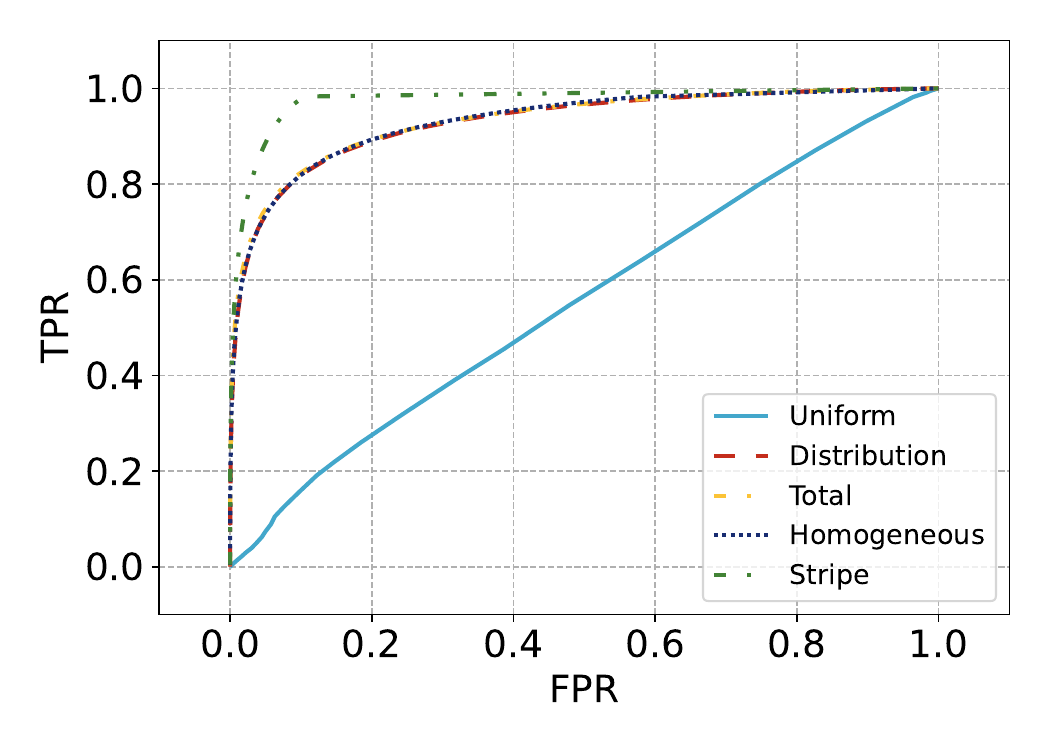}
    }\hfil
    \subfloat[\label{fig:info_infl}]{%
    \includegraphics[width=0.49\textwidth]{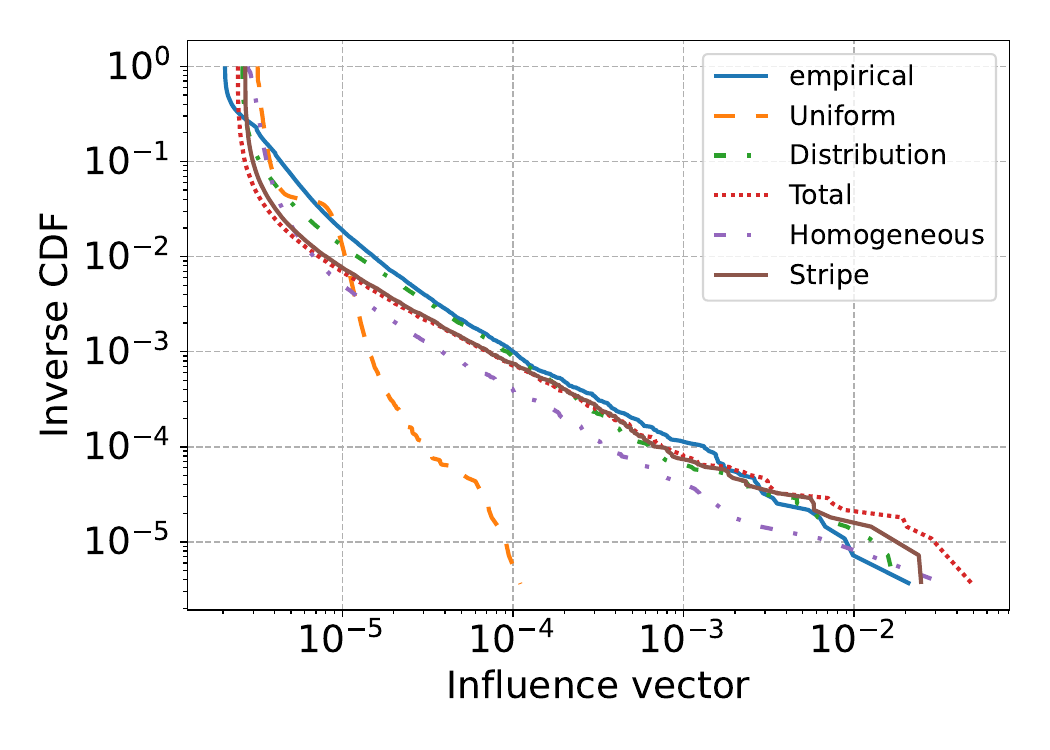}
    }
    \caption{\scriptsize Comparison of reconstruction accuracy for the different cases of firm-level information outlined in the Working with aggregate data section of the main text: complementary cumulative distribution of the in (a) and out (b) degrees; average nearest neighbour out-degree vs out-degree (c) and average nearest neighbour in-degree vs in-degree (d), the shaded area represents the interquartile range for the binned degrees; relative effect of firm level information on the reconstruction accuracy in terms of ROC (e) and Influence vector complementary cumulative distribution (f). }
\end{figure} 

In the main text we have briefly discussed the effect of firm-level information on the reconstruction accuracy of the model. We report here in figures \ref{fig:info_dout_icdf} and \ref{fig:info_din_icdf} the effect on reconstructing the degree distribution. We can see that the uniform case performs poorly as expected, while the distribution case performs adequately in the in-degree but worse in the out-degree tail. Similarly for the average nearest neighbour degree in figures \ref{fig:info_knn_oo} and \ref{fig:info_knn_ii}, we see that firm-level information plays an important role in estimating the correct neighbourhood of the nodes. Note here that these plots have been generated by sampling multiple times from the ensemble and pooling the results to obtain the distribution. The confidence intervals are obtained by binning the degree in a logarithmic fashion. Furthermore, we can see from the ROC in figure \ref{fig:info_roc} that the stripe model outperforms all other models, while the total, the distribution and the homogeneous case all perform similarly. This does not imply that unless the true stripe are known then the model performs poorly, rather this will depend on the quantity of interest. In figure \ref{fig:info_infl} we have shown for example the distribution of the influence vector presented in \cite{carvalho2014micro}. We can clearly see that other than the uniform case all others perform adequately in reproducing the empirical distribution. Note that in the distribution case, since we are assuming the the fitnesses are not known we could perform much worse that this by assigning a high fitness to an empirically low fitness node. To avoid this we assume here to know the true ranking of nodes in terms of their strength, this ensures that we are in the best case scenario. 

\begin{figure}
    \centering
    \includegraphics[width=0.6\textwidth]{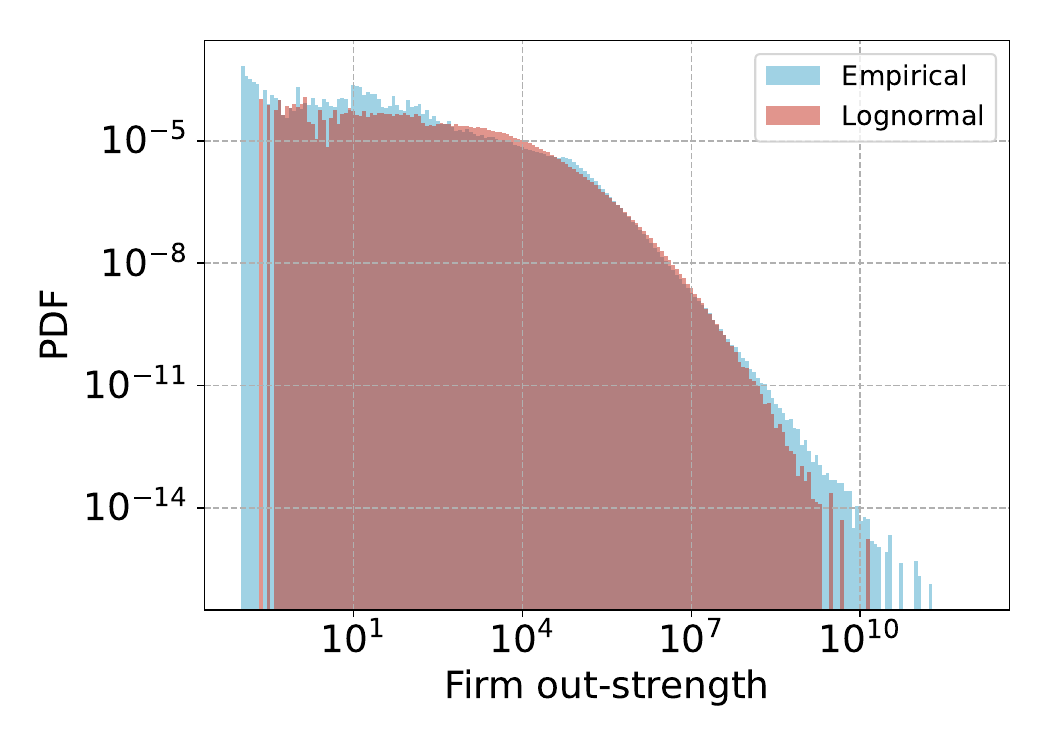}
    \caption{Comparison between the empirical and fitted log-normal probability distributions.}\label{fig:lgn_fit}
\end{figure}

We report in figure \ref{fig:lgn_fit} the comparison of the empirical probability distribution function and the one fitted using a log-normal distribution. The fitted parameters are then used to generate the distribution case in the plots above and in the main text. 

\newpage

\subsection{Undirected networks and accounting for self-loops}
The multi-scale model allows for connection of a node with itself. Even if these do not exist at the lowest level, when no aggregation is present, they can arise through aggregation. We can see that in the model we have presented in the main text the aggregation rule remains the same also for self-loops as all of the $a_{i_l k_l}$ terms in the following equation are independent events:
\begin{align} \label{eq:self_agg_edge}
    a_{i_{l+1}i_{l+1}}^{(l+1)} &= 1 - \prod_{i_l \in i_{l+1}}\prod_{k_l \in i_{l+1}} (1 - a_{i_lk_l}^{(l)})\\
    &= 1 - \left(\prod_{i_l \in i_{l+1}} (1 - a_{i_l i_l}^{(l)})\right) \left( \prod_{i_l \in i_{l+1}}\prod_{\substack{k_l \in i_{l+1} \\  k_l \neq i_l} } (1 - a_{i_lk_l}^{(l)})\right)  \quad .
\end{align}
The functional form of the multi-scale model satisfies the requirement above. In the case of undirected networks however the functional form differs slightly. The undirected stripe multi-scale model is obtained as in the directed case by selecting our parameter vector as $\theta_i^{(l)} \coloneqq \bm{s}_i$ where $\bm{s}_i$ is a column vector with elements $\alpha$ given by $s_{i, \alpha}$ which is a suitable proxy of the size of trading of product $\alpha$ by firm $i$. The value of $s_{i, \alpha}$ could for example be defined as $s_{i, \alpha} \coloneqq s^{\text{out}}_{i_l, \alpha} + s^{\text{in}}_{i_l, \alpha}$. We can now set the matrix $\bm{B} \coloneqq \text{diag}(\bm{\delta})$ and obtain the following functional form:
\begin{equation} \label{eq:p_inv_stripe_undir_pre}
    p_{i_{l}j_{l}} (\bm{\delta}) = p_{j_{l}i_{l}} (\bm{\delta}) = 1 - e^{-\sum_{\alpha} \delta_\alpha s_{i_l, \alpha} s_{j_l, \alpha}}
\end{equation}
We note however that in this undirected case the aggregation rule is now different if we are talking about self-loops or not. Indeed for self-loops we have that the only independent events are now given by
\begin{align} \label{eq:self_agg_edge_undir}
    a_{i_{l+1}i_{l+1}}^{(l+1)} &= 1 - \prod_{i_l \in i_{l+1}}\prod_{\substack{k_l \in i_{l+1} \\  k_l \leq i_l} } (1 - a_{i_lk_l}^{(l)}) \\
    &= 1 - \left(\prod_{i_l \in i_{l+1}} (1 - a_{i_l i_l}^{(l)}) \right) \left(\prod_{i_l \in i_{l+1}}\prod_{\substack{k_l \in i_{l+1} \\  k_l < i_l} } (1 - a_{i_lk_l}^{(l)}) \right) \quad .
\end{align}
We can see that the functional form \eqref{eq:p_inv_stripe_undir_pre} in this case would give the following issue
\begin{align}
    1 - p_{i_{l+1} i_{l+1}} &= e^{-\sum_{\alpha} \delta_\alpha \left( \sum_{i_l \in i_{l+1}} s_{i_l, \alpha} \right) \left(\sum_{k_l \in i_{l+1}} s_{k_l, \alpha}\right)} \\
    &= \prod_{i_l \in i_{l+1}}\prod_{k_l \in i_{l+1}} e^{-\sum_{\alpha} \delta_\alpha s_{i_l, \alpha} s_{k_l, \alpha}} \\
    &= \prod_{i_l \in i_{l+1}}\prod_{k_l \in i_{l+1}} (1 - p_{i_{l} k_{l}}) \\
    &\neq \prod_{i_l \in i_{l+1}}\prod_{\substack{k_l \in i_{l+1} \\  k_l \leq i_l}} (1 - p_{i_{l} k_{l}})
\end{align}
We can fix this by adjusting slightly the functional form to be
\begin{equation} \label{eq:p_inv_stripe_undir}
    p_{i_{l}j_{l}} (\bm{\delta}) = \left\{\begin{array}{ll} 1 - e^{-\sum_{\alpha} \frac{1}{2} \delta_\alpha s_{i_l, \alpha}^2} \quad & \text{if} \ i_{l} = j_{l} \ , \\ 
    1 - e^{-\sum_{\alpha} \delta_\alpha s_{i_l, \alpha} s_{j_l, \alpha}} \quad &\text{otherwise.}
  \end{array} \right. 
\end{equation}
We can now see that we obtain the correct result:
\begin{align}
    1 - p_{i_{l+1} i_{l+1}} &= e^{-\sum_{\alpha} \frac{1}{2} \delta_\alpha \left( \sum_{i_l \in i_{l+1}} s_{i_l, \alpha} \right)^2} \\
    &= e^{-\sum_{\alpha} \frac{1}{2} \delta_\alpha \left(\sum_{i_l \in i_{l+1}} s_{i_l, \alpha}^2 +  \sum_{i_l \in i_{l+1}} \sum_{\substack{k_l \in i_{l+1} \\  k_l \neq i_l}} s_{i_l, \alpha} s_{k_l, \alpha} \right) } \\
    &= e^{-\sum_{\alpha} \frac{1}{2} \delta_\alpha \left(\sum_{i_l \in i_{l+1}} s_{i_l, \alpha}^2 + 2 \sum_{i_l \in i_{l+1}} \sum_{\substack{k_l \in i_{l+1} \\  k_l < i_l}} s_{i_l, \alpha} s_{k_l, \alpha} \right) } \\
    &= \left(\prod_{i_l \in i_{l+1}} e^{-\sum_{\alpha} \frac{1}{2} \delta_\alpha s_{i_l, \alpha}^2} \right) \left( \prod_{i_l \in i_{l+1}}\prod_{\substack{k_l \in i_{l+1} \\  k_l < i_l}} e^{-\sum_{\alpha} \delta_\alpha s_{i_l, \alpha} s_{k_l, \alpha}} \right) \\
    &= \left(\prod_{i_l \in i_{l+1}} 1 - p_{i_{l} i_{l}} \right) \left( \prod_{i_l \in i_{l+1}}\prod_{\substack{k_l \in i_{l+1} \\  k_l < i_l}} 1 - p_{i_{l} k_{l}} \right) \\
    &= \prod_{i_l \in i_{l+1}}\prod_{\substack{k_l \in i_{l+1} \\  k_l \leq i_l}} (1 - p_{i_{l} i_{l}}) \quad .
\end{align}

\subsection{Aggregating product layers}
We note that the functional form of the multi-scale model is not only invariant to node aggregation but also to product aggregation under certain conditions. To see this we must first generalize our model to include the possibility of an edge existing between layers. We define $a_{i_{l}j_{l}} (\alpha_{k}, \beta_{k})$ as the link going from node $i_l$ in layer $\alpha_k$ to node $j_l$ in layer $\beta_k$. Then by using the matrix $\bm{B} \coloneqq \begin{bmatrix} \bm{0} & \bm{\Delta}_k \\ \bm{0} & \bm{0} \end{bmatrix}$ where now $\bm{\Delta}_k$ is an $S_k \times S_k$ matrix with elements $\delta_{\alpha_k, \beta_k}$, we obtain that the functional form for the link probability is given by 
\begin{equation} \label{eq:p_stripe_gen}
    p_{i_{l}j_{l}}^{(k)} (\bm{\delta}) = 1 - e^{-\sum_{\alpha_k} \sum_{\beta_k} \delta_{\alpha_k, \beta_k} s^{\text{out}}_{i_l, \alpha_k} s^{\text{in}}_{j_l, \beta_k}}
\end{equation}
where $k$ denotes the product aggregation level. We have now defined the edges to be dependent on the couple $(\alpha_{k}, \beta_{k})$, such that multiple edges might exist between the same nodes. We can define the independent edge probability as
\begin{equation}\label{eq:p_layer_gen}
    p_{i_{l}j_{l}} (\delta_{\alpha_k, \beta_k}) \coloneqq 1 - e^{- \delta_{\alpha_k, \beta_k} s^{\text{out}}_{i_l, \alpha_k} s^{\text{in}}_{j_l, \beta_k}}
\end{equation}
such that $p_{i_{l}j_{l}}^{(k)} (\bm{\delta}) = 1 - \prod_{\alpha_k} \prod_{\beta_k} (1 - p_{i_{l}j_{l}} (\delta_{\alpha_k, \beta_k}) )$.

If we define an aggregation of products such that $s^{\text{out}}_{i, \alpha_{k+1}} = \sum_{\alpha_{k} \in \alpha_{k+1}} s^{\text{out}}_{i, \alpha_k}$ and $s^{\text{in}}_{i, \alpha_{k+1}} = \sum_{\alpha_{k} \in \alpha_{k+1}} s^{\text{in}}_{i, \alpha_k}$, and we require that
\begin{equation} \label{eq:inv_req_prod}
    1 - p_{i_{l}j_{l}}^{\alpha_{k+1},  \beta_{k+1}} (\delta_{\alpha_{k+1}, \beta_{k+1}}) = \prod_{\alpha_{k} \in \alpha_{k+1}} \prod_{\beta_{k} \in \beta_{k+1}} (1 - p_{i_{l}j_{l}}^{\alpha_k,  \beta_k} (\delta_{\alpha_k, \beta_k})) \quad \forall i_l, j_l
\end{equation}
then the functional form in equation \eqref{eq:p_stripe_gen} respects the invariance provided 
\begin{equation} \label{eq:delta_prod_agg}
    \delta_{\alpha_{k+1}, \beta_{k+1}} = \frac{ \sum_{\alpha_{k} \in \alpha_{k+1}} \sum_{\beta_{k} \in \beta_{k+1}} s^{\text{out}}_{i_l, \alpha_{k}} s^{\text{in}}_{j_l, \beta_{k}} \delta_{\alpha_{k}, \beta_{k}}}{\left(\sum_{\alpha_{k} \in \alpha_{k+1}} s^{\text{out}}_{i_l, \alpha_{k}} \right)\left(\sum_{\beta_{k} \in \beta_{k+1}} s^{\text{in}}_{j_l, \beta_{k}}\right)} \quad \forall i_l, j_l \ .
\end{equation}
We can show this by substituting equation \eqref{eq:p_layer_gen} into the right-hand side of equation \eqref{eq:inv_req_prod}:
\begin{align}
    \text{R.H.S} &= \prod_{\alpha_{k} \in \alpha_{k+1}} \prod_{\beta_{k} \in \beta_{k+1}} (1 - p_{i_{l}j_{l}} (\delta_{\alpha_k, \beta_k})) \\
    &= \prod_{\alpha_{k} \in \alpha_{k+1}} \prod_{\beta_{k} \in \beta_{k+1}} e^{- \delta_{\alpha_k, \beta_k} s^{\text{out}}_{i_l, \alpha_k} s^{\text{in}}_{j_l, \beta_k}} \\
    &= e^{- \sum_{\alpha_{k} \in \alpha_{k+1}} \sum_{\beta_{k} \in \beta_{k+1}} \delta_{\alpha_k, \beta_k} s^{\text{out}}_{i_l, \alpha_k} s^{\text{in}}_{j_l, \beta_k}} \\
    &=  e^{- \frac{\sum_{\alpha_{k} \in \alpha_{k+1}} \sum_{\beta_{k} \in \beta_{k+1}} \delta_{\alpha_k, \beta_k} s^{\text{out}}_{i_l, \alpha_k} s^{\text{in}}_{j_l, \beta_k}}{\left(\sum_{\alpha_{k} \in \alpha_{k+1}} s^{\text{out}}_{i_l, \alpha_{k}}\right) \left( \sum_{\beta_{k} \in \beta_{k+1}} s^{\text{in}}_{j_l, \beta_{k}} \right)} \left(\sum_{\alpha_{k} \in \alpha_{k+1}} s^{\text{out}}_{i_l, \alpha_{k}}\right) \left( \sum_{\beta_{k} \in \beta_{k+1}} s^{\text{in}}_{j_l, \beta_{k}} \right) } \\
    &=  e^{- \delta_{\alpha_{k+1}, \beta_{k+1}} \left(\sum_{\alpha_{k} \in \alpha_{k+1}} s^{\text{out}}_{i_l, \alpha_{k}}\right) \left( \sum_{\beta_{k} \in \beta_{k+1}} s^{\text{in}}_{j_l, \beta_{k}} \right) } = \text{L.H.S} \ .
\end{align}
The added complexity here is that we require equations \eqref{eq:inv_req_prod} and therefore \eqref{eq:delta_prod_agg} hold for all $(i_l, j_l)$ pairs. An obvious case in which this is true is if the parameter is independent of the layer meaning that $\delta_{\alpha_{k+1}, \beta_{k+1}} = \delta$ holds. This is however also a pretty uninteresting as it is essentially the case in which these layer structures play no role in determining the link probability as all layers contribute equally. 

A more interesting case is the one in which we retain the product layer structure and only allow links on the layer rather than between them. In this case we have that the link probability is given by:
\begin{equation}\label{eq:p_layer_stripe}
    p_{i_{l}j_{l}} (\delta_{\alpha_k}) \coloneqq 1 - e^{- \delta_{\alpha_k} s^{\text{out}}_{i_l, \alpha_k} s^{\text{in}}_{j_l, \alpha_k}}
\end{equation}
such that the probability of observing at least one link on any layer between two nodes is given by:
\begin{align}
    p_{i_{l}j_{l}}^{(k)} (\bm{\delta}) &= 1 - \prod_{\alpha_k} (1 - p_{i_{l}j_{l}} (\delta_{\alpha_k}) ) \\
    &= 1 - e^{- \sum_{\alpha_k} \delta_{\alpha_k} s^{\text{out}}_{i_l, \alpha_k} s^{\text{in}}_{j_l, \alpha_k}} \\
    &= 1 - e^{- {\bm{s}_{i_l}^{\text{out}}}^T \bm{D} \bm{s}_{j_l}^\text{in}}
\end{align}
where $\bm{D}$ is a diagonal matrix with elements $\delta_{\alpha_k}$. Given that equation \eqref{eq:p_layer_stripe} is the same as equation \eqref{eq:p_layer_gen} for the case $\alpha_k = \beta_k$, what is the relation between these two models and how do the parameters compare? We can find this by specifying that no link can exist between layers such that $p_{i_{l}j_{l}} (\delta_{\alpha_k, \beta_k}) = 0$,  $\forall \alpha_k \neq \beta_k$ which is easily achieved by setting $\delta_{\alpha_k, \beta_k} = 0$ $\forall \alpha_k \neq \beta_k$ and $\delta_{\alpha_k, \alpha_k} = \delta_{\alpha_k}$ otherwise. Under these conditions the models are equivalent so the requirement for product aggregation now becomes: 
\begin{equation} \label{eq:delta_prod_agg_stripe}
    \delta_{\alpha_{k+1}} = \frac{ \sum_{\alpha_{k} \in \alpha_{k+1}} s^{\text{out}}_{i_l, \alpha_{k}} s^{\text{in}}_{j_l, \alpha_{k}} \delta_{\alpha_{k}}}{\left(\sum_{\alpha_{k} \in \alpha_{k+1}} s^{\text{out}}_{i_l, \alpha_{k}} \right)\left(\sum_{\alpha_{k} \in \alpha_{k+1}} s^{\text{in}}_{j_l, \alpha_{k}}\right)} \quad \forall i_l, j_l \ .
\end{equation}
This relation is, of course, not always true but it can be computed and it allows us to fit the model at different scales. 

\subsection{Estimating the parameters from aggregate data}
When estimating the parameters from an input-output table we usually have that the number of layers and aggregated nodes are the same. This is due to the fact that we have industries as nodes but we are also using the industrial classification of the source vertex of an edge to determine the product layer. As such the in-strength vector of node $i$ will be of dimension equal to the number of industries and be different from zero only when there exist a connection with that sector. As such the matrix of the in-strengths is identical to the weighted adjacency matrix, while the out-strength matrix is a matrix with the total output of each industry on the diagonal and zero otherwise. This construction unfortunately means that for any $\delta_{\alpha_{k}} > 0$ then $p_{i_{l}j_{l}}^{(k)} (\bm{\delta}) > 0$ if there is a node in the aggregated graph and $0$ otherwise. Estimating the model parameters directly is therefore not possible since what maximizes the likelihood and gives the correct density value is setting $\delta_{\alpha_{k}}$ to infinity, giving a likelihood for the observed graph of one. This is not per se wrong, since it is indeed desired that the structure of the input-output table is clearly returned with probability approaching one, but the parameters estimated in this way do not have any information of the density of the firm-level graph.

One possible solution to this issue is to use the aggregation rule described by equation \eqref{eq:delta_prod_agg_stripe}. This then allows us to fit the model using the total strength of each industry but then re-scale the parameter to ensure consistency. We do this by first fitting the model using a single global parameter $\delta$ with the functional $p_{i_l, j_l} = 1 - e^{- \delta \left(\sum_{\alpha_{k} \in \alpha_{k+1}} s^{\text{out}}_{i_l, \alpha_{k}}\right) \left( \sum_{\beta_{k} \in \beta_{k+1}} s^{\text{in}}_{j_l, \beta_{k}} \right)}$. We can then use equation \eqref{eq:delta_prod_agg_stripe} to scale the parameter to get a global $\delta_k$ defined at $k^\text{th}$ level. This is quite simply given by
\begin{equation} \label{eq:delta_disagg_rule}
    \delta_k = \frac{\left(\sum_{\alpha_{k}} s^{\text{out}}_{i_l, \alpha_{k}} \right)\left(\sum_{\alpha_{k}} s^{\text{in}}_{j_l, \alpha_{k}}\right)} {\sum_{\alpha_{k}} s^{\text{out}}_{i_l, \alpha_{k}} s^{\text{in}}_{j_l, \alpha_{k}}} \delta \quad \forall i_l, j_l \ .
\end{equation}
The difficulty with this approach is that we have to ensure this holds for all $(i_l, j_l)$ pairs which can be difficult. One could imagine finding various strategies to computationally find a optimal solution. For the purposes of this work we only highlight the issue, as in the main text we have avoided this problem by fitting the stripe model at a more disaggregated scale.

\subsection{Conditional probability under fine-graining}\label{sec:conditional}
In many cases of practical interest we will not only have access to the fitness variables of the node and the global density for calibration but to some coarse-grained graph as well. In this case, when trying to find the probability distribution over all fined-grained graphs, it is reasonable to require that we only consider graphs compatible with the observed one. This implies we want to reject any configuration that does not coarse-grain to the observed one. We want therefore express the conditional probability of each link given the observed coarse-grained network.

The conditional probability of having a link between $i_l, j_l$, i.e. $ a_{i_l j_l} = 1$, will depend on the existence of a link between the macro-nodes that contain $i_l$ and $j_l$, that is $i_{l+1}$ and $j_{l+1}$ respectively. We then have that 
\begin{align}
    \label{eq:cond_prob}
    \bar{p}_{i_l j_l} &:= P(a_{i_l j_l} = 1 | a_{i_{l+1} j_{l+1}} = 1) 
    = \frac{P(a_{i_l j_l} = 1 \cap a_{i_{l+1} j_{l+1}} = 1)}{P(a_{i_{l+1} j_{l+1}} = 1)} \\
    &=\frac{P(a_{i_l j_l} = 1)\overbrace{P\left(a_{i_{l+1} j_{l+1}} = 1|a_{i_l j_l} = 1\right)}^{=1}}{P(a_{i_{l+1} j_{l+1}} = 1)} = \frac{p_{i_l j_l}}{p_{i_{l+1} j_{l+1}}} \ .
\end{align}
It should be clear that this follows from the fact that we need a single link between nodes of each partition for the link to exist in the coarse-grained one. Of course, having an aggregate link does not imply that we must observe one for each pair that could compose it. We have summarised the various possibilities in table \ref{tb:cond_p}. It should be clear from the table that conditioning on the coarse-grained graph can greatly reduce the entropy of the model at finer partitions if the observed graph is very sparse. 

It is important to note here that the solution we have found in \eqref{eq:cond_prob} is not unique to this model. However what is unique to this model is that the denominator of the expression can be computed very efficiently. Indeed computing this for a general ERGM requires $M\times N$ operations to compute the probability of not having any connections between nodes in the two groups $i_{l+1}$ and $j_{l+1}$ where $M$ and $N$ are the number of nodes in $i_{l+1}$ and $j_{l+1}$ respectively. For the multi-scale model this translates to a complexity of $M + N + 1$ as we only have to perform the addition of the parameters in each group and one probability computation. 

\begin{table}[h!]
\centering
{\setlength{\tabcolsep}{1em}
\begin{tabular}{lcc}
    & \multicolumn{2}{c}{$a_{i_{l+1} j_{l+1}}$} \\\cline{2-3}
                                          & 0 & 1 \\ \hline
$P(a_{i_l j_l} = 0| a_{i_{l+1} j_{l+1}})$ & 1 & $1 - \bar{p}_{i_l j_l}$ \\
$P(a_{i_l j_l} = 1| a_{i_{l+1} j_{l+1}})$ & 0 & $\bar{p}_{i_l j_l}$      
\end{tabular}} \vspace{1em}
\caption{Conditional probability table for a fine-grained link $a_{i_l j_l}$ given the observed relevant coarse-grained edge $a_{i_{l+1} j_{l+1}}$.}\label{tb:cond_p}
\end{table}

In computing expected properties of the ensemble after conditioning we must be careful to consider the various cases as the probabilities might depend on the same macro edges. To this end we report here the various cases of expected values of pairs of edges. The difference between these cases is given by how many independent macro links the pair depends on. For a pair of edges $(a_{i_l j_l}, a_{r_l s_l})$ where $(i_{l+1}, j_{l+1}) \neq (r_{l+1}, s_{l+1})$, which implies $(i_l, j_l) \neq (r_l, s_l)$, we have that the conditional probabilities of all possible events are given by
\begin{align}
    P &(a_{i_l j_l} = w, a_{r_l s_l} = x| a_{i_{l+1} j_{l+1}} = y, a_{r_{l+1} s_{l+1}} = z) \\
    &= \frac{P(a_{i_l j_l} = w \cap a_{r_l s_l} = x \cap a_{i_{l+1} j_{l+1}} = y \cap a_{r_{l+1} s_{l+1}} = z)}{P(a_{i_{l+1} j_{l+1}} = y \cap a_{r_{l+1} s_{l+1}} = z)} \\
    &= \frac{P(a_{i_l j_l} = x \cap a_{i_{l+1} j_{l+1}} = y)}{P(a_{i_{l+1} j_{l+1}} = y)} \frac{P(a_{r_l s_l} = x \cap a_{r_{l+1} s_{l+1}} = z)}{P(a_{r_{l+1} s_{l+1}} = z)} \ . \label{eq:cond_pair}
\end{align}
Equation \eqref{eq:cond_pair} highlights that the conditional events $P(a_{i_l j_l} = w |a_{i_{l+1} j_{l+1}} = y)$ and $ P(a_{r_l s_l} = x| a_{r_{l+1} s_{l+1}} = z)$ are independent provided $(i_{l+1}, j_{l+1}) \neq (r_{l+1}, s_{l+1})$. We have of course two special cases: if $(i_{l+1}, j_{l+1}) = (r_{l+1}, s_{l+1})$ with $(i_l, j_l) \neq (r_l, s_l)$, and the case $(i_l, j_l) = (r_l, s_l)$. In the latter instance, it is clear that we have only one event conditional on its coarse-grained edge, so we are back in the simple conditional probability detailed in table \ref{tb:cond_p}. In the first special case however we have that there are two distinct edges that compose the same macro one. Here we have that $a_{i_{l+1} j_{l+1}} = a_{r_{l+1} s_{l+1}} = y$ giving us 
\begin{align}
    P &(a_{i_l j_l} = w, a_{r_l s_l} = x| a_{i_{l+1} j_{l+1}} = y) \\
    &= \frac{P(a_{i_l j_l} = w \cap a_{r_l s_l} = x \cap a_{i_{l+1} j_{l+1}} = y )}{P(a_{i_{l+1} j_{l+1}} = y)} \\
    &= \frac{P(a_{i_l j_l} = w \cap a_{r_l s_l} = x) P(a_{i_{l+1} j_{l+1}} = y | a_{i_l j_l} = w, a_{r_l s_l} = x)}{P(a_{i_{l+1} j_{l+1}} = y)} \\
    &= \frac{P(a_{i_l j_l} = w) P(a_{r_l s_l} = x) P(a_{i_{l+1} j_{l+1}} = y | a_{i_l j_l} = w, a_{r_l s_l} = x)}{P(a_{i_{l+1} j_{l+1}} = y)}  \ . \label{eq:cond_pair_spec}
\end{align}
Equation \eqref{eq:cond_pair_spec} has three important cases: if at least one of $w$ or $x$ is one then $a_{i_{l+1} j_{l+1}} = 1$ with probability one and as such we have that this conditional probability will be either one or zero depending on $y$. The last case is if both $w$ or $x$ are zero. In this scenario the value of $a_{i_{l+1} j_{l+1}}$ is not certain but will depend on all the other links that compose it. We now have that 
\begin{equation}
    P(a_{i_{l+1} j_{l+1}} = 0 | a_{i_l j_l} = 0, a_{r_l s_l} = 0) = {\prod_{m \in i_{l+1}} \prod_{n \in j_{l+1}}}_{(m, n) \neq (i, j), (r, s)} (1 - p_{m_l n_l})
\end{equation}
and
\begin{equation}
    P(a_{i_{l+1} j_{l+1}} = 1 | a_{i_l j_l} = 0, a_{r_l s_l} = 0) = 1 - {\prod_{m \in i_{l+1}} \prod_{n \in j_{l+1}}}_{(m, n) \neq (i, j), (r, s)} (1 - p_{m_l n_l}) \ .
\end{equation}
We can now summarise all possible cases depending on the values of $w$, $x$, $y$ and $z$ in table \ref{tb:cond_pairs}. 

\begin{table}[ht]
\centering
{\setlength{\tabcolsep}{0.5em}
\begin{tabular}{lcccc}
    & \multicolumn{4}{c}{$a_{i_{l+1} j_{l+1}}, a_{r_{l+1} s_{l+1}}$} \\ \cline{2-5}
    & 0,0 & 0,1 & 1,0 & 1,1 \\ \hline \hline
If $(i_{l+1}, j_{l+1}) \neq (r_{l+1}, s_{l+1})$ & & & & \\
$P(a_{i_l j_l} = 0, a_{r_l s_l} = 0| a_{i_{l+1} j_{l+1}}, a_{r_{l+1} s_{l+1}})$ 
    & 1 & $1 - \bar{p}_{r_l s_l}$ & $1 - \bar{p}_{i_l j_l}$ & $(1 - \bar{p}_{i_l j_l})(1 - \bar{p}_{r_l s_l})$ \\
$P(a_{i_l j_l} = 0, a_{r_l s_l} = 1| a_{i_{l+1} j_{l+1}}, a_{r_{l+1} s_{l+1}})$ 
    & 0 & $\bar{p}_{r_l s_l}$ & 0 &  $(1 - \bar{p}_{i_l j_l})\bar{p}_{r_l s_l}$ \\
$P(a_{i_l j_l} = 1, a_{r_l s_l} = 0| a_{i_{l+1} j_{l+1}}, a_{r_{l+1} s_{l+1}})$ 
    & 0 & 0 & $\bar{p}_{i_l j_l}$ & $\bar{p}_{i_l j_l}(1 - \bar{p}_{r_l s_l})$ \\
$P(a_{i_l j_l} = 1, a_{r_l s_l} = 1| a_{i_{l+1} j_{l+1}}, a_{r_{l+1} s_{l+1}})$ 
    & 0 & 0 & 0 & $\bar{p}_{i_l j_l}\bar{p}_{r_l s_l}$ \\ \hline
If $(i_{l+1}, j_{l+1}) = (r_{l+1}, s_{l+1}), a_{i_l j_l} \neq a_{r_l s_l}$ & & & & \\ 
$P(a_{i_l j_l} = 0, a_{r_l s_l} = 0| a_{i_{l+1} j_{l+1}})$ 
    & 1 & & & $\frac{(1 - p_{i_l j_l})(1 - p_{r_l s_l})}{p_{i_{l+1} j_{l+1}}} - \frac{1 - p_{i_{l+1} j_{l+1}}}{p_{i_{l+1} j_{l+1}}}$ \\
$P(a_{i_l j_l} = 0, a_{r_l s_l} = 1| a_{i_{l+1} j_{l+1}})$ 
    & 0 & & & $\frac{(1 - p_{i_l j_l})p_{r_l s_l}}{p_{i_{l+1} j_{l+1}}}$ \\
$P(a_{i_l j_l} = 1, a_{r_l s_l} = 0| a_{i_{l+1} j_{l+1}})$ 
    & 0 & & & $\frac{p_{i_l j_l}(1 - p_{r_l s_l})}{p_{i_{l+1} j_{l+1}}}$ \\
$P(a_{i_l j_l} = 1, a_{r_l s_l} = 1| a_{i_{l+1} j_{l+1}})$ 
    & 0 & & & $\frac{p_{i_l j_l}p_{r_l s_l}}{p_{i_{l+1} j_{l+1}}}$ \\ \hline
If $ a_{i_l j_l} = a_{r_l s_l}$ & & & & \\ 
$P(a_{i_l j_l} = 0| a_{i_{l+1} j_{l+1}})$ 
    & 1 & & & $1 - \bar{p}_{i_l j_l}$ \\
$P(a_{i_l j_l} = 1| a_{i_{l+1} j_{l+1}})$ 
    & 0 & & & $\bar{p}_{i_l j_l}$ 
\end{tabular}} \vspace{1em}
\caption{Conditional probability table for the fine-grained links $a_{i_l j_l}$ and $a_{r_l s_l}$ given the observed relevant coarse-grained edges $a_{i_{l+1} j_{l+1}}$ and $a_{r_{l+1} s_{l+1}}$.}\label{tb:cond_pairs}
\end{table}

\newpage
Based on the tables above we can now compute the conditional expected values of the degree sequence and average nearest neighbour degree. For the out-degree we simply have that 
\begin{equation}
    \bar{k}_{i_l}^{\text{out}} \coloneqq E\left(\left. k_{i_l}^{\text{out}} \right| \bm{A}^{(l+1)} \right) = E\left(\left. \sum_{j_l} a_{i_l j_l} \right| \bm{A}^{(l+1)} \right) = \sum_{j_l} E\left(\left. a_{i_l j_l} \right| \bm{A}^{(l+1)} \right) = \sum_{j_l} \bar{p}_{i_l j_l}
\end{equation}
and similarly for the in-degree we have $ \bar{k}_{i_l}^{\text{in}} \coloneqq E\left(\left. k_{i_l}^{\text{in}} \right| \bm{A}^{(l+1)} \right) = \sum_{j_l} \bar{p}_{j_l i_l}$ . 

We report here the derivation for the average out-nearest neighbour out-degree:
\begin{align}
    E &\left(\left. k_{nn_{\text{out}}, i_l}^{\text{out}} \right| \bm{A}^{(l+1)} \right) 
    \coloneqq E\left(\left. \frac{1}{k_{i_l}^\text{out}}\sum_{j_l \in nn_{\text{out}, i_l}} k_{j_l}^\text{out} \right| \bm{A}^{(l+1)} \right) \\  \label{eq:knn_approx}
    &\approx \frac{1}{E\left(\left. k_{i_l}^\text{out}\right| \bm{A}^{(l+1)} \right)}\sum_{j_l \in nn_{\text{out}, i_l}} E\left(\left. k_{j_l}^\text{out} \right| \bm{A}^{(l+1)} \right) \\
    &= \frac{1} {\bar{k}_{nn_{\text{out}}, i_l}^{\text{out}}} \sum_{j_l \neq i_l} E\left(\left. a_{i_l j_l} k_{j_l}^\text{out} \right| \bm{A}^{(l+1)} \right)
    = \frac{1} {\bar{k}_{nn_{\text{out}}, i_l}^{\text{out}}} \sum_{j_l \neq i_l} \sum_{k_l \neq j_l} E\left(\left. a_{i_l j_l} a_{j_l k_l} \right| \bm{A}^{(l+1)} \right) \\
    &= \frac{1} {\bar{k}_{nn_{\text{out}}, i_l}^{\text{out}}} \left[ \sum_{j_l \neq i_l } \sum_{\substack{k_l \neq j_l \\ j_l \notin i_{l+1} \lor k_l \notin i_{l+1}}} \bar{p}_{i_l j_l} \bar{p}_{j_l k_l} + \sum_{\substack{j_l \neq i_l \\ j_l \in i_{l+1}}} \sum_{\substack{k_l \neq j_l \\ k_l \in i_{l+1}}} \frac{p_{i_l j_l} p_{j_l k_l}}{p_{i_{l+1} i_{l+1}}} \right]\ . \label{eq:knn_out_out}
\end{align}

\begin{figure}[t]
    \centering
    \includegraphics[width=0.6\linewidth]{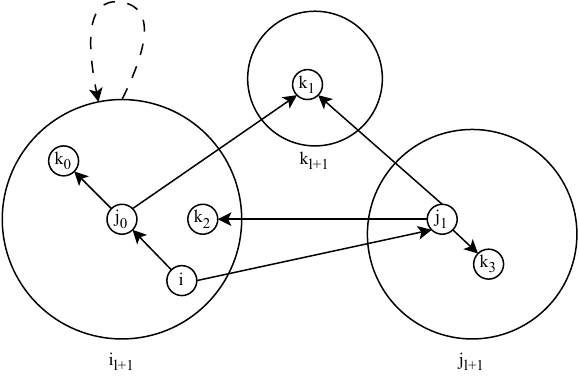}
    \caption{Illustration of possible dependencies in computing the conditional average nearest neighbours degree.}
    \label{fig:knn_diag}
\end{figure}

Note that in equation \eqref{eq:knn_approx} we have used the first order approximation of the Taylor expansion of $E\left[ \frac{X}{Y} \right]$. We also note that as we have summarized in table \ref{tb:cond_pairs} we have three possible cases for the conditional probability of $a_{i_l j_l}$ and $a_{j_l k_l}$: if the connection is the same, if the links are different but belong to the same macro edge $a_{i_{l+1} j_{l+1}}$, or finally if they belong to different ones. In figure \ref{fig:knn_diag} we have highlighted all the possible cases for the out-out case. We note that the connection being the same is ruled out by construction with the directed average nearest neighbour degree as $a_{i_l j_l} \neq a_{j_l k_l} \forall k_l$ if $j_l \neq i_l$ but self-loops are excluded from the computation. We further note that the macro edge being the same is only possible in two scenarios: first if $i_l$, $j_l$ and $k_l$ all belong to the same coarse-grained node such that $a_{i_l j_l}$ and $a_{j_l k_l}$ all belong to the self-loop $a_{i_{l+1} i_{l+1}}$; the second case happens only for the in-out and out-in average nearest neighbour degrees. This can be seen from the diagram in figure \ref{fig:knn_diag} by letting the connection go from $k_2$ to $j_1$, then we would have that $a_{i j_1}$ and $a_{k_2 j_1}$ both belong to $a_{i_{l+1} i_{l+1}}$. These considerations are what give us the two distinct sums in equation \eqref{eq:knn_out_out}.

\subsection{Additional figures}
We report for completeness additional figures generated for this analysis.

\begin{figure}
    \centering
    \subfloat[\label{fig:d_levels_b}]{%
    \includegraphics[width=0.49\linewidth]{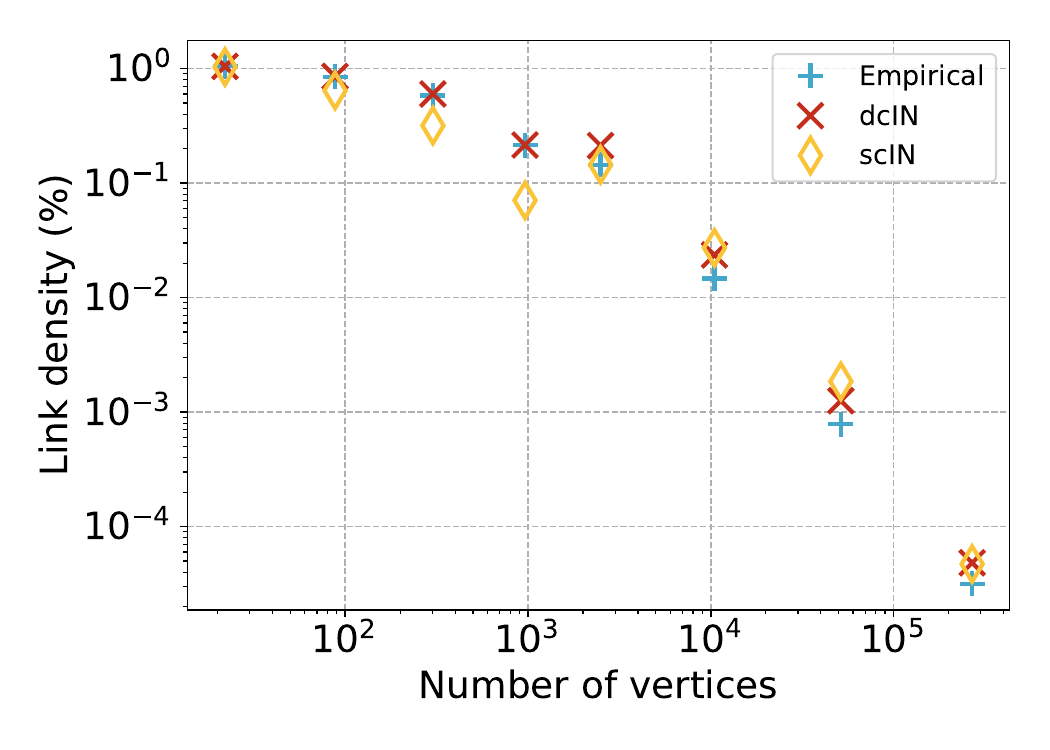}
    }\hfil
    \subfloat[\label{fig:sc3_params_b}]{%
    \includegraphics[width=0.49\linewidth]{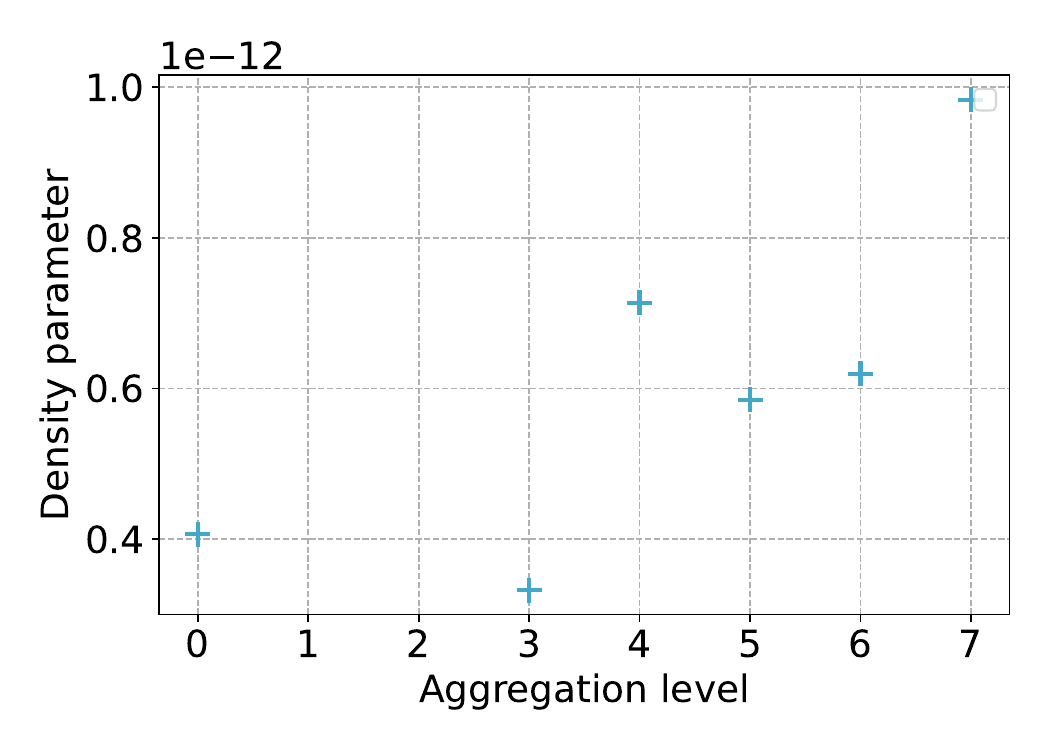}
    }\hfil
    \subfloat[\label{fig:sbi_digits_b}]{%
    \includegraphics[width=0.49\textwidth]{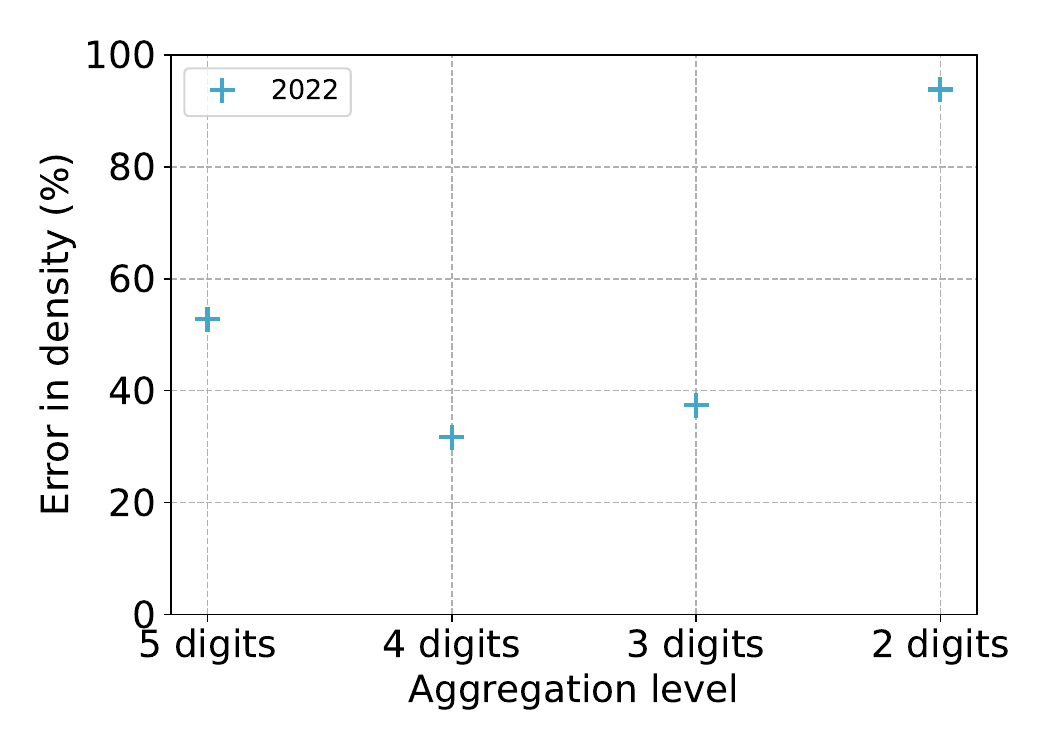}
    }
    \caption{We report here the results for the ING dataset in the case of the aggregation scenario. In panel (a) we observe a similar result for expected density under coarse and fine graining. In panel (b) we instead plot the value of the estimated parameter at the various aggregation levels and in panel (c) the percentage error in density as a function of NACE digits. We note that differently from the ABN results the error here is more consistent across scales. }
\end{figure} 

\begin{figure}
    \centering
    \subfloat[Level 0]{%
    \includegraphics[width=0.49\textwidth]{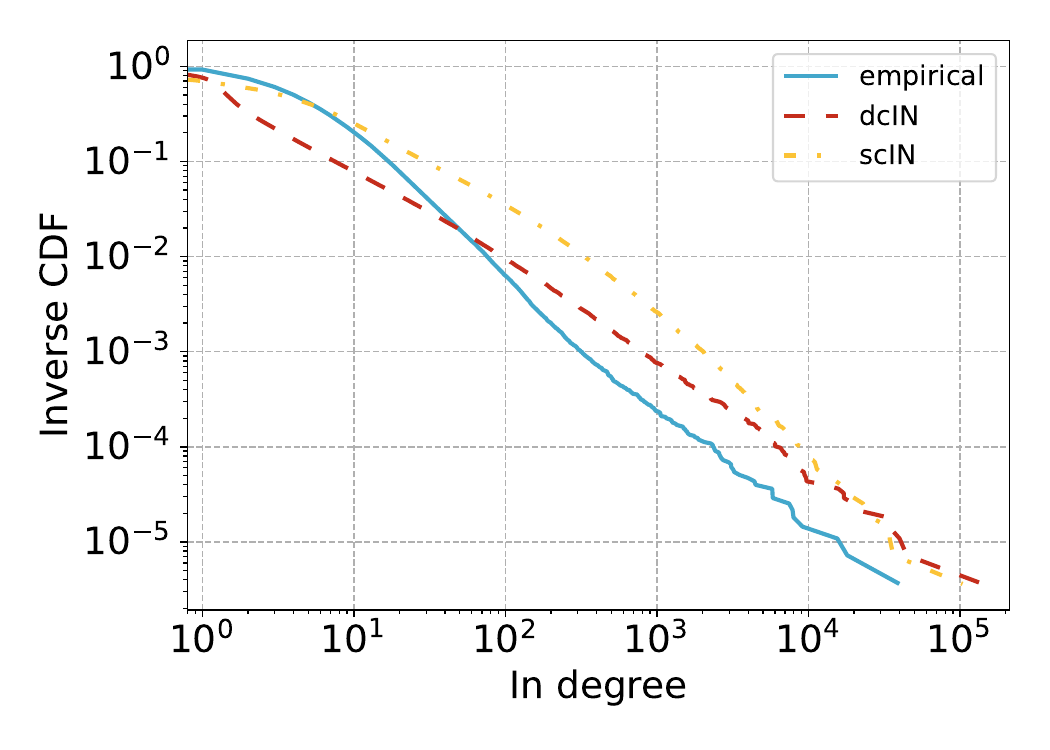}
    }\hfil
    \subfloat[Level 0]{%
    \includegraphics[width=0.49\textwidth]{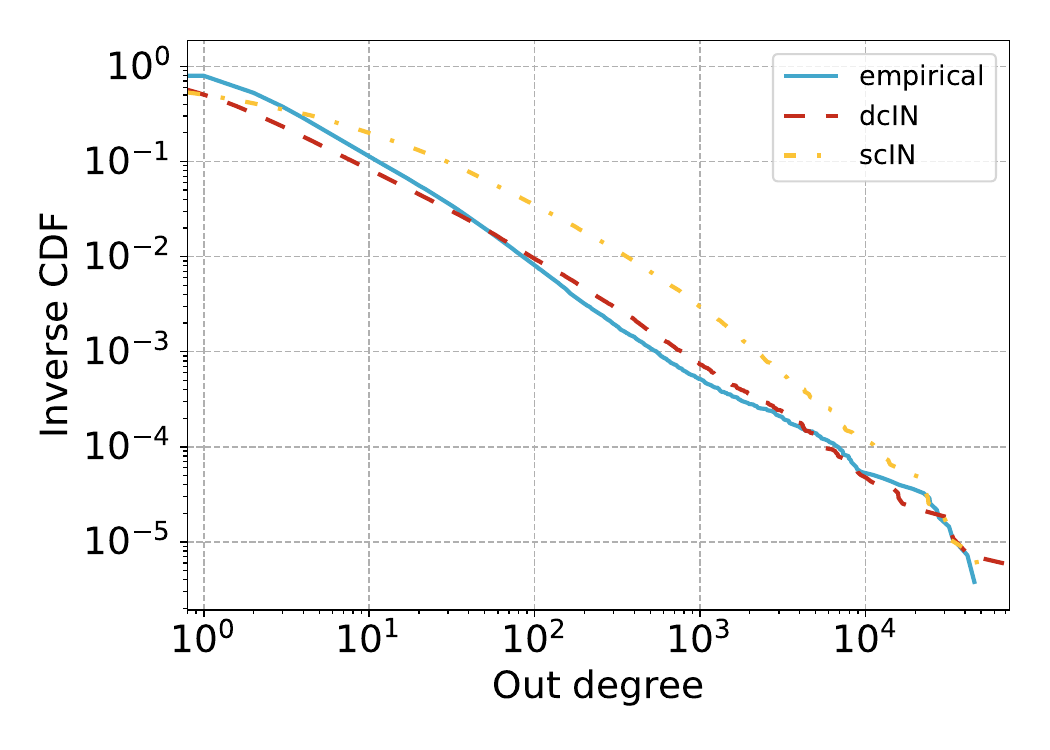}%
    }\vfil
    \subfloat[Level 2]{%
    \includegraphics[width=0.49\textwidth]{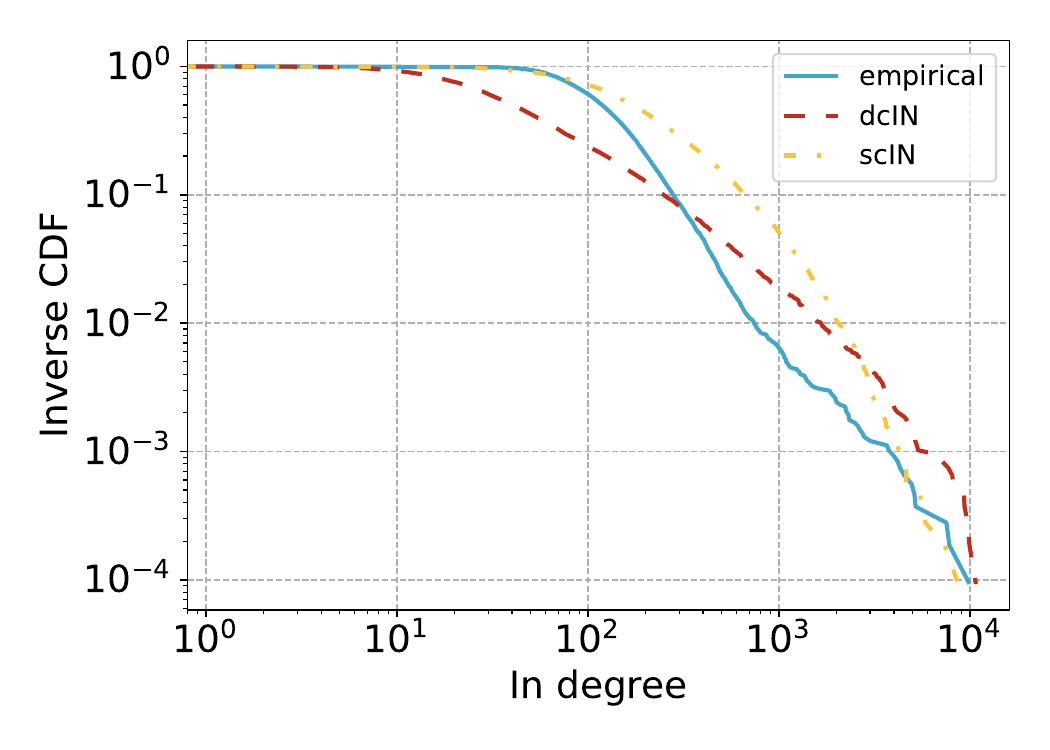}
    }\hfil
    \subfloat[Level 2]{%
    \includegraphics[width=0.49\textwidth]{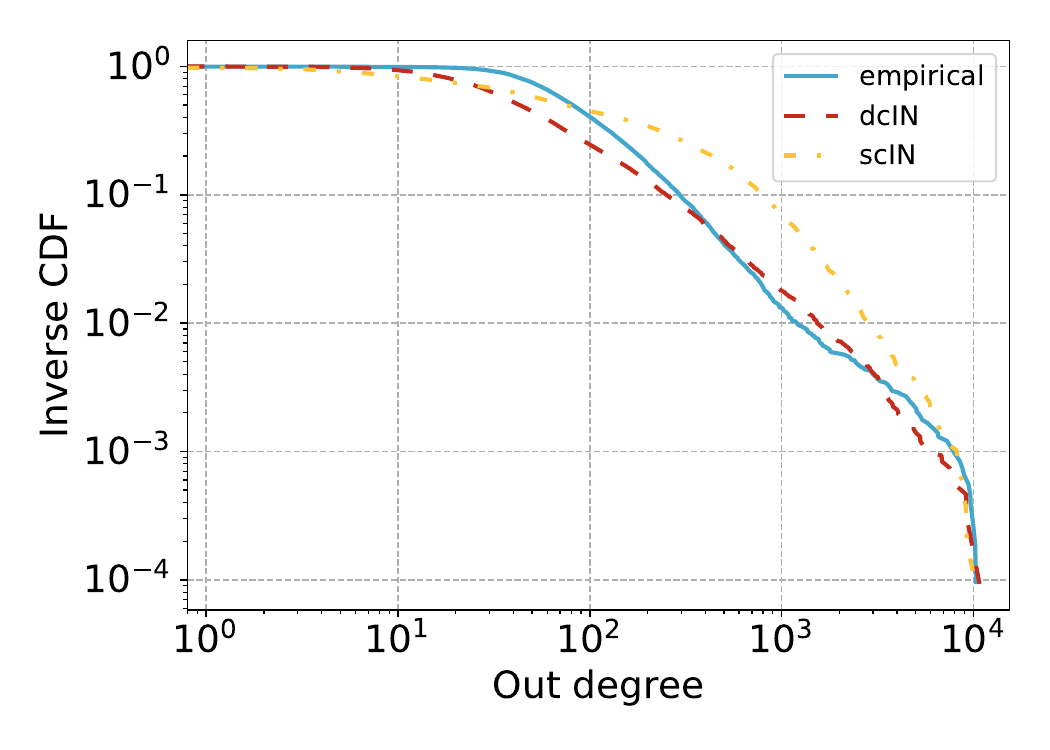}%
    }\vfil
    \subfloat[Level 4]{%
    \includegraphics[width=0.49\textwidth]{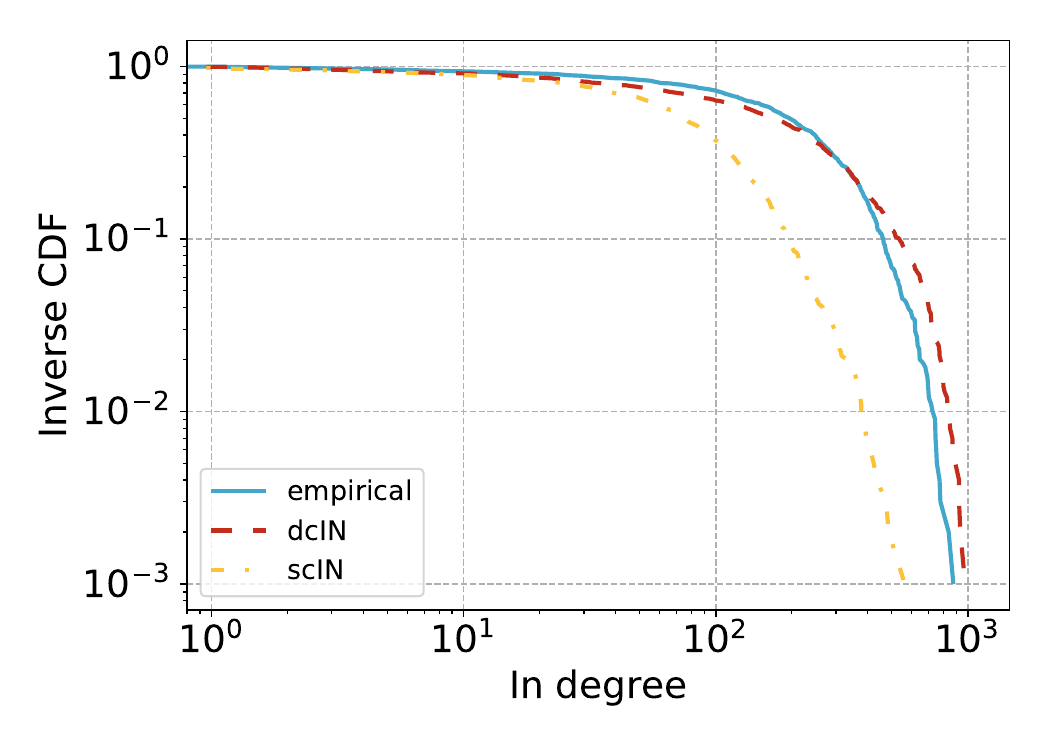}
    }\hfil
    \subfloat[Level 4]{%
    \includegraphics[width=0.49\textwidth]{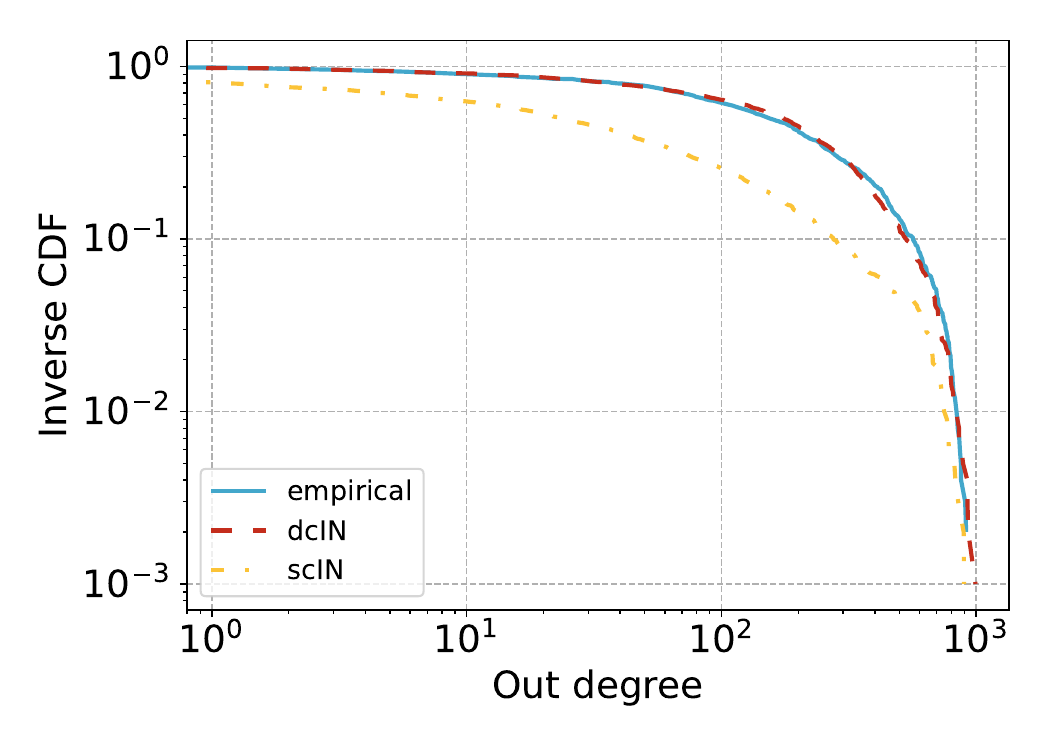}%
    }\vfil
    \caption{In and out degree distribution at different aggregation levels compared with the ensemble average for the multi-scale models.}
\end{figure} 

\begin{figure}
    \centering
    \subfloat[Level 0]{%
    \includegraphics[width=0.49\textwidth]{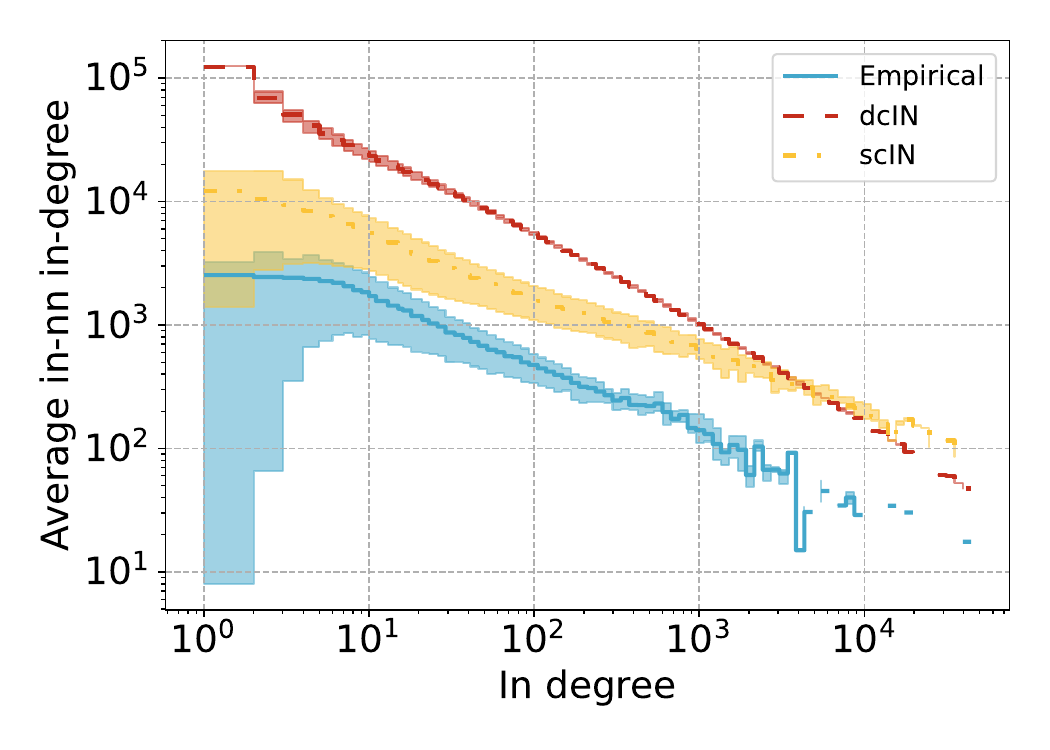}
    }\hfil
    \subfloat[Level 0]{%
    \includegraphics[width=0.49\textwidth]{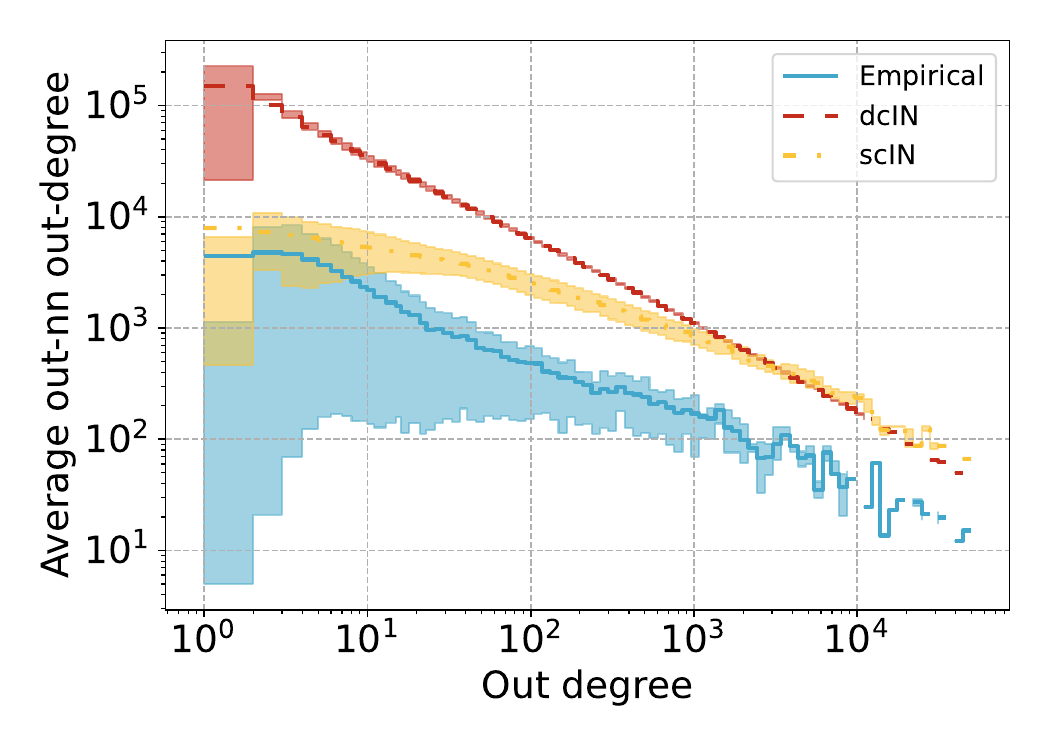}%
    }\vfil
    \subfloat[Level 2]{%
    \includegraphics[width=0.49\textwidth]{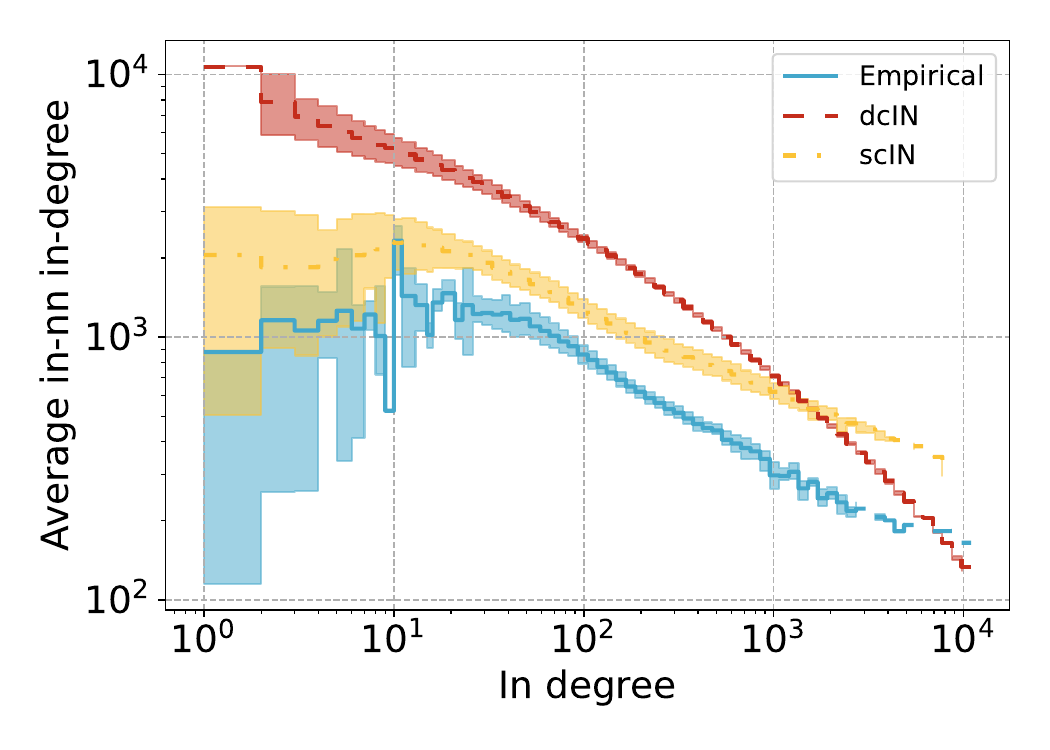}
    }\hfil
    \subfloat[Level 2]{%
    \includegraphics[width=0.49\textwidth]{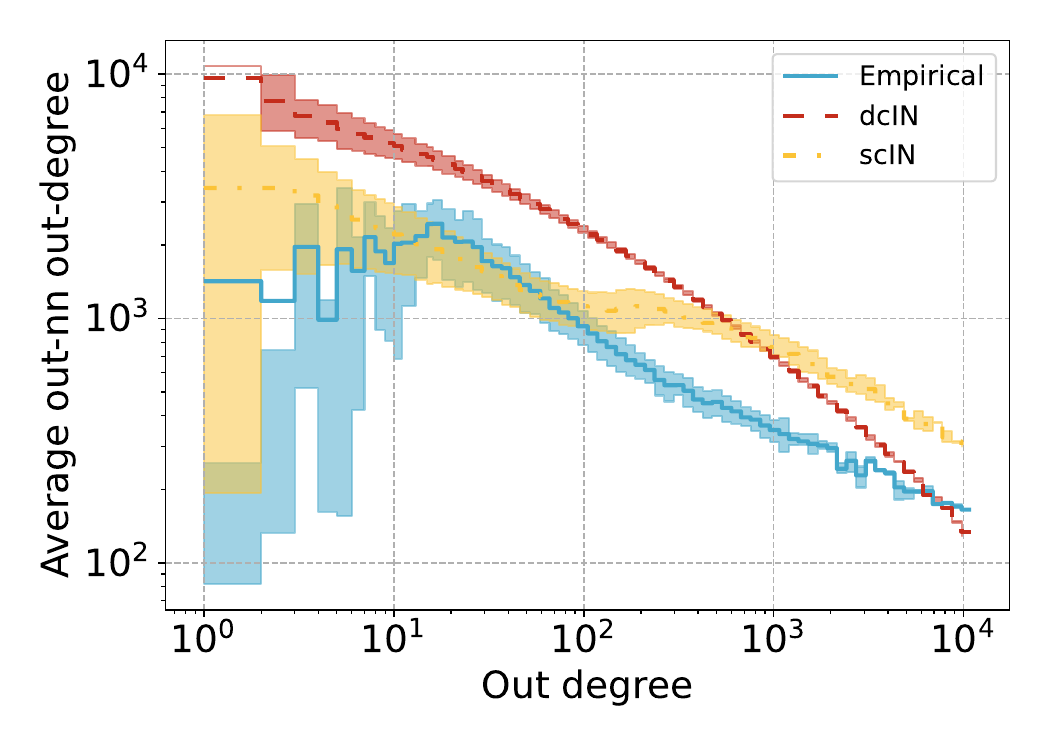}%
    }\vfil
    \subfloat[Level 4]{%
    \includegraphics[width=0.49\textwidth]{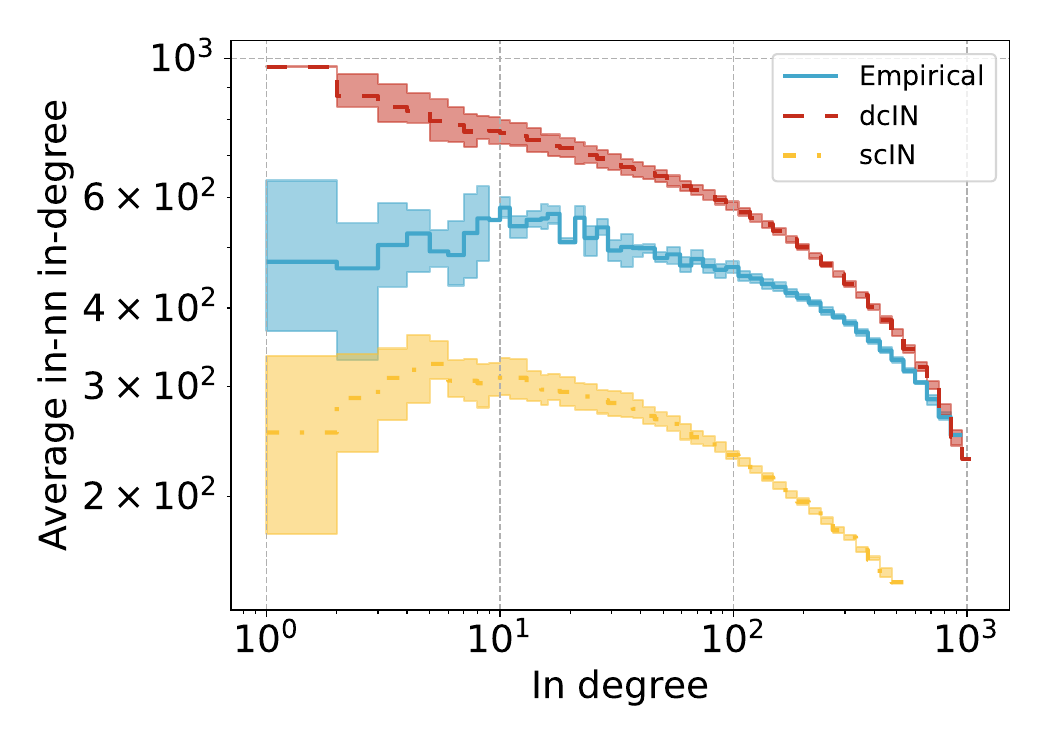}
    }\hfil
    \subfloat[Level 4]{%
    \includegraphics[width=0.49\textwidth]{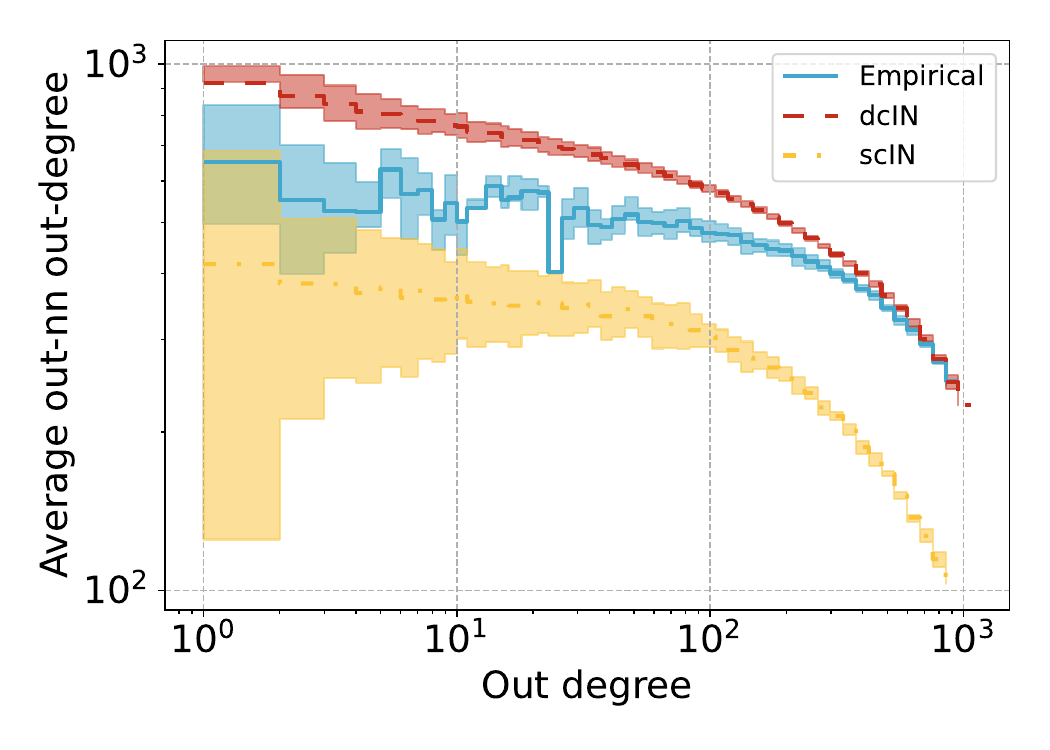}%
    }\vfil
    \caption{Average nearest neighbour degree at different aggregation levels compared with the ensemble average for the multi-scale models. The shaded area represents the interquartile range.}
\end{figure} 

\end{document}